\documentclass[%
 reprint,
superscriptaddress,
 amsmath,amssymb,
 aps, 
prb,
floatfix,
]{revtex4-2}
\usepackage{amsmath,amssymb}
\usepackage{graphicx}
\usepackage{dcolumn}
\usepackage{bm}
\usepackage{multirow}
\usepackage{float}
\usepackage{hhline}
\usepackage{color}
\usepackage{afterpage}
\usepackage{tabularx}
\usepackage{comment}
\usepackage{float}
\usepackage{mathtools}
\usepackage[utf8]{inputenc}
\usepackage[T1]{fontenc}
\usepackage{mathptmx}
\usepackage[normalem]{ulem}
\usepackage{physics}
\usepackage[caption=false]{subfig}

\newcommand{\RomanNumeralCaps}[1]{\MakeUppercase{\romannumeral #1}}
\newcommand{\Sp}{0.3cm}
\newcommand{\drhodt}[1]{\frac{\mathrm{d} #1}{\mathrm{d} \tau}}

\newcolumntype{C}[1]{>{\hsize=#1\hsize\centering\arraybackslash}X}%
\newcolumntype{L}[1]{>{\hsize=#1\hsize\raggedright\arraybackslash}X}%

\graphicspath{{Figures/}}

\begin{document}

\preprint{AIP/123-QED}

\title{Many-Body Coherence in Quantum Transport}

\author{Ching-Chi Hang}
\affiliation{Institute of Atomic and Molecular Sciences, Academia Sinica, Taipei, Taiwan}
\author{Liang-Yan Hsu}
\email{lyhsu@gate.sinica.edu.tw}
\affiliation{Institute of Atomic and Molecular Sciences, Academia Sinica, Taipei, Taiwan}
\affiliation{Department of Chemistry, National Taiwan University, Taipei, Taiwan}
\affiliation{National Center for Theoretical Sciences, Taipei, Taiwan}

\begin{abstract}
In this study, we propose the concept of harnessing quantum coherence to control electron transport in a many-body system. Combining an open quantum system technique based on Hubbard operators, we show that many-body coherence can eliminate the well-known Coulomb staircase and cause strong negative differential resistance. To explore the mechanism, we analytically derive the current-coherence relationship in the zero electron-phonon coupling limit. Furthermore, by incorporating a gate field, we demonstrate the possibility of constructing a coherence-controlled transistor. This development opens up a new direction for exploring quantum electronic devices based on many-body coherence. 
\end{abstract}

\maketitle

\section{Introduction} 
\begin{figure}[b!]
    \centering
    \includegraphics[width=0.5\textwidth]{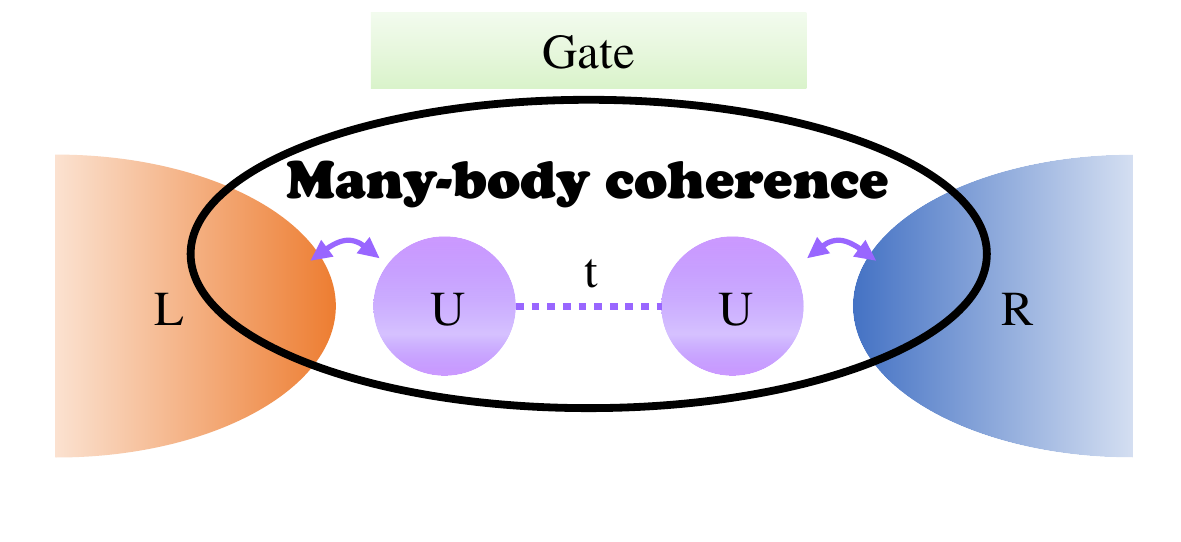} 
    \caption{(a) Illustration of a quantum electronic device in a many-body system. The system coupled with one gate and two leads L and R contains on-site Coulomb repulsion $U$, and intersite electron hopping $t$. \label{fig:Schematic}}
\end{figure}
Quantum coherence is a fundamental concept in quantum mechanics that sets it apart from classical physics. The unique properties of quantum coherence have been applied in a diverse range of fields across various disciplines. For instance, quantum coherence has been utilized to enhance the energy transfer efficiency in quantum biology~\cite{Engel2007,Panitchayangkoon2010,Scholes2017,Bredas2017} and the performance of nanoscale heat engines in quantum thermodynamics~\cite{Scully2011,Samuelsson2017,Saryal2021,Tajima2021,Kamimura2022}. Moreover, quantum coherence can be exploited to store and transfer information for quantum communication~\cite{Wu2018,Nguyen2019,Bhaskar2020,Zhai2022}. In nanoelectronics, the importance of quantum coherence is manifested in the interference of an single electron passing through a junction with multiple tunneling pathways, e.g., a quantum interference transistor~\cite{Guedon2012,Ballmann2012,Hsu2012,Li2019,Bai2019,Greenwald2021}. Despite extensive studies on quantum interference in quantum transport, how to directly connect quantum coherence and transport properties, particularly a current-coherence relationship in many-body systems, remains an open question. 
\newline \indent
Many-body effects in quantum transport have attracted considerable attention due to their critical significance in open quantum systems and their potential applications in nanoelectronics~\cite{Galperin2007,Mitchell2017,Yu2017,Fu2018, Kimura2019, Fetherolf2020,Shein-Lumbroso2022}. Numerous intriguing many-body quantum transport phenomena, including  Coulomb blockade~\cite{Park2002,Brotons-Gisbert2019}, Kondo resonance~\cite{Liang2002,Kurzmann2021}, Franck-Condon blockade~\cite{Koch2005,Burzuri2014,Du2021}, and current hysteresis~\cite{Galperin2005,Schwarz2016}, have been extensively explored in semiconductor nanostructures, 2D materials, and single-molecule junctions. However, the concept of many-body coherence, which refers to quantum coherence between two many-body states, has not received enough attention in the field of quantum transport. 
\newline \indent
In this paper, inspired by the Bloch-Redfield formalism~\cite{Brandes2005,Landi2022,Dani2022,Segal2000,Hsu2014,Agarwalla2015,Anto-Sztrikacs2023}, we introduce quantum coherence from a Redfield-type fermionic quantum master equation and study quantum transport in a minimal model that incorporates many-body effects such as electron-electron interactions. 
Based on the minimal model, we aim to clarify the role of many-body coherence in quantum transport, thus shedding light on how to harness many-body coherence to design quantum electronic devices.
\section{Model Hamiltonian}
To demonstrate the effect of many-body coherence on quantum transport, we consider a quantum electronic device shown in Fig.~\ref{fig:Schematic}. The device is described by the total Hamiltonian
\begin{equation}
    \hat{H}=\hat{H}_{\mathrm{sys}}+\hat{H}_{\mathrm{lead}}+\hat{H}_{\mathrm{sys-lead}}+\hat{H}_{\mathrm{gate}},
\end{equation}
which is composed of the system Hamiltonian $\hat{H}_{\mathrm{sys}}$, the lead Hamiltonian $\hat{H}_{\mathrm{lead}}$, the system-lead coupling term $\hat{H}_{\mathrm{sys-lead}}$, and the gate Hamiltonian $\hat{H}_{\mathrm{gate}}$. Furthermore, to simplify the complexity of a many-body system while retaining electron-electron interactions, we consider the two-site Hubbard model to be the system, including on-site energy $\varepsilon$, on-site Coulomb repulsion $U$, and intersite electron hopping $t$. The system Hamiltonian has the form
\begin{equation}
    \hat{H}_{\mathrm{sys}} = \varepsilon \sum_{i,\sigma} \hat{c}_{i\sigma}^{\dagger} \hat{c}_{i\sigma} + U \sum_{i} \hat{c}_{i\uparrow}^{\dagger} \hat{c}_{i\uparrow}\hat{c}_{i\downarrow}^{\dagger} \hat{c}_{i\downarrow} - t \sum_{\sigma} (\hat{c}_{2\sigma}^{\dagger}\hat{c}_{1\sigma} + \mathrm{H.c.}),
\end{equation}
where $\hat{c}_{i\sigma}^{\dagger}$ ($\hat{c}_{i\sigma}$) is the fermionic operator which creates (annihilates) an electron on site $i=1,2$ with spin $\sigma = \uparrow,\downarrow$. The model can accommodate at most 4 electrons and generate 16 different many-body electronic states in total~\cite{Li2014}. To properly describe many-body states, we denote the many-body states of the system as $\ket{N_a,a}$ with energy $\varepsilon_a$ as shown in Table~\ref{tab:state}, where $N_a$ represents the number of electrons of state $a$.
According to the previous study~\cite{Koole2016}, we believe that a two-site system, such as thiolated arylethynylene with 9,10-dihydroanthracene core (AH), is experimentally feasible for the demonstration of the effect of many-body coherence on quantum transport. 
\newline \indent
The two leads and the gate are modeled as follows. For the gate, we model its Hamiltonian as 
\begin{equation}
    \hat{H}_{\mathrm{gate}}=-\mathrm{e}V_{\mathrm{g}} \sum_{i,\sigma} \hat{c}_{i\sigma}^{\dagger} \hat{c}_{i\sigma},
\end{equation}
where the gate voltage $V_{\mathrm{g}}$ shifts the on-site energy $\varepsilon$ by $-\mathrm{e}V_{\mathrm{g}}$. The two leads are described by a noninteracting electron gas model, 
\begin{equation}
    \hat{H}_{\mathrm{lead}} = \sum_{l,k,\sigma} \xi_{k\sigma} \hat{d}_{lk\sigma}^{\dagger} \hat{d}_{lk\sigma},
\end{equation}
where $\hat{d}_{lk\sigma}^{\dagger}(\hat{d}_{lk\sigma})$ creates (annihilates) an electron in the state $\ket{lk\sigma}$ with energy $\xi_{lk\sigma}$ in the lead $l$, and $l = \mathrm{L}$ and $\mathrm{R}$ represents the left and the right leads. Assuming that the electrons in the leads stay at equilibrium, we express the average occupation number as $\ev*{\hat{d}_{lk\sigma}^{\dagger} \hat{d}_{l'k'\sigma'}} = \delta_{l,l'}\delta_{k,k'}\delta_{\sigma,\sigma'} f_l(\xi_{k\sigma})$, where $f_l(\xi_{k\sigma}) = (1+e^{(\xi_{k\sigma}-\mu_l)/k_{\mathrm{B}}T})^{-1}$ is the Fermi function of lead $l$ with chemical potential $\mu_l$ at temperature $T$. In this work, we consider the symmetric bias condition $\mu_l = \mu_0 + \zeta_l \mathrm{e}V_\mathrm{sd}/2$ with $\zeta_\mathrm{L} = 1$ and $\zeta_\mathrm{R} = -1$, where $V_\mathrm{sd}$ is the bias voltage, and $\mu_0$ is the equilibrium chemical potential for the electrodes.
The system-lead coupling is modeled as 
\begin{equation}
    \hat{H}_{\mathrm{sys-lead}} = \sum_{k,\sigma} (T_{\mathrm{L}k,1}\hat{c}_{1\sigma}^{\dagger}\hat{d}_{\mathrm{L}k\sigma} + T_{\mathrm{R}k,2} \hat{c}_{2\sigma}^{\dagger} \hat{d}_{\mathrm{R}k\sigma} + \mathrm{H.c.}),
\end{equation}
where we assume the left (right) lead is only coupled to the first (second) site of the system.  Furthermore, we specify the transitions between many-body states using Hubbard operators $\hat{X}^{a,b} \equiv \ketbra{N_a,a}{N_b,b}$; see Appendix~\ref{sec:HO_coeff} for more details. The advantage of using Hubbard operators is to provide a convenient way to describe many-body state transitions and incorporate characteristics of fermions in the coefficient of each operator~\cite{Esposito2009,Li2014}. As a result, we rewrite the coupling Hamiltonian as 
\begin{equation}
    \hat{H}_{\mathrm{sys-lead}} = \sum_{ab,k,\sigma} ( V_{Lk\sigma,ab}^* \hat{X}^{b,a} \hat{d}_{\mathrm{L}k\sigma} + V_{Rk\sigma,ab}^* \hat{X}^{b,a} \hat{d}_{\mathrm{R}k\sigma} + \mathrm{H.c.} )
\end{equation}
based on the Hubbard operator techniques, where the transformed coupling becomes $V_{lk\sigma,ab} = T_{lk,i}^* \cdot \mel{N_a,a}{\hat{c}_{i\sigma}}{N_b,b}$. The index $i$ is neglected in $V_{lk\sigma,ab}$ because $i$ is uniquely determined by $l$, i.e., $i=1~(2)$ when $l=\mathrm{L}~(\mathrm{R})$. 
Here we do not consider the effect of the external potential exerted by the bias, i.e., the on-site energy $\varepsilon$ does not vary with the source-drain voltage $V_\mathrm{sd}$. This effect can lead to level renormalization and slightly modify the pattern of Coulomb staircase~\cite{Wunsch2005,Luo2011}. 
\def\arraystretch{1.3}%
\begin{table}[ht]
    \centering
    \begin{ruledtabular}
    \begin{tabularx}{1\linewidth}{p{2cm}<{\centering} C{0.3} L{0.3}}
    Hilbert space & Energy $\varepsilon_a$ & Eigenstate $\ket*{N_a,a}$ \\
    \colrule
    Zero-electron & 0 & $\ket{0,S^0}$ \\
    One-electron & $\varepsilon-t$ & $\ket*{1,D^1_{+,\uparrow}}$, $\ket*{1,D^1_{+,\downarrow}}$ \\
                 & $\varepsilon+t$ & $\ket*{1,D^1_{-,\uparrow}}$, $\ket*{1,D^1_{-,\downarrow}}$ \\
    Two-electron & $2\varepsilon-(x-U)/2$ & $\ket*{2,S^2_+}$ \\
                 & $2\varepsilon$ & $\ket{2,T^2_0}$, $\ket*{2,T^2_{+1}}$, $\ket*{2,T^2_{-1}}$ \\
                 & $2\varepsilon+U$ & $\ket*{2,S^2_{\mathrm{CS}}}$\\
                 & $2\varepsilon+(x+U)/2$ & $\ket*{2,S^2_-}$\\
    Three-electron & $3\varepsilon+U-t$ & $\ket*{3,D^3_{-,\uparrow}}$, $\ket*{3,D^3_{-,\downarrow}}$ \\
                   & $3\varepsilon+U+t$ & $\ket*{3,D^3_{+,\uparrow}}$, $\ket*{3,D^3_{+,\downarrow}}$ \\
    Four-electron & $4\varepsilon+2U$ & $\ket*{4,S^4}$ \\            
    \end{tabularx}
    \end{ruledtabular}
    \caption{The 16 eigenstates of the system Hamiltonian and their corresponding energies~\cite{Thomas2021}, $x \equiv \sqrt{U^2+16t^2}$.}
    \label{tab:state}
\end{table}
\section{Quantum master equation analysis}
To incorporate the effect of many-body coherence into quantum transport, instead of using the Pauli master equation (PME) or the Lindblad quantum master equation, we adopt the Redfield formalism, which has been used extensively to describe electronic bath in the electrodes~\cite{Brandes2005,Dani2022,Landi2022} or phonon effects on quantum transport~\cite{Segal2000,Hsu2014,Agarwalla2015,Anto-Sztrikacs2023}. 
We start from the quantum Liouville equation, treat the two leads $\hat{H}_{\mathrm{lead}}$ as bath, make the Born-Markov approximation, and finally derive a Redfield-type fermionic quantum master equation based on Hubbard operators. A detailed derivation and discussion may be found in Appendix~\ref{sec:deriv_FR} and the final result is as follows,
\begin{align}
\label{eq:RME}
    \frac{\mathrm{d} \hat{\rho}_{\mathrm{sys}}(t)}{\mathrm{d} t} = -\frac{i}{\hbar} [\hat{H}_{\mathrm{sys}},\hat{\rho}_{\mathrm{sys}}(t)] + \mathcal{R}_{\mathrm{lead}} \hat{\rho}_{\mathrm{sys}}(t), 
\end{align}
where $\hat{\rho}_{\mathrm{sys}}(t)$ is the electronic density matrix, $\mathcal{R}_{\mathrm{lead}}$ is the lead Redfield superoperator which describes the electron transport processes between the system and electrodes. It is well-known that the phonon bath can lead to electronic state relaxation and decoherence in the electronic density matrix~\cite{Ueda2010, Hartle2011, Kilgour2015}, but the effect of the electronic bath (associated with the lead Redfield superoperator $\mathcal{R}_{\mathrm{lead}}$) on electron transport is quite vague. In order to focus on many-body electronic coherence due to electronic bath, we neglect the effect of the phonon bath on coherence in the main text. 
\newline \indent
The operation of the lead Redfield superoperator on the electronic density matrix can be expressed as 
\begin{align}
\label{eq:Rtensor}
    \bra{N_a,a} &\mathcal{R}_{\mathrm{lead}} \hat{\rho}_{\mathrm{sys}}(t) \ket{N_b,b} = \sum_{cd} \mathcal{R}_{ab,cd} \rho_{cd},
\end{align}
where states $(a,b,c,d)$ serve as the eigenstates of $\hat{H}_{\mathrm{sys}}$. Several remarks are listed below. $\mathcal{R}_{ab,cd}$ in Eq.~\eqref{eq:Rtensor} can be decomposed into four mechanisms $\mathcal{R}^{\RomanNumeralCaps{1}}$, $\mathcal{R}^{\RomanNumeralCaps{2}}$, $\mathcal{R}^{\RomanNumeralCaps{3}}$, and $\mathcal{R}^{\RomanNumeralCaps{4}}$. The first mechanism $\mathcal{R}^{\RomanNumeralCaps{1}}$ and the second mechanism  $\mathcal{R}^{\RomanNumeralCaps{2}}$ correspond to the two-path quantum inference formed of state-to-state transitions caused by electron and hole injections, respectively. The third mechanism $\mathcal{R}^{\RomanNumeralCaps{3}}$ and the fourth mechanism  $\mathcal{R}^{\RomanNumeralCaps{4}}$ correspond to the indirect interference caused by electron and hole injections, respectively. For example, 
$\mathcal{R}^{\RomanNumeralCaps{1}}_{ab,cd}\rho_{cd} = - \frac{i}{\hbar} \sum_{l} \left[ \Sigma^{(l),<}_{db,ca}(\varepsilon_{db}) - \left(\Sigma^{(l),<}_{ca,db}(\varepsilon_{ca})\right)^*  \right]\rho_{cd}$ represents the two-path quantum interference formed of $\ket{N-1,c} \rightarrow \ket{N,a}$ and $\ket{N-1,d} \rightarrow \ket{N,b}$ caused by electron injections, where lesser self-energy $\Sigma^{(l),<}_{db,ca}(\varepsilon_{db})$ describes the state-to-state transition accompanied by a single-electron injection with energy $\varepsilon_{db}\equiv \varepsilon_b-\varepsilon_d$ (see Appendix~\ref{sec:deriv_FR} for more details).
\newline\indent 
For simplicity, we consider the wideband approximation~\cite{Covito2018}, and the lesser self-energy can be expressed in terms of Hubbard operator $\hat{X}^{a,b}$ as 
\begin{align}
    \label{eq:self_wideband}
    \Sigma^{(l),<}_{db,ca}(\varepsilon_{db}) = i\frac{\Gamma}{2} f_l(\varepsilon_{db}) \sum_{\sigma} \mathrm{Tr}\left[ \hat{c}_{i\sigma}^{\dagger}\hat{X}^{d,b} \right]^* \mathrm{Tr}\left[ \hat{c}_{i\sigma}^{\dagger}\hat{X}^{c,a} \right],  
\end{align}
which is composed of a coupling constant $\Gamma$, the occupation of electrons $f_l(\varepsilon_{db})$, and the transition amplitude between many-body states of the system due to an injected electron. Similarly, the greater self-energy in the wideband approximation comprises a coupling constant $\Gamma$, the occupation of holes $1-f_l(\varepsilon_{db})$, and the transition amplitude between many-body states of the system due to a hole entering the system.
\newline \indent
To explore the correlation between the steady-state electric current and many-body coherence, we compute the electric current~\cite{haug2008quantum} from the steady-state density matrix $\hat{\rho}_{\mathrm{sys}}(t)$ as (see Appendix~\ref{sec:SS_current})
\begin{align}
\label{eq:current}
    I = \frac{2\mathrm{e}}{\hbar} \sum_{acd} \mathrm{Im} \bigg\{ &\Big[ \Sigma^{(\mathrm{L}),<}_{da,ca}(\varepsilon_{da}) - \left(\Sigma^{(\mathrm{L}),>}_{ac,ad}(\varepsilon_{ad})\right)^* \Big]  \rho_{cd} \bigg\},
\end{align}
where $\Sigma^{(\mathrm{L}),<}_{da,ca}(\varepsilon_{da})$ corresponds to a transition from $N$-electron to $(N+1)$-electron state due to an injected electron from the left electrode, while $\Sigma^{(\mathrm{L}),>}_{ac,ad}(\varepsilon_{ad})$ corresponds to a transition from $N$-electron to $(N-1)$-electron state caused by an injected hole. 
\section{Many-body coherence and current blockade}
\begin{figure}[!t]
    \centering
    {\includegraphics[width=0.5\textwidth]{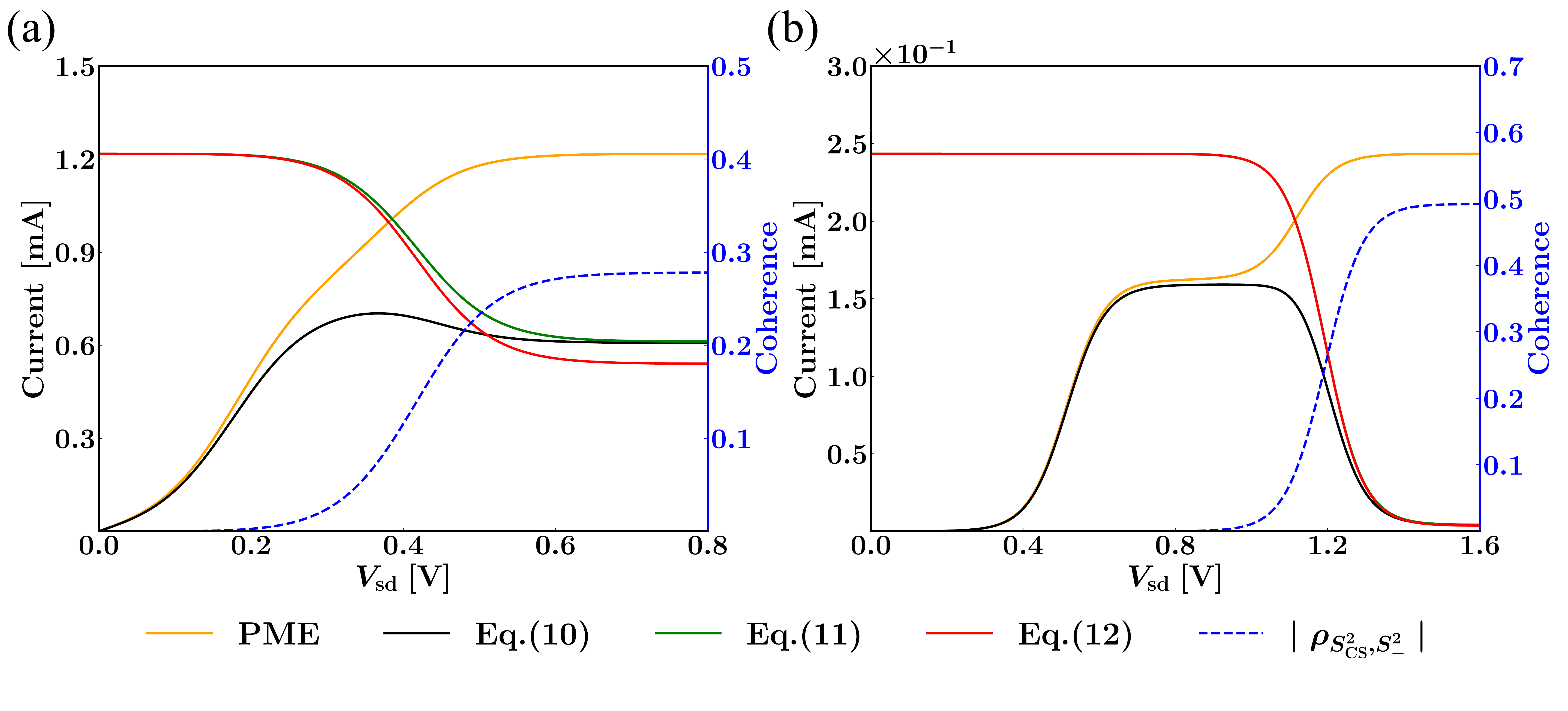}}
    \caption{Current blockade induced by many-body coherence in an AH system~\cite{Koole2016} for (a) $\varepsilon = 0.1\ \mathrm{eV}$, $t = 0.01\ \mathrm{eV}$, $U = 0.08\ \mathrm{eV}$, $\Gamma = 0.005\ \mathrm{eV}$ and in a model system for (b) $\varepsilon = -0.25\ \mathrm{eV}$, $t = 0.005\ \mathrm{eV}$, $U = 0.8\ \mathrm{eV}$, $\Gamma = 0.001\ \mathrm{eV}$. Other parameters are $T=300\ \mathrm{K}$, and $V_\mathrm{g}=0\ \mathrm{V}$. The orange, black, green, and red solid lines correspond to steady-state currents derived from PME, Eq.~\eqref{eq:current}, Eq.~\eqref{eq:analytical}, and Eq.~\eqref{eq:analytical2}, respectively. The dashed blue line describes the magnitude of coherence $\rho_{S^2_{\mathrm{CS}}, S^2_{-}}$.\label{fig:2}}
\end{figure}

To demonstrate that the effect of many-body coherence on quantum transport can be experimentally observed in a realistic system, we consider AH with experimental parameters~\cite{Koole2016}. As shown in Fig.~\ref{fig:2}a, the electric current (the black solid line) decreases as many-body coherence between eigenstates $\ket{2,S^2_\mathrm{CS}}$ and $\ket{2,S^2_-}$ (the blue dashed line) increases with bias. Furthermore, we find that, for a model system with large Coulomb repulsion and weak intersite electron hopping, many-body coherence can reach a maximum, and the electric current can be completely blocked to zero, as shown in Fig.~\ref{fig:2}b. It is worth mentioning that the current blockade phenomenon in Fig.~\ref{fig:2}a and \ref{fig:2}b is completely different from the well-known ``Coulomb blockade''. In Coulomb blockade, the electric current exhibits ``Coulomb staircase'' with the increasing bias voltage (the orange solid lines in Fig.~\ref{fig:2}a and \ref{fig:2}b), whereas Fig.~\ref{fig:2}a and \ref{fig:2}b show that the electric current decreases with the increasing bias voltage, similar to the behavior of a negative difference resistance. Here, we would like to emphasize that Coulomb staircase can be fully understood by the PME approach, and this approach is extensively employed to study nanodevices~\cite{Boyle2019,Vyas2020,Thomas2021}. However, the PME approach does not account for the effect of ``coherence'' induced by the interaction between many-body states and electron baths. Note that the current suppression is found to be robust against electron-phonon couplings (see Appendix~\ref{sec:vib}). Our numerical simulations clearly demonstrate that coherence between many-body states cannot be neglected and is directly associated with electric current.    
\newline\indent 
To quantitatively understand the current blockade in Fig.~\ref{fig:2}a and \ref{fig:2}b, we derive a current-coherence relationship for a system with weak hopping and strong Coulomb repulsion. The relationship is established based on two assumptions. First, to include the effect of Coulomb repulsion $U$ on currents, we consider that $\mathrm{e}V_\mathrm{sd} > U$ in the zero temperature limit. Furthermore, for the simplicity of derivation, we neglect the influence of $\varepsilon$ and $t$ on the Fermi function. In this condition, we can approximate  $f_{\rm{L}}(\varepsilon_{ca})=1$ and $f_{\rm{R}}(\varepsilon_{ca})=0$ in Eq.~\eqref{eq:self_wideband}. Second, we only keep many-body coherence $\rho_{S_{+}^2, T_0^2}$, $\rho_{S_{+}^2, T_{+1}^2}$, $\rho_{S_{+}^2, T_{-1}^2}$ and $\rho_{S_{\mathrm{CS}}^2, S_-^2}$ when solving Eq.~\eqref{eq:RME}. It is well-known that coherence can be neglected while there is a large energy gap between two states, i.e., the secular approximation for the derivation of the PME approach. When $t/U$ is small, the energy gap between $\ket*{2,S^2_\mathrm{CS}}$ and $\ket*{2,S^2_-}$ and the energy gap between $\ket*{2,S^2_+}$ and triplet states $\ket*{2,T^2_0}$, $\ket*{2,T^2_{+1}}$, $\ket*{2,T^2_{-1}}$ are the smallest. As a result, we consider these coherence terms when solving Eq.~\eqref{eq:RME} and find that only $\rho_{S_{\mathrm{CS}}^2, S_-^2}$ is associated with current. 
Finally, we obtain a current-coherence relationship as (see Appendix~\ref{sec:ana})
\begin{equation}
\label{eq:analytical}
     I = \frac{\mathrm{e} \Gamma}{\hbar}  \bigg\{ 1 - 2  \big[ 1 + \frac{1}{4}  (\frac{4t^2}{U\Gamma})^2\big]^{-1/2}  \abs{\rho_{S_{\mathrm{CS}}^2, S_-^2}} \bigg\},
\end{equation}
showing that the electric current can be expressed in terms of many-body coherence $ \rho_{S^2_{\mathrm{CS}}, S^2_{-}}$ and the kinetic exchange $4t^2/U$ in the unit of system-lead coupling $\Gamma$. In Fig.~\ref{fig:2}a and \ref{fig:2}b, the green lines almost coincide with the black lines when current blockade occurs, which reveals that Eq.~\eqref{eq:analytical} has successfully captured the physics behind the current blockade and elucidated the influence of many-body coherence on quantum transport. Furthermore, the kinetic exchange $4t^2/U$, resulting from the interplay between hopping and many-body interactions, describes the intersite delocalization of electrons. Therefore, when the kinetic exchange is small, electrons accumulate on a single site, and the current is blockaded. Note that $4t^2/U$ corresponds to the energy gap $\Delta E_{S^2_\mathrm{CS}, S^2_-} = (\sqrt{U^2+16t^2}-U)/2$ when $t/U \ll 1$. If the energy gap $\Delta E_{S^2_\mathrm{CS}, S^2_-}$ is small enough, i.e, $4t^2/U\Gamma$ is negligible, then Eq.~\eqref{eq:analytical} can be further simplified as 
\begin{equation}
\label{eq:analytical2}
     I = \frac{\mathrm{e} \Gamma}{\hbar}  \bigg\{ 1 - 2  \abs{\rho_{S_{\mathrm{CS}}^2, S_-^2}} \bigg\}, 
\end{equation}
indicating that many-body coherence $ \rho_{S^2_{\mathrm{CS}}, S^2_{-}}$ becomes a dominant factor in current blockade. 
When $t/U$ is not small enough, e.g., $t/U=0.125$ in Fig.~\ref{fig:2}a, Eq.~\eqref{eq:analytical2} (the red line) slightly underestimates the electric current in the current blockade region due to the neglect of the kinetic exchange effect. On the other hand, when $t/U \ll 1$, e.g., $t/U=0.00625$ in Fig.~\ref{fig:2}b, the red line matches the black line in the current blockade region, testifying that many-body coherence predominates the current suppression. 
\begin{figure}[b!]
    \centering
    {\includegraphics[width=0.5\textwidth]{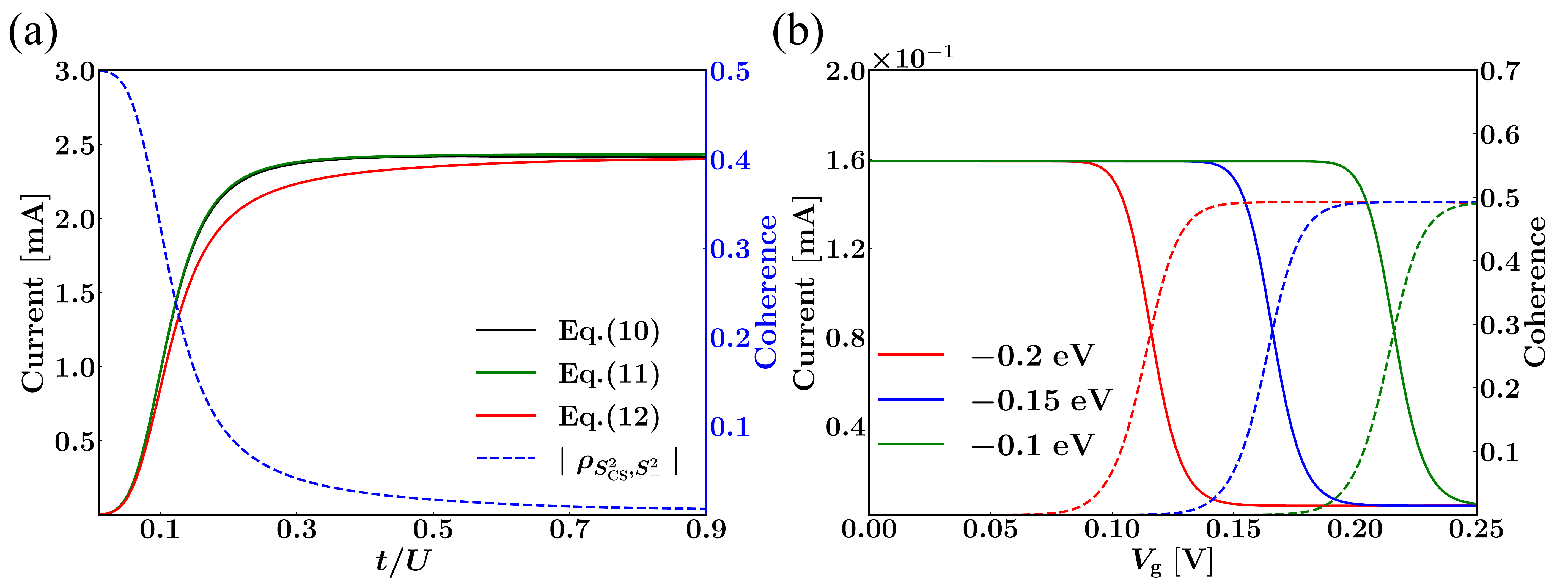}}
    \caption{Coherence-controlled current blockade of a model system tuned by (a) the Hamiltonian design $t/U$ for $\varepsilon = -0.1\ \mathrm{eV}$, $t = 0.002\text{ to }0.18\ \mathrm{eV}$, $U = 0.2\ \mathrm{eV}$, $\Gamma = 0.01\ \mathrm{eV}$ under bias $V_{\mathrm{sd}}=0.8\ \mathrm{V}$, gate  $V_\mathrm{g} = 0\ \mathrm{V}$ and (b) the gate voltage $V_\mathrm{g}$ for $t = 0.005\ \mathrm{eV}$, $U = 0.8\ \mathrm{eV}$, $\Gamma = 0.001\ \mathrm{eV}$ under bias $V_{\mathrm{sd}}=1.0\ \mathrm{V}$, gate $V_\mathrm{g} = 0\text{ to }0.25\ \mathrm{V}$. The black, green, and red solid lines in (a) correspond to currents from Eq.~\eqref{eq:current}, Eq.~\eqref{eq:analytical}, and Eq.~\eqref{eq:analytical2}, and the dashed blue line denotes the magnitude of coherence $\rho_{S^2_{\mathrm{CS}}, S^2_{-}}$. The red, blue, and green lines in (b) correspond to $\varepsilon = -0.2,\ -0.15,\ -0.1\ \mathrm{eV}$, and the solid and dashed lines denote currents and the magnitude of coherence $\rho_{S^2_{\mathrm{CS}}, S^2_{-}}$ respectively. Other parameters used here are $T=300\ \mathrm{K}$ for (a) and $T=77\ \mathrm{K}$ for (b).\label{fig:3}}
\end{figure}
\section{Control of current blockade}
Control of electric current is a key issue in quantum transport~\cite{Grifoni1998,Hsu2012,Hsu2013,White2013,Damanet2019}. Here, we demonstrate that it is feasible to operate many-body coherence and current blockade via internal Hamiltonian design and an external gate voltage. 
\newline \indent
First, for Hamiltonian design, the relative magnitudes of intersite coupling $t$ and Coulomb repulsion $U$ are directly related to many-body coherence and current blockade. As shown in Fig.~\ref{fig:3}a, when $t/U \ll 0.1$, the current decreases to almost zero, and many-body coherence $ \rho_{S^2_{\mathrm{CS}}, S^2_{-}}$ approaches its maximum $0.5$. In brief, the maximum value of coherence can be understood by the fact that small $t/U$ reduces the energy gap between $\ket*{2,S^2_\mathrm{CS}}$ and $\ket*{2,S^2_-}$ to almost zero and thus leads to the maximum of $ \rho_{S^2_{\mathrm{CS}}, S^2_{-}}=0.5$. The origin of strong current blockade results mainly from many-body coherence $\rho_{S^2_\mathrm{CS},S^2_-}$, i.e., when $t/U \ll 0.1$, the current calculated from Eq.~\eqref{eq:analytical2} (the red line), which neglects the kinetic exchange effect, coincides with the current calculated from Eq.~\eqref{eq:analytical} (the green line). The small deviation between the green line and the red line in the region $t/U \approx 0.1 \sim 0.7$ indicates that the kinetic exchange $4t^2/U$ can affect the electric current, but many-body coherence is still the main mechanism for the current blockade. When $t/U \gg 0.7$, many-body coherence $ \rho_{S^2_\mathrm{CS},S^2_-}$ reaches zero, so current blockade disappears. Fig.~\ref{fig:3}a clearly shows that one can control electric current and many-body coherence via the modification of $t/U$. 
\newline \indent
Second, we find that many-body coherence of a system can be significantly influenced by an external gate field. Fig.~\ref{fig:3}b shows that, with an increasing gate voltage $V_\mathrm{g}$, many-body coherence $ \rho_{S^2_\mathrm{CS},S^2_-}$ transitions from zero to its maximum and the current drops to zero. Moreover, the transition gate voltage increases with the increasing on-site energy $\varepsilon$, where  the red, blue, and green line correspond to $\varepsilon=-0.2$, $-0.15$, and $-0.1$ eV, respectively. Control of the gate voltage $V_\mathrm{g}$ and the Hamiltonian design correspond to different mechanisms of forming the current blockade because control of $V_\mathrm{g}$ does not change the kinetic exchange $4t^2/U$. To explain the gate dependence of many-body coherence $ \rho_{S^2_{\mathrm{CS}}, S^2_{-}}$, we derive an analytical expression for the coherence-gate relationship, $ \mid \rho_{S^2_{\mathrm{CS}}, S^2_{-}} \mid = 1/2 \times \left[2-\Theta(\mu_\mathrm{L}-\varepsilon-U+\mathrm{e}V_\mathrm{g})\right]/\left[8-7\Theta(\mu_\mathrm{L}-\varepsilon-U+\mathrm{e}V_\mathrm{g})\right],$ by making the approximation $f_\mathrm{L}(\varepsilon+U-\mathrm{e}V_\mathrm{g}) = \Theta(\mu_\mathrm{L}-\varepsilon-U+\mathrm{e}V_\mathrm{g})$ and $t\approx 0$ (see Appendix~\ref{sec:ana}), where $\Theta(\mu_\mathrm{L}-\varepsilon-U+\mathrm{e}V_\mathrm{g})$ is the Heaviside step function. According to the coherence-gate relation, when $\varepsilon=-0.2~\mathrm{eV}$, $U=0.8~\mathrm{eV}$, and $V_\mathrm{sd}=1.0~\mathrm{V}$, many-body coherence $  \rho_{S^2_{\mathrm{CS}}, S^2_{-}} $ has a maximum value $0.5$ while $V_\mathrm{g}$ exceeds $0.1~\mathrm{V}$, which is consistent with our simulation result (the red line). Fig.~\ref{fig:3}b also indicates that, with lower on-site energies, the electric current and many-body coherence can be operated with smaller gate voltages, showing potential as  transistors.
\section{Conclusions}
We have demonstrated the significance of many-body coherence in quantum transport and established a current-coherence relationship Eq.~\eqref{eq:analytical} for a model system using the Redfield-type fermionic quantum master equation. The results imply that many-body coherence can eliminate the well-known Coulomb staircase and lead to the negative differential resistance, which cannot be described by the PME approach~\cite{Boyle2019,Vyas2020,Thomas2021} due to the lack of coherence. Furthermore, it is shown that many-body coherence can be manipulated through modifying the internal system Hamiltonian or applying an external gate voltage. Finally, we find that the electric current can be switched based on many-body coherence at a low gate voltage, indicating potential as coherence-controlled transistors. The results here open a new class of electronic devices in quantum electronics, which will motive further experimental and theoretical investigations on the effects of many-body coherence in condensed matter physics and quantum technology.

\begin{acknowledgments}
We thank Chih‐En Shen, Hung‐Sheng Tsai, Ming-Wei Lee, Yi-Ting Chuang, Qian-Rui Huang, Michitoshi Hayashi, and Yang-Hao Chan for useful discussions. This research was supported by Academia Sinica (AS-CDA-111-M02) and National Science and Technology Council (Grant Nos. 110-2113-M-001-053 and 111-2113-M-001-027-MY4).
\end{acknowledgments}

\appendix
\counterwithin{figure}{section}
\counterwithin{table}{section}
\section{Expression for Single-Electron Operators by Hubbard Operators \label{sec:HO_coeff}}
\indent
In this section, we show how to adopt Hubbard operators to express the single-electron creation and annihilation operators. First, we introduce the occupation number (ON) vector representation $\ket*{n_{1\uparrow}n_{\downarrow}n_{2\uparrow}n_{2\downarrow}}$ to denote states spanned on the site basis depicting the distribution of electrons. These states are arranged as $\{ \ket*{0}, \ket*{1} \dots \ket*{15}\}$, with each state $\ket*{p},\ p = (n_{1\uparrow})_p\cdot2^0+(n_{1\downarrow})_p\cdot2^1+(n_{2\uparrow})_p\cdot2^2+(n_{2\downarrow})_p\cdot2^3$, and $(n_{i\sigma})_p$ represents the occupation of an electron on site $i$ with spin $\sigma$ in state $\ket*{p}$. Note that the definition of $\ket{0}$ to $\ket{15}$ that we adopt is slightly different from the definition in the previous study~\cite{Li2014}. The Hubbard operators defined as $\hat{X}^{p,p'} \equiv \ket*{p}\bra*{p'}$ can be used to describe the transition from state $\ket*{p'}$ to state $\ket*{p}$ \cite{Esposito2009,Li2014}, and the single-electron creation (annihilation) operators spanned on the site basis can be expressed by the aforementioned Hubbard operators. Take $\hat{c}_{2\uparrow}^\dagger$ for instance. The operations of $\hat{c}_{2\uparrow}^\dagger$ on states $\ket*{0}$ to $\ket*{15}$:
\begin{align*}
    \footnotesize
    \begin{array}{rlrlrlrlrlr}
        \hat{c}_{2\uparrow}^{\dagger} \ket{0}
        &=& \hat{c}_{2\uparrow}^{\dagger} \ket{0000} &=& \ket{0010} &=& \ket{4} &\rightarrow& \hat{X}^{4,0} && \\
        \hat{c}_{2\uparrow}^{\dagger} \ket{1} &=& \hat{c}_{2\uparrow}^{\dagger} \ket{1000}  &=& -\ket{1010} &=& -\ket{5} &\rightarrow& -\hat{X}^{5,1} && \\
        \hat{c}_{2\uparrow}^{\dagger} \ket{2} &=& \hat{c}_{2\uparrow}^{\dagger} \ket{0100} &=& -\ket{0110} &=& -\ket{6} &\rightarrow& -\hat{X}^{6,2} && \\
        \hat{c}_{2\uparrow}^{\dagger} \ket{8} &=& \hat{c}_{2\uparrow}^{\dagger} \ket{0001} &=& \ket{0011} &=& \ket{12} &\rightarrow& \hat{X}^{12,8} && \\
        \hat{c}_{2\uparrow}^{\dagger} \ket{3} &=& \hat{c}_{2\uparrow}^{\dagger} \ket{1100} &=& \ket{1110} &=& \ket{7} &\rightarrow& \hat{X}^{7,3} && \\
        \hat{c}_{2\uparrow}^{\dagger} \ket{9} &=& \hat{c}_{2\uparrow}^{\dagger} \ket{1001} &=& -\ket{1011} &=& -\ket{13} &\rightarrow& -\hat{X}^{13,9} && \\
        \hat{c}_{2\uparrow}^{\dagger} \ket{10} &=& \hat{c}_{2\uparrow}^{\dagger} \ket{0101} &=& -\ket{0111} &=& -\ket{14} &\rightarrow& -\hat{X}^{14,10} && \\
        \hat{c}_{2\uparrow}^{\dagger} \ket{11} &=& \hat{c}_{2\uparrow}^{\dagger} \ket{1101} &=& \ket{1111} &=& \ket{15} &\rightarrow& \hat{X}^{15,11}. &&
    \end{array}
\end{align*}
The operation of $\hat{c}_{2\uparrow}^\dagger$ on any of the other many-body states equals 0. Thus,
\begin{equation*}
    \hat{c}_{2\uparrow}^{\dagger} = \hat{X}^{4,0} - \hat{X}^{5,1} - \hat{X}^{6,2} + \hat{X}^{7,3} + \hat{X}^{12,8} - \hat{X}^{13,9} - \hat{X}^{14,10} + \hat{X}^{15,11}.
\end{equation*}
\indent
Following the similar procedures, other single-electron operators can also be expressed in terms of the Hubbard operators spanned on the site basis as
\begin{align*}
\footnotesize
    \begin{array}{rlrlrlrlrlrlrlrlr}
        \hat{c}_{1\uparrow}^{\dagger} &=& \hat{X}^{1,0} &+& \hat{X}^{3,2} &+& \hat{X}^{5,4} &+& \hat{X}^{7,6} \\
        &+& \hat{X}^{9,8} &+& \hat{X}^{11,10} &+& \hat{X}^{13,12} &+& \hat{X}^{15,14} \\
        \hat{c}_{1\downarrow}^{\dagger} &=& \hat{X}^{2,0} &-& \hat{X}^{3,1} &+& \hat{X}^{6,4} &-& \hat{X}^{7,5} \\
        &+& \hat{X}^{10,8} &-& \hat{X}^{11,9} &+& \hat{X}^{14,12} &-& \hat{X}^{15,13} \\
        \hat{c}_{2\uparrow}^{\dagger} &=& \hat{X}^{4,0} &-& \hat{X}^{5,1} &-& \hat{X}^{6,2} &+& \hat{X}^{7,3} \\
        &+& \hat{X}^{12,8} &-& \hat{X}^{13,9} &-& \hat{X}^{14,10} &+& \hat{X}^{15,11} \\
        \hat{c}_{2\downarrow}^{\dagger} &=& \hat{X}^{8,0} &-& \hat{X}^{9,1} &-& \hat{X}^{10,2} &+& \hat{X}^{11,3} \\
        &-& \hat{X}^{12,4} &+& \hat{X}^{13,5} &+& \hat{X}^{14,6} &-& \hat{X}^{15,7} \\
        \hat{c}_{1\uparrow} &=& \hat{X}^{0,1} &+& \hat{X}^{2,3} &+& \hat{X}^{4,5} &+& \hat{X}^{6,7} \\
        &+& \hat{X}^{8,9} &+& \hat{X}^{10,11} &+& \hat{X}^{12,13} &+& \hat{X}^{14,15} \\
        \hat{c}_{1\downarrow} &=& \hat{X}^{0,2} &-& \hat{X}^{1,3} &+& \hat{X}^{4,6} &-& \hat{X}^{5,7} \\
        &+& \hat{X}^{8,10} &-& \hat{X}^{9,11} &+& \hat{X}^{12,14} &-& \hat{X}^{13,15} \\
        \hat{c}_{2\uparrow} &=& \hat{X}^{0,4} &-& \hat{X}^{1,5} &-& \hat{X}^{2,6} &+& \hat{X}^{3,7} \\
        &+& \hat{X}^{8,12} &-& \hat{X}^{9,13} &-& \hat{X}^{10,14} &+& \hat{X}^{11,15} \\
        \hat{c}_{2\downarrow} &=& \hat{X}^{0,8} &-& \hat{X}^{1,9} &-& \hat{X}^{2,10} &+& \hat{X}^{3,11} \\
        &-& \hat{X}^{4,12} &+& \hat{X}^{5,13} &+& \hat{X}^{6,14} &-& \hat{X}^{7,15}.
    \end{array} 
\end{align*}
\indent
In Appendix~\ref{sec:deriv_FR}, we will apply a shorthand notation $\hat{c}_{i\sigma}=\sum_{p<q}(r_{i\sigma})_{p,q}\hat{X}^{p,q}$, $(r_{i\sigma})_{p,q}\in\{\pm 1,0\}$, $ (r_{i\sigma})_{p,q}=(r_{i\sigma})_{q,p}$ to denote the above operators (see Eq.~\eqref{eq:RT_site} to Eq.~\eqref{eq:redef_coup}). The non-zero elements of $(r_{i\sigma})_{p,q}$ are listed below,
\begin{align*}
    \begin{array}{rlrlrlrlrl}
        (r_{1\uparrow})_{0,1} &=& (r_{1\uparrow})_{2,3} &=& (r_{1\uparrow})_{4,5} &=& (r_{1\uparrow})_{6,7} &=& 1 &\\
        (r_{1\uparrow})_{8,9} &=& (r_{1\uparrow})_{10,11} &=& (r_{1\uparrow})_{12,13} &=& (r_{1\uparrow})_{14,15} &=& 1 &\\
        (r_{1\downarrow})_{0,2} &=& (r_{1\downarrow})_{4,6} &=& (r_{1\downarrow})_{8,10} &=& (r_{1\downarrow})_{12,14} &=& 1 &\\
        (r_{1\downarrow})_{1,3} &=& (r_{1\downarrow})_{5,7} &=& (r_{1\downarrow})_{9,11} &=& (r_{1\downarrow})_{13,15} &=& -1 &\\
        (r_{2\uparrow})_{0,4} &=& (r_{2\uparrow})_{3,7} &=& (r_{2\uparrow})_{8,12} &=& (r_{2\uparrow})_{11,15} &=& 1 &\\
        (r_{2\uparrow})_{1,5} &=& (r_{2\uparrow})_{2,6} &=& (r_{2\uparrow})_{9,13} &=& (r_{2\uparrow})_{10,14} &=& -1 &\\
        (r_{2\downarrow})_{0,8} &=& (r_{2\downarrow})_{3,11} &=& (r_{2\downarrow})_{5,13} &=& (r_{2\downarrow})_{6,14} &=& 1 &\\
        (r_{2\downarrow})_{1,9} &=& (r_{2\downarrow})_{2,10} &=& (r_{2\downarrow})_{4,12} &=& (r_{2\downarrow})_{7,15} &=& -1 &.   
    \end{array}
\end{align*}

\section{Derivation of Redfield-Type Fermionic Quantum Master Equation\label{sec:deriv_FR}}
\indent
In this section, we outline the derivation of Eq.~(2) in the main text. We start from the quantum Liouville equation,
\begin{align}
\label{eq:liouville}
    \frac{\mathrm{d} \hat{\rho}(t)}{\mathrm{d} t} = -\frac{i}{\hbar} \big[ \hat{H}, \hat{\rho}(t) \big],
\end{align}
where $\hat{H}$ and $\hat{\rho}$ denote the total Hamiltonian and density matrix. In the interaction picture, the quantum Liouville equation can be written as an integro-differential equation as follows
\begin{align}
\label{eq:dyson_series}
    \frac{\mathrm{d} \hat{\tilde{\rho}}(t)}{\mathrm{d} t} = &-\frac{i}{\hbar} \big[ \hat{\tilde{H}}_\mathrm{sys-lead}(t), \hat{\tilde{\rho}}(t_0) \big] \nonumber \\
    &- \frac{1}{\hbar^2} \int_{t_0}^{t} \mathrm{d}t_1 \Big[ \hat{\tilde{H}}_\mathrm{sys-lead}(t),\ \big[ \hat{\tilde{H}}_\mathrm{sys-lead}(t_1),\ \hat{\tilde{\rho}}(t_1)\big] \Big],
\end{align}
in which $\hat{H} = \hat{H}_\mathrm{sys}+\hat{H}_\mathrm{lead}+\hat{H}_\mathrm{sys-lead}$, with $\hat{\tilde{H}}$ and $\hat{\tilde{\rho}}$ as the total Hamiltonian and the total density matrix in the interaction picture respectively. Due to weak coupling between the system and the leads, the dynamics of the system and the dynamics of
the bath occur at different time scales, and we apply the Born approximation. Under the Born approximation, the density matrix of the total system is
approximated as the direct product of the electronic density matrix of the system $\hat{\rho}_\mathrm{sys}(t)$ and the density matrix of the lead $\hat{\rho}_\mathrm{lead}(t)$, i.e., $\hat{\rho}(t) = \hat{\rho}_\mathrm{sys}(t) \otimes \hat{\rho}_\mathrm{lead}(t)$. In the interaction picture, one can derive
\begin{align}
\label{eq:DM_BA}
    \hat{\tilde{\rho}}(t) = \hat{\tilde{\rho}}_\mathrm{sys}(t) \otimes \hat{\tilde{\rho}}_\mathrm{lead}(t),
\end{align}
with $\hat{\tilde{\rho}}_\mathrm{sys}(t) = e^{i\hat{H}_\mathrm{sys}t/\hbar} \hat{\rho}_\mathrm{sys}(t) e^{-i\hat{H}_\mathrm{sys}t/\hbar}$ and $\hat{\tilde{\rho}}_\mathrm{lead}(t) = e^{i\hat{H}_\mathrm{lead}t/\hbar} \hat{\rho}_\mathrm{lead}(t) e^{-i\hat{H}_\mathrm{lead}t/\hbar}$. Since the leads are weakly coupled to the system and relax rapidly, we assume that the leads do not change with time and always stay in thermal equilibrium. As a result, we have the relation
\begin{align}
\label{eq:lead_TE}
    \hat{\rho}_\mathrm{lead}(t) = \hat{\rho}_\mathrm{lead}(t_0) = \hat{\bar{\sigma}}_\mathrm{lead} = \frac{e^{-\beta_\mathrm{lead} \hat{H}_\mathrm{lead}}}{\mathrm{Tr}_\mathrm{lead}\big(e^{-\beta_\mathrm{lead} \hat{H}_\mathrm{lead}}\big)},
\end{align}
where $\beta_\mathrm{lead} = 1/kT_\mathrm{lead}$ represents the reciprocal of the thermodynamic temperature of the leads.
By $\big[ \hat{\rho}_\mathrm{lead}(t), \hat{H}_\mathrm{lead} \big] = 0$, the density matrix in the interaction picture can be obtained as 
\begin{align}
\label{eq:lead_TE_int}
    \hat{\tilde{\rho}}_\mathrm{lead}(t) = \hat{\bar{\sigma}}_\mathrm{lead}.
\end{align}
\indent
The system degrees of freedom and lead degrees of freedom can be further separated in the coupling terms. The system-lead coupling $\hat{H}_\mathrm{sys-lead} = \sum_{k,\sigma} (T_{\mathrm{L}k,1}\hat{c}_{1\sigma}^{\dagger}\hat{d}_{\mathrm{L}k,\sigma} + T_{\mathrm{R}k,2} \hat{c}_{2\sigma}^{\dagger} \hat{d}_{\mathrm{R}k,\sigma} + \mathrm{H.c.})$ can be rewritten as
\begin{align}
\label{eq:coup_DP}
    \hat{H}_\mathrm{sys-lead} &= \hat{c}_{1\uparrow}^{\dagger} \otimes \sum_k T_{\mathrm{L}k,1}\hat{d}_{\mathrm{L}k\uparrow} + \hat{c}_{2\uparrow}^{\dagger} \otimes \sum_k T_{\mathrm{R}k,2}\hat{d}_{\mathrm{R}k\uparrow} 
    \nonumber \\
    &+~ \hat{c}_{1\downarrow}^{\dagger} \otimes \sum_k T_{\mathrm{L}k,1}\hat{d}_{\mathrm{L}k\downarrow} + \hat{c}_{2\downarrow}^{\dagger} \otimes \sum_k T_{\mathrm{R}k,2}\hat{d}_{\mathrm{R}k\downarrow} 
    \nonumber \\
    &+~ \mathrm{H.c.},
\end{align}
in which each coupling element is expressed as the direct product of operators acting on the system and the bath. One can verify that the average of each lead operator is zero, e.g., $\expval{\hat{d}_{\mathrm{L}k\uparrow}} = 0$; therefore, the non-Markovian master equation can be derived as
\begin{align}
\label{eq:nonMarkovian}
    \frac{\mathrm{d} \hat{\tilde{\rho}}_\mathrm{sys}(t)}{\mathrm{d} t} = &\mathrm{Tr}_\mathrm{lead} \bigg\{ \frac{\mathrm{d} \hat{\tilde{\rho}}(t)}{\mathrm{d} t}\bigg\} \\
    = &- \frac{1}{\hbar^2} \int_{0}^{t-t_0} \mathrm{d}\tau \mathrm{Tr}_\mathrm{lead} \bigg\{ \Big[ \hat{\tilde{H}}_\mathrm{sys-lead}(t), \nonumber \\
    &\hspace{0.3cm}\big[ \hat{\tilde{H}}_\mathrm{sys-lead}(t-\tau),\ \hat{\tilde{\rho}}_\mathrm{sys}(t-\tau)\otimes \hat{\bar{\sigma}}_\mathrm{lead} \big] \Big] \bigg\}.
\end{align}
\indent
Next, we assume that the density matrix varies slower than the decay time of the bath (lead) correlation. Therefore, we apply the first Markov approximation, which assumes that  $\hat{\tilde{\rho}}_\mathrm{sys}(t-\tau) = \hat{\tilde{\rho}}_\mathrm{sys}(t)$ in Eq.~\eqref{eq:nonMarkovian}, and the second Markov approximation, which considers the long-time limit $t-t_0 \rightarrow \infty$. After applying the Markov approximations to Eq.~\eqref{eq:nonMarkovian}, we obtain the Redfield equation in the Schr\"{o}dinger picture:
\begin{align}
\label{eq:REquation}
    \frac{\mathrm{d} \hat{\rho}_{\mathrm{sys}}(t)}{\mathrm{d} t} = 
    -&\frac{i}{\hbar} [\hat{H}_{\mathrm{sys}},\hat{\rho}_{\mathrm{sys}}(t)] + \mathcal{R}_\mathrm{lead} \hat{\rho}_{\mathrm{sys}}(t) \\
\label{eq:RTensor}
    \mathcal{R}_\mathrm{lead} \hat{\rho}_{\mathrm{sys}}(t) = -&\frac{1}{\hbar^2} \int_{0}^{\infty} \mathrm{d}\tau \mathrm{Tr}_\mathrm{lead} \bigg\{ \Big[ \hat{H}_\mathrm{sys-lead}(0), \nonumber \\
    &\big[ \hat{H}_\mathrm{sys-lead}(-\tau),\ \hat{\rho}_\mathrm{sys}(t)\otimes \hat{\bar{\sigma}}_\mathrm{lead} \big] \Big] \bigg\}.
\end{align}
\indent
We further simply Eq.~\eqref{eq:RTensor} by tracing out the lead degrees of freedom because both the system-lead coupling $\hat{H}_\mathrm{sys-lead}$ and the density matrix $\hat{\rho}(t)$ can be divided into the system part and the lead part. Take the term $\hat{H}_\mathrm{sys-lead}(0)\hat{H}_\mathrm{sys-lead}(-\tau)\hat{\rho}_\mathrm{sys}\otimes \hat{\bar{\sigma}}_\mathrm{lead}$ in Eq.~\eqref{eq:RTensor} and $\hat{c}_{1\uparrow}^\dagger \otimes \sum_k T_{\mathrm{L}k,1}\hat{d}_{\mathrm{L}k \uparrow}$ in $\hat{H}_\mathrm{sys-lead}$ as an example,
\begin{align}
    &\mathrm{Tr}_\mathrm{lead} \Big\{ \hat{c}_{1\uparrow}^\dagger(0) \otimes \sum_k T_{\mathrm{L}k,1}\hat{d}_{\mathrm{L}k \uparrow}(0) \nonumber \\
    &\hspace{1cm}\times \hat{c}_{1\uparrow}^\dagger(-\tau) \otimes \sum_{k'} T_{\mathrm{L}k',1}\hat{d}_{\mathrm{L}k' \uparrow}(-\tau) \times \hat{\rho}_\mathrm{sys}(t)\otimes\hat{\bar{\sigma}}_\mathrm{lead} \Big\} \nonumber \\ 
    =~&\hat{c}_{1\uparrow}^\dagger(0) \hat{c}_{1\uparrow}^\dagger(-\tau) \hat{\rho}_\mathrm{sys}(t) \otimes \nonumber \\
    &\hspace{1.05cm}\sum_k \sum_{k'} T_{\mathrm{L}k,1} T_{\mathrm{L}k',1} \times \mathrm{Tr}_\mathrm{lead} \Big\{ \hat{d}_{\mathrm{L}k \uparrow}(0)  \hat{d}_{\mathrm{L}k' \uparrow}(-\tau) \hat{\bar{\sigma}}_\mathrm{lead} \Big\}. \nonumber
\end{align}
\indent
Following the similar procedures, we can obtain bath correlation functions and then classify them into four types, i.e., 
\begin{subequations}
    \begin{align}
        \mathrm{Tr}_\mathrm{lead} \{ \hat{d}_{l_1 k_1 \sigma_1}^{\dagger}(0) \hat{d}_{l_2 k_2 \sigma_2}^{\dagger}(-\tau)\hat{\bar{\sigma}}_\mathrm{lead} \} \\
        \mathrm{Tr}_\mathrm{lead} \{ \hat{d}_{l_1 k_1 \sigma_1}^{\dagger}(0) \hat{d}_{l_2 k_2 \sigma_2}(-\tau)\hat{\bar{\sigma}}_\mathrm{lead} \} \\
        \mathrm{Tr}_\mathrm{lead} \{ \hat{d}_{l_1 k_1 \sigma_1}(0) \hat{d}_{l_2 k_2 \sigma_2}^{\dagger}(-\tau)\hat{\bar{\sigma}}_\mathrm{lead} \} \\
        \mathrm{Tr}_\mathrm{lead} \{ \hat{d}_{l_1 k_1 \sigma_1}(0) \hat{d}_{l_2 k_2 \sigma_2}(-\tau)\hat{\bar{\sigma}}_\mathrm{lead} \}
    \end{align}
\end{subequations}
in Eq.~\eqref{eq:RTensor}. Two of the correlation functions are non-zero and can be calculated as
\begin{subequations}
\begin{align}
\label{eq:lesser_green}
    \mathrm{Tr}_\mathrm{lead} \Big\{ \hat{d}_{l_1 k_1 \sigma_1}^{\dagger}(0) &\hat{d}_{l_2 k_2 \sigma_2}(-\tau)\hat{\bar{\sigma}}_\mathrm{lead} \Big\} \nonumber \\
    &= -i\hbar g^<_{l_1 k_1 \sigma_1}(-\tau)  \delta_{l_1,l_2}\delta_{k_1,k_2}\delta_{\sigma_1,\sigma_2} \\
\label{eq:greater_green}
    \mathrm{Tr}_\mathrm{lead} \Big\{\hat{d}_{l_1 k_1 \sigma_1}(0) &\hat{d}_{l_2 k_2 \sigma_2}^{\dagger}(-\tau)\hat{\bar{\sigma}}_\mathrm{lead} \Big\} \nonumber \\
    &= i\hbar g^>_{l_1 k_1 \sigma_1}(\tau)  \delta_{l_1,l_2}\delta_{k_1,k_2}\delta_{\sigma_1,\sigma_2},
\end{align}
\end{subequations}
with the lesser Green's function of free electrons $g^<_{lk\sigma}(t) \equiv (i/\hbar) f_{l}(\xi_{k\sigma}) e^{-i \xi_{k\sigma} t / \hbar}$ and the greater Green's function of free electrons  $g^>_{lk\sigma}(t) \equiv -(i/\hbar) [1-f_{l}(\xi_{k\sigma})] e^{-i \xi_{k\sigma} t / \hbar}$ \cite{Esposito2009}, where $f_l(\xi_{k\sigma}) = (1+e^{\beta_\mathrm{lead}(\xi_{k\sigma}-\mu_l)})^{-1}$ is the Fermi function of lead $l$ with chemical potential $\mu_l$ at temperature $T_\mathrm{lead}$. 
By means of Hubbard operators introduced in section~\ref{sec:HO_coeff}, the Redfield tensor for electrons spanned on the site basis becomes
\begin{widetext}
    \begin{align}
    \label{eq:RT_site}
    \mathcal{R}_\mathrm{lead} \hat{\rho}_{\mathrm{sys}}(t) = 
    &\frac{i}{\hbar} \sum_{p<q} \sum_{p'<q'} \sum_{(l,i)} \sum_{k} \sum_{\sigma} \int_{0}^{\infty} \mathrm{d}\tau \nonumber \\
    &\hspace{1cm} \Big\{ \;\;\, g_{lk\sigma}^{<}(-\tau) \abs{T_{lk,i}}^2 (r_{i\sigma})_{p,q} (r_{i\sigma})_{p',q'} \big[ \hat{X}^{p,q}, \hat{X}^{q',p'}(-\tau)\hat{\rho}_{\mathrm{el}}(t)\big] \nonumber \\
    &\hspace{1cm} -\ g_{lk\sigma}^{>}(\tau) \abs{T_{lk,i}}^2 (r_{i\sigma})_{p,q} (r_{i\sigma})_{p',q'} \big[ \hat{X}^{q,p}, \hat{X}^{p',q'}(-\tau)\hat{\rho}_{\mathrm{el}}(t)\big] \nonumber \\   
    &\hspace{1cm} +\ \mathrm{H.c.} \Big\},
    \end{align}
\end{widetext}
where $(l,i) \in \{(\mathrm{L},1),(\mathrm{R},2)\}$, and $(r_{i\sigma})_{p,q}$ serve as the coefficients of single-electron operators spanned on the site basis $\hat{c}_{i\sigma}=\sum_{p<q}(r_{i\sigma})_{p,q} \hat{X}^{p,q}$, $ (r_{i\sigma})_{p,q}\in\{\pm 1,0\}$, $ (r_{i\sigma})_{p,q}=(r_{i\sigma})_{q,p}$. 
\begin{table*}[t!]
\centering
\renewcommand*{\arraystretch}{1.9}
\begin{tabular}{ ccccc  }
\hline\hline
\multicolumn{5}{c}{One-electron states} \\
\hline
& $\ket{1000}$ & $\ket{0100}$ & $\ket{0010}$ & $\ket{0001}$ \\
$\ket*{1,D^1_{+,\uparrow}}$ & $\frac{1}{\sqrt{2}}$ & 0 & $\frac{1}{\sqrt{2}}$ & 0 \\
$\ket*{1,D^1_{+,\downarrow}}$ & 0 & $\frac{1}{\sqrt{2}}$ & 0 & $\frac{1}{\sqrt{2}}$ \\
$\ket*{1,D^1_{-,\uparrow}}$ & $\frac{1}{\sqrt{2}}$ & 0 & $-\frac{1}{\sqrt{2}}$ & 0 \\
$\ket*{1,D^1_{-,\downarrow}}$ & 0 & $\frac{1}{\sqrt{2}}$ & 0 & $-\frac{1}{\sqrt{2}}$ \\
\hline\hline
\end{tabular} 
\hspace{\Sp}
\begin{tabular}{ ccccc  }
\hline\hline
\multicolumn{5}{c}{Two-electron states} \\
\hline
& $\ket{1100}$ & $\ket{1001}$ & $\ket{0110}$ & $\ket{0011}$ \\
$\ket*{2,S^2_+}$ & $c_2$ & $c_1$ & $-c_1$ & $c_2$ \\
$\ket*{2,T^2_0}$ & 0 & $\frac{1}{\sqrt{2}}$ & $\frac{1}{\sqrt{2}}$ & 0  \\
$\ket*{2,S^2_{\mathrm{CS}}}$ & $-\frac{1}{\sqrt{2}}$ & 0 & 0 & $\frac{1}{\sqrt{2}}$ \\
$\ket*{2,S^2_-}$ & $c_1$ & $-c_2$ & $c_2$ & $c_1$ \\
\hline\hline
\end{tabular}
\hspace{\Sp}
\begin{tabular}{ ccccc  }
\hline\hline
\multicolumn{5}{c}{Three-electron states} \\
\hline
& $\ket{1110}$ & $\ket{1101}$ & $\ket{1011}$ & $\ket{0111}$ \\
$\ket*{3,D^3_{-,\uparrow}}$ & $\frac{1}{\sqrt{2}}$ & 0 & $-\frac{1}{\sqrt{2}}$ & 0 \\
$\ket*{3,D^3_{-,\downarrow}}$ & 0 & $\frac{1}{\sqrt{2}}$ & 0 & $-\frac{1}{\sqrt{2}}$ \\
$\ket*{3,D^3_{+,\uparrow}}$ & $\frac{1}{\sqrt{2}}$ & 0 & $\frac{1}{\sqrt{2}}$ & 0 \\
$\ket*{3,D^3_{+,\downarrow}}$ & 0 & $\frac{1}{\sqrt{2}}$ & 0 & $\frac{1}{\sqrt{2}}$ \\
\hline\hline
\end{tabular}
\caption{Basis transformation between site basis and eigenbasis, with $c_1=\frac{1}{2}\sqrt{1+\frac{U}{x}}$, $c_2=\frac{1}{2}\sqrt{1-\frac{U}{x}}$, and $x=\sqrt{U^2+16t^2}$.}
\label{tab:basis_trans}
\end{table*}
Using a basis transformation from the site basis to eigenbasis (Table~\ref{tab:basis_trans}), we can derive the time evolution of the Hubbard operators as
\begin{align}
\label{eq:basis_trans}
    \ket{N_a,a} &= \sum_{p} U_{p,a} \ket{p} \\
\label{eq:HO}
    \hat{X}^{q',p'}(-\tau) &= e^{-i\hat{H}_{\mathrm{sys}}\tau/\hbar} \ket{q'}\bra{p'} e^{i\hat{H}_{\mathrm{sys}}\tau/\hbar} \nonumber \\
    &= \sum_{ab} U_{a,q'}^{\dagger}U_{p',b} e^{-i\varepsilon_{ba}\tau/\hbar} \ket{N_a,a}\bra{N_b,b} \nonumber \\
    &= \sum_{ab} U_{a,q'}^{\dagger}U_{p',b} e^{-i\varepsilon_{ba}\tau/\hbar} \hat{X}^{a,b},
\end{align} 
where $\ket*{N_a,a}$ is the eigenstate for the system Hamiltonian $\hat{H}_{\mathrm{sys}}$. For simplicity, we use the notation $\hat{X}^{a,b} = \ket*{N_a,a}\bra*{N_b,b}$ and $\varepsilon_{ba}=\varepsilon_a-\varepsilon_b$, where $\varepsilon_a$ is the energy for state $\ket*{N_a,a}$. We derive the Redfield equation spanned on eigenbasis,
\begin{widetext}
    \begin{align}
    \label{eq:RT_energy}
    \mathcal{R}_{\mathrm{lead}} \hat{\rho}_{\mathrm{sys}}(t) = &\;\frac{i}{\hbar} \sum_{p<q} \sum_{p'<q'} \sum_{(l,i)} \sum_{k} \sum_{\sigma} \sum_{abcd} \int_{0}^{\infty} \mathrm{d}\tau \hspace{0.1cm} \hat{X}^{a,b} \nonumber \\
    &\Big\{ \;\; e^{-i\varepsilon_{cd}\tau/\hbar} g^<_{lk\sigma}(-\tau) \abs{T_{lk,i}}^2 (r_{i\sigma})_{p,q} (r_{i\sigma})_{p',q'} U_{a,p}^{\dagger}U_{q,d}U_{d,q'}^{\dagger}U_{p',c} \times \rho_{cb} \nonumber \\
    &- e^{-i\varepsilon_{ca}\tau/\hbar} g^<_{lk\sigma}(-\tau) \abs{T_{lk,i}}^2 (r_{i\sigma})_{p,q} (r_{i\sigma})_{p',q'} U_{d,p}^{\dagger}U_{q,b}U_{a,q'}^{\dagger}U_{p',c} \times \rho_{cd} \nonumber \\
    &- e^{-i\varepsilon_{cd}\tau/\hbar} g^>_{lk\sigma}(\tau) \abs{T_{lk,i}}^2 (r_{i\sigma})_{p,q} (r_{i\sigma})_{p',q'} U_{a,q}^{\dagger}U_{p,d}U_{d,p'}^{\dagger}U_{q',c} \times \rho_{cb} \nonumber \\
    &+ e^{-i\varepsilon_{ca}\tau/\hbar} g^>_{lk\sigma}(\tau) \abs{T_{lk,i}}^2 (r_{i\sigma})_{p,q} (r_{i\sigma})_{p',q'} U_{d,q}^{\dagger}U_{p,b}U_{a,p'}^{\dagger}U_{q',c} \times  \rho_{cd} \;\; \Big\} \nonumber \\    
    + &\;\mathrm{H.c.}.
    \end{align}
\end{widetext}

\indent
Next, we do the Laplace transform of the lesser (greater) Green's functions in Eq.~\eqref{eq:RT_energy}, i.e.,
\begin{subequations}
\label{eq:green}
\begin{align}
\label{eq:lesser_green_LT}
    g^{<}_{lk\sigma}(\epsilon) &\equiv \int_0^\infty e^{i\epsilon \tau/\hbar} g^{<}_{lk\sigma}(\tau) \mathrm{d}\tau \nonumber \\
    &= \frac{i}{\hbar} f_{l}(\xi_{k\sigma})  \Big[i\hbar\mathbf{P}\frac{1}{\epsilon-\xi_{k\sigma}}+\pi\hbar\delta(\epsilon-\xi_{k\sigma}) \Big] \\
\label{eq:greater_green_LT}
    g^{>}_{lk\sigma}(\epsilon) &\equiv \int_0^\infty e^{i\epsilon \tau/\hbar} g^{>}_{lk\sigma}(\tau) \mathrm{d}\tau \nonumber \\
    &= -\frac{i}{\hbar} [1-f_{l}(\xi_{k\sigma})]  \Big[i\hbar\mathbf{P}\frac{1}{\epsilon-\xi_{k\sigma}}+\pi\hbar\delta(\epsilon-\xi_{k\sigma}) \Big],
\end{align}    
\end{subequations}
where $\mathbf{P}$ represents the Cauchy principal value. The lesser (greater) Green's function contains both electron (hole) injection and energy shift of the system due to the leads. Then, we organize Eq.~\eqref{eq:RT_energy} by utilizing a redefined coupling,
\begin{subequations}
\label{eq:redef_coup}
\begin{align}
    V_{lk\sigma,ab} &\equiv T_{lk,i}^* \mel{N_a,a}{\hat{c}_{i\sigma}}{N_b,b} \\
    &= T_{lk,i}^* \sum_{p<q} (r_{i\sigma})_{p,q} U_{a,p}^{\dagger}U_{q,b} \\
    &= T_{lk,i}^* ~ \mathrm{Tr}[\hat{c}_{i\sigma}\hat{X}^{b,a}],
\end{align}
\end{subequations}
and the lesser (greater) self-energy $\Sigma^{(l),\lessgtr}_{ca,db}(\varepsilon_{ac}) = \sum_{k,\sigma} V_{lk\sigma,ca}^* \times g^{\lessgtr}_{lk\sigma}(\varepsilon_{ac}) \times V_{lk\sigma,db}$. Note that the order of the subscript of the self-energy represents the transition from the $N$-electron state to the $(N+1)$-electron state. Finally, we obtain the Redfield-type fermionic quantum master equation,
\begin{alignat}{5}
    &\bra{N_a,a}\mathcal{R}_{\mathrm{lead}} \hat{\rho}_{\mathrm{el}}(t)\ket{N_b,b} \nonumber \\
\label{eq:RTensor_Ebasis}
    = &\sum_{cd} \mathcal{R}_{ab,cd} \rho_{cd} \\
\label{eq:RT_Lead}
    = &- \frac{i}{\hbar} \sum_{l} 
    \nonumber \\
    &\hspace{0.3cm} \bigg\{ \sum_{cd} \Big[ \Sigma^{(l),<}_{db,ca}(\varepsilon_{db})~&-&\hspace{0.15cm} \left(\Sigma^{(l),<}_{ca,db}(\varepsilon_{ca})\right)^*  \Big]\rho_{cd}&&  
    \nonumber \\
    &\hspace{0.3cm}+ \sum_{cd} \Big[ -\Sigma^{(l),>}_{bd,ac}(\varepsilon_{ac})~&+&\hspace{0.15cm} \left(\Sigma^{(l),>}_{ac,bd}(\varepsilon_{bd})\right)^*  \Big]\rho_{cd}&& 
    \nonumber \\
    &\hspace{0.3cm}- \sum_{cde} \Big[ \Sigma^{(l),<}_{de,be}(\varepsilon_{de}) \delta_{a,c}~&-&\hspace{0.15cm} \left(\Sigma^{(l),<}_{ce,ae}(\varepsilon_{ce})\right)^* \delta_{b,d} \Big]\rho_{cd}&&  
    \nonumber \\
    &\hspace{0.3cm}- \sum_{cde} \Big[ -\Sigma^{(l),>}_{ea,ec}(\varepsilon_{ec}) \delta_{b,d}~& +&\hspace{0.15cm} \left(\Sigma^{(l),>}_{eb,ed}(\varepsilon_{ed})\right)^* \delta_{a,c} \Big]\rho_{cd}&& \bigg\}     
\end{alignat}
\begin{figure*}[!t]
    \centering
    {\includegraphics[width=0.9\textwidth]{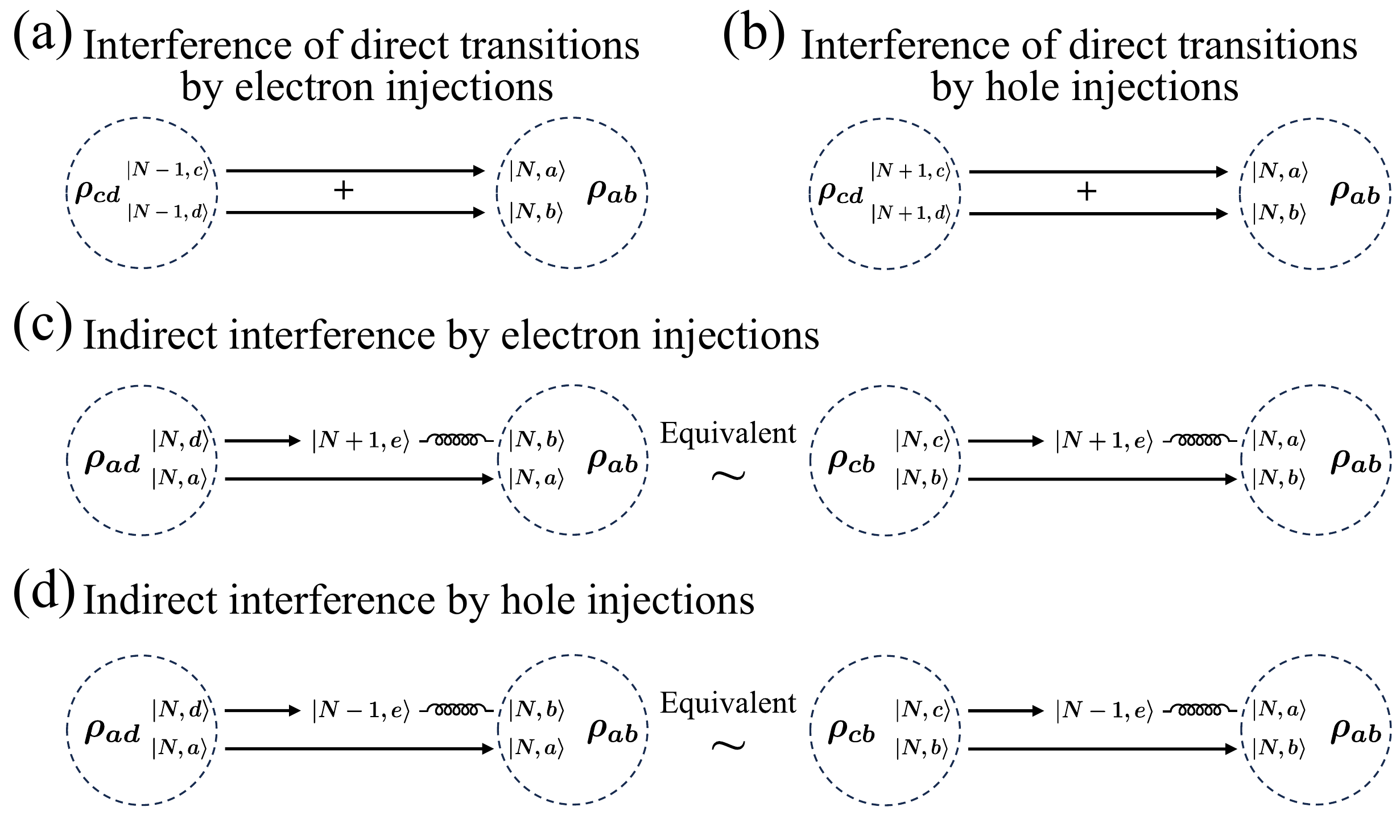}}
    \caption{Illustration of four main processes in Eq.~\eqref{eq:RTensor_Ebasis}. (a) Interference of direct transitions by electron injections $\mathcal{R}_{ab,cd}^{\RomanNumeralCaps{1}}\rho_{cd} = - \frac{i}{\hbar} \sum_{l} \left[ \Sigma^{(l),<}_{db,ca}(\varepsilon_{db}) - \left(\Sigma^{(l),<}_{ca,db}(\varepsilon_{ca})\right)^*  \right]\rho_{cd}$ (b) Interference of direct transitions by hole injections $\mathcal{R}_{ab,cd}^{\RomanNumeralCaps{2}}\rho_{cd} = - \frac{i}{\hbar} \sum_{l} \left[ -\Sigma^{(l),>}_{bd,ac}(\varepsilon_{ac})+ \left(\Sigma^{(l),>}_{ac,bd}(\varepsilon_{bd})\right)^*  \right]\rho_{cd}$ (c) Indirect interference by electron injections $\mathcal{R}_{ab,cd}^{\RomanNumeralCaps{3}}\rho_{cd} = - \frac{i}{\hbar} \sum_{l}\sum_{cde} \Big[ \Sigma^{(l),<}_{de,be}(\varepsilon_{de}) \delta_{a,c} - \left(\Sigma^{(l),<}_{ce,ae}(\varepsilon_{ce})\right)^* \delta_{b,d} \Big]\rho_{cd}$ (d) Indirect interference by hole injections $\mathcal{R}_{ab,cd}^{\RomanNumeralCaps{4}}\rho_{cd} = -\frac{i}{\hbar} \sum_{l} \sum_{cde} \Big[ -\Sigma^{(l),>}_{ea,ec}(\varepsilon_{ec}) \delta_{b,d} + \left(\Sigma^{(l),>}_{eb,ed}(\varepsilon_{ed})\right)^* \delta_{a,c} \Big]\rho_{cd}$ \label{fig:RT}}
\end{figure*}
\indent
In Fig.~\ref{fig:RT}, $\mathcal{R}_{ab,cd}$ in Eq.~\eqref{eq:RTensor_Ebasis} can be decomposed into four mechanisms $\mathcal{R}^{\RomanNumeralCaps{1}}$, $\mathcal{R}^{\RomanNumeralCaps{2}}$, $\mathcal{R}^{\RomanNumeralCaps{3}}$, and $\mathcal{R}^{\RomanNumeralCaps{4}}$.
The first mechanism $\mathcal{R}^{\RomanNumeralCaps{1}}$ (Fig.~\ref{fig:RT}a) is the quantum interference of state-to-state transitions $\ket{N_c=N-1,c} \rightarrow \ket{N_a=N,a}$ and $\ket{N_d=N-1,d} \rightarrow \ket{N_b=N,b}$ by one electron injection from the electrodes, and the second mechanism $\mathcal{R}^{\RomanNumeralCaps{2}}$ (Fig.~\ref{fig:RT}b) is the quantum interference of state-to-state transitions $\ket{N_c=N+1,c} \rightarrow \ket{N_a=N,a}$ and $\ket{N_d=N+1,d} \rightarrow \ket{N_b=N,b}$ by one hole injection. When $c=d$, $a=b$, these processes correspond to population transfer from $P_c \equiv \rho_{cc}$ to $P_a$ that causes an increase of population $P_a$ in PME~\cite{Datta2005}. The third mechanism $\mathcal{R}^{\RomanNumeralCaps{3}}$ (Fig.~\ref{fig:RT}c) is the indirect interference of state-to-state transition $\ket{N_a=N,a}/\ket{N_b=N,b} \rightarrow \ket{N_e=N+1,e}$ and state-to-state transition $\ket{N_c=N,c}/\ket{N_d=N,d} \rightarrow \ket{N_e=N+1,e}$ by one electron injection. The fourth mechanism  $\mathcal{R}^{\RomanNumeralCaps{4}}$ (Fig.~\ref{fig:RT}d) is the indirect interference of state-to-state transition $\ket{N_a=N,a}/\ket{N_b=N,b} \rightarrow \ket{N_e=N-1,e}$ and state-to-state transition $\ket{N_c=N,c}/\ket{N_d=N,d} \rightarrow \ket{N_e=N-1,e}$ by one hole injection. When $a=b=d$ ($a=b=c$) on the left (right) hand side of Fig.~\ref{fig:RT}c and \ref{fig:RT}d, these processes correspond to population decay of $P_a$ that decreases population $P_a$ in PME~\cite{Datta2005}.
\section{Expression for Steady-State Electric Current \label{sec:SS_current}}
\indent
In this section, we derive the steady-state electric current expression, i.e., Eq.~~(4), in the main text. From the definition of steady-state electric current \cite{haug2008quantum},
\begin{align}
\label{eq:I_def}
    I &\equiv \mathrm{e} \times \big( -\frac{\mathrm{d} \expval{\hat{N}_{\mathrm{L}}}}{\mathrm{d}t} \big) 
    \nonumber \\
    &= \mathrm{e} \times \frac{\mathrm{d}\expval{\hat{N}_{\mathrm{el}}}_{\mathrm{L}}}{\mathrm{d}t}
    \nonumber \\
    &= \mathrm{e} \times \frac{\mathrm{d}}{\mathrm{d}t} \mathrm{Tr} \left\{ \hat{N}_{\mathrm{el}}\hat{\rho}_{\mathrm{sys}} \right\}_{\mathrm{L}},
\end{align}
where $\expval{\hat{N}_\mathrm{L}}$ denotes the average number of electrons in the left electrode, $\expval{\hat{N}_\mathrm{el}}$ denotes the average number of electrons in the system, and the subscript $\mathrm{L}$ represents that we focus on change of the system due to electron (hole) injections from the left electrode. The second equality in Eq.~\eqref{eq:I_def} comes from the condition that all of the electrons leaving the left electrode enter the system. Since the number operator commutes with electronic Hamiltonian $\hat{H}_\mathrm{el}$, we derive a current expression spanned on the eigenbasis of $\hat{H}_\mathrm{el}$,
\begin{align}
\label{eq:I_energy}
    I &= \mathrm{e} \cdot \sum_{ab} \frac{\mathrm{d}}{\mathrm{d}t} \big[ \bra{b}\hat{N}_{\mathrm{el}}\ket{a}\bra{a}\hat{\rho}_{\mathrm{sys}}\ket{b} \big]_{\mathrm{L}} 
    \nonumber \\
    &= \mathrm{e} \cdot \sum_{ab} \frac{\mathrm{d}}{\mathrm{d}t} \big[ N_{a}\bra{b}\ket{a}\bra{a}\hat{\rho}_{\mathrm{sys}}\ket{b} \big]_{\mathrm{L}} 
    \nonumber \\
    &= \mathrm{e} \cdot \sum_{a} N_{a} \left( \frac{\mathrm{d}P_a}{\mathrm{d}t} \right)_{\mathrm{L}},
\end{align}
in which $P_a=\rho_{aa}$ denotes the population of state $\ket*{N_a,a}$. From Eq.~\eqref{eq:RT_Lead}, we obtain the dynamic equations of populations as
\begin{align}
\label{eq:population}
    \frac{\mathrm{d} P_a}{\mathrm{d}t} = \frac{2}{\hbar} \sum_{cd} &\sum_{l} \nonumber \\
    \mathrm{Im}\bigg\{ \hspace{0.1cm} &\Sigma^{(l),<}_{da,ca}(\varepsilon_{da}) \rho_{cd} \hspace{0.1cm}+\hspace{0.1cm} 
    \left(\Sigma^{(l),>}_{ac,ad}(\varepsilon_{ad})\right)^* \rho_{cd} \nonumber \\
    -\ \hspace{0.1cm}&\Sigma^{(l),<}_{cd,ad}(\varepsilon_{cd}) \rho_{ac} \hspace{0.1cm}-\hspace{0.1cm} \left(\Sigma^{(l),>}_{da,dc}(\varepsilon_{dc})\right)^* \rho_{ac} \bigg\}.
\end{align}
\indent
From Eq.~\eqref{eq:I_energy} and Eq.~\eqref{eq:population}, we derive an expression for steady-state electric current as
\begin{align}
\label{eq:I_form}
    I 
    = \frac{2\mathrm{e}}{\hbar} \sum_{acd}& \nonumber \\
    \mathrm{Im}\bigg\{ \hspace{0.1cm} &N_a \Sigma^{(\mathrm{L}),<}_{da,ca}(\varepsilon_{da}) \rho_{cd} \hspace{0.1cm}+\hspace{0.1cm} 
    N_a \left(\Sigma^{(\mathrm{L}),>}_{ac,ad}(\varepsilon_{ad})\right)^* \rho_{cd} \nonumber \\
    -\ \hspace{0.1cm} &N_a \Sigma^{(\mathrm{L}),<}_{cd,ad}(\varepsilon_{cd}) \rho_{ac} \hspace{0.1cm}-\hspace{0.1cm} N_a \left(\Sigma^{(\mathrm{L}),>}_{da,dc}(\varepsilon_{dc})\right)^* \rho_{ac} \bigg\} \nonumber \\
    = \frac{2\mathrm{e}}{\hbar} \sum_{acd}& \nonumber \\
    \mathrm{Im}\bigg\{ \hspace{0.1cm} &N_a \Sigma^{(\mathrm{L}),<}_{da,ca}(\varepsilon_{da}) \rho_{cd} \hspace{0.1cm}+\hspace{0.1cm} 
    N_a \left(\Sigma^{(\mathrm{L}),>}_{ac,ad}(\varepsilon_{ad})\right)^* \rho_{cd} \nonumber \\
    -\ \hspace{0.1cm} &N_c \Sigma^{(\mathrm{L}),<}_{da,ca}(\varepsilon_{da}) \rho_{cd} \hspace{0.1cm}-\hspace{0.1cm} N_c \left(\Sigma^{(\mathrm{L}),>}_{ac,ad}(\varepsilon_{ad})\right)^* \rho_{cd} \bigg\} \nonumber \\
    = \frac{2\mathrm{e}}{\hbar} \sum_{acd}& \mathrm{Im} \bigg\{ \Big[ \Sigma^{(\mathrm{L}),<}_{da,ca}(\varepsilon_{da}) - \left(\Sigma^{(\mathrm{L}),>}_{ac,ad}(\varepsilon_{ad})\right)^* \Big] \cdot \rho_{cd} \bigg\} 
\end{align}

\section{Effect of Phonons on Many-Body Coherence \label{sec:vib}}
\indent
In this section, we explore phonon effects on electronic coherence in a transport system. The two-site Hubbard model with the phonon bath~\cite{Toyozawa1981,Hsu2010} can be written as
\begin{equation}
    \hat{H}_{\mathrm{sys}} = \hat{H}_{\mathrm{el}}+\hat{H}_{\mathrm{ph}}+\hat{H}_{\mathrm{el-ph}}, 
\end{equation}
which is composed of the electronic Hamiltonian $\hat{H}_{\mathrm{el}}$ that represents the two-site Hubbard model, the phonon Hamiltonian $\hat{H}_{\mathrm{ph}}$, and the electron-phonon coupling $\hat{H}_{\mathrm{el-ph}}$. We consider the phonon Hamiltonian and the electron-phonon coupling as
\begin{subequations}
\begin{align}
    \hat{H}_{\mathrm{ph}} &= \sum_{\alpha} \hbar \omega_{\alpha} (\hat{b}_{\alpha}^{\dagger}\hat{b}_{\alpha} + \frac{1}{2}) \\
    \hat{H}_{\mathrm{el-ph}} &= g\sum_{i,\sigma,\alpha} \hat{c}_{i\sigma}^{\dagger} \hat{c}_{i\sigma} (\hat{b}_{\alpha}^{\dagger} + \hat{b}_{\alpha}),
\end{align}
\end{subequations}
where $\omega_{\alpha}$ and $\hat{b}_\alpha^\dagger$ ($\hat{b}_\alpha$) stand for phonon frequency and bosonic creation (annihilation) operators of the phonon mode $\alpha$, respectively.
\newline \indent
To derive the dynamic equation for electron transport with the effect of phonons, we begin from the quantum Liouville equation for the system in Eq.~\eqref{eq:REquation}. In the interaction picture, the dynamic equation becomes
\begin{align}
    \label{eq:inter_vib}
    \frac{\mathrm{d} \hat{\tilde{\rho}}_{\mathrm{sys}}(t)}{\mathrm{d} t} 
    = &-\frac{i}{\hbar} [\hat{\tilde{H}}_{\mathrm{el-ph}}(t),\hat{\tilde{\rho}}_{\mathrm{sys}}(t)] \nonumber \\
    &+ e^{i(\hat{H}_\mathrm{el}+\hat{H}_\mathrm{ph})t/\hbar} \mathcal{R}_\mathrm{lead} \hat{\rho}_{\mathrm{sys}}(t) e^{-i(\hat{H}_\mathrm{el}+\hat{H}_\mathrm{ph})t/\hbar}.
\end{align}
\noindent
The system density matrix can be divided into the electronic part and phonon part, i.e., $\hat{\rho}_\mathrm{sys}(t) = \hat{\rho}_{\mathrm{el}}(t) \otimes \hat{\rho}_{\mathrm{ph}}(t)$. Following the similar procedures in Eqs.~\eqref{eq:dyson_series}, \eqref{eq:DM_BA}, \eqref{eq:lead_TE_int}, \eqref{eq:coup_DP}, and \eqref{eq:nonMarkovian}, we can obtain the dynamic equation as
\begin{widetext}
    \begin{align}
    \label{eq:nonMarkovian_vib}
    \frac{\mathrm{d} \hat{\tilde{\rho}}_{\mathrm{el}}(t)}{\mathrm{d} t} = \mathrm{Tr}_\mathrm{ph} \bigg\{ \frac{\mathrm{d} \hat{\tilde{\rho}}_{\mathrm{sys}}(t)}{\mathrm{d} t} \bigg\} = &- \frac{1}{\hbar^2} \int_{0}^{t-t_0} \mathrm{d}\tau_2 \mathrm{Tr}_\mathrm{ph} \bigg\{ \Big[ \hat{\tilde{H}}_{\mathrm{el-ph}}(t),\ \left[ \hat{\tilde{H}}_{\mathrm{el-ph}}(t-\tau_2),\ \hat{\tilde{\rho}}_{\mathrm{el}}(t-\tau_2) \otimes \hat{\bar{\sigma}}_\mathrm{ph} \right]\Big] \bigg\} \nonumber \\
    &- \frac{i}{\hbar} \int_{t_0}^t \mathrm{d}t_2 \mathrm{Tr}_\mathrm{ph} \bigg\{ \left[ \hat{\tilde{H}}_{\mathrm{el-ph}}(t), e^{i(\hat{H}_\mathrm{el}+\hat{H}_\mathrm{ph})t_2/\hbar} \mathcal{R}_\mathrm{lead} \hat{\rho}_{\mathrm{el}}(t_2) \otimes \hat{\bar{\sigma}}_\mathrm{ph} e^{-i(\hat{H}_\mathrm{el}+\hat{H}_\mathrm{ph})t_2/\hbar}\right] \bigg\} \nonumber \\
    &+ \mathrm{Tr}_\mathrm{ph} \bigg\{e^{i(\hat{H}_\mathrm{el}+\hat{H}_\mathrm{ph})t/\hbar} \mathcal{R}_\mathrm{lead} \hat{\rho}_{\mathrm{el}}(t) \otimes \hat{\bar{\sigma}}_\mathrm{ph} e^{-i(\hat{H}_\mathrm{el}+\hat{H}_\mathrm{ph})t/\hbar} \bigg\}.
    \end{align}
\end{widetext}
\indent
The first line in Eq.~\eqref{eq:nonMarkovian_vib} is a typical term in a non-Markovian master equation for the description of electron-phonon coupling. By tracing out the phonon degrees of freedom, the second line can be simplified as
\begin{align}
    &\mathrm{Tr}_\mathrm{ph} \bigg\{ \Big[ \hat{\tilde{H}}_{\mathrm{el-ph}}(t), \nonumber \\
    &\hspace{1.2cm} e^{i(\hat{H}_\mathrm{el}+\hat{H}_\mathrm{ph})t_2/\hbar} \mathcal{R}_\mathrm{lead} \hat{\rho}_{\mathrm{el}}(t_2) \otimes \hat{\bar{\sigma}}_\mathrm{ph} e^{-i(\hat{H}_\mathrm{el}+\hat{H}_\mathrm{ph})t_2/\hbar}\Big] \bigg\} \nonumber \\
    =~ &\mathrm{Tr}_\mathrm{ph} \bigg\{ \Big[ \hat{H}_{\mathrm{el-ph}},  \mathcal{R}_\mathrm{lead} \hat{\rho}_{\mathrm{el}}(t_2) \otimes \hat{\bar{\sigma}}_\mathrm{ph} \Big] \bigg\}.
\end{align}
\indent
When we trace out the phonon degrees of freedom, the second line in Eq.~\eqref{eq:nonMarkovian_vib} becomes zero because $\mathrm{Tr}_\mathrm{ph}\{ \hat{H}_\mathrm{el-ph} \hat{\rho}_{\mathrm{el}}(t_2) \otimes \hat{\bar{\sigma}}_\mathrm{ph} \} = 0$ and  $\mathcal{R}_\mathrm{lead}$ is independent of the phonon degrees of freedom since the system-lead coupling does not influence the phonon degrees of freedom under the weak coupling condition, i.e., $\hat{H}_\mathrm{sys-lead} = \sum_{\sigma} (\hat{c}_{1\sigma}^{\dagger} \otimes \hat{\mathbb{I}}_\mathrm{ph} \otimes \sum_k T_{\mathrm{L}k,1}\hat{d}_{\mathrm{L}k\sigma} + \hat{c}_{2\sigma}^{\dagger} \otimes \hat{\mathbb{I}}_\mathrm{ph} \otimes \sum_k T_{\mathrm{R}k,2} \hat{d}_{\mathrm{R}k\sigma} + \mathrm{H.c.})$
The last term in Eq.~\eqref{eq:nonMarkovian_vib} is equivalent to $\mathcal{R}_\mathrm{lead} \hat{\rho}_{\mathrm{el}}(t)$. 
After applying the Markov approximations, we derive the dynamic equation in the Schr\"{o}dinger picture as follows,
\begin{widetext}
    \begin{align}
    \label{eq:REtotal}
    \frac{\mathrm{d} \hat{\rho}_{\mathrm{el}}(t)}{\mathrm{d} t} &= -\frac{i}{\hbar} [\hat{H}_{\mathrm{el}},\hat{\rho}_{\mathrm{el}}(t)] + \mathcal{R}_\mathrm{lead} \hat{\rho}_{\mathrm{el}}(t) + \mathcal{R}_\mathrm{ph} \hat{\rho}_{\mathrm{el}}(t) \\
    \label{eq:RTlead_el}
    \mathcal{R}_\mathrm{lead} \hat{\rho}_{\mathrm{el}}(t) &= -\frac{1}{\hbar^2} \int_{0}^{\infty} \mathrm{d}\tau \mathrm{Tr}_\mathrm{lead} \bigg\{ \Big[ \hat{H}_\mathrm{sys-lead}(0),\ \big[ \hat{H}_\mathrm{sys-lead}(-\tau),\ \hat{\rho}_\mathrm{el}(t)\otimes\hat{\bar{\sigma}}_\mathrm{lead} \big] \Big] \bigg\} \\
    \label{eq:RTph_el}
    \mathcal{R}_\mathrm{ph} \hat{\rho}_{\mathrm{el}}(t) &= - \frac{1}{\hbar^2} \int_{0}^{\infty} \mathrm{d}\tau_2 \mathrm{Tr}_\mathrm{ph} \bigg\{ \Big[ \hat{H}_{\mathrm{el-ph}}(0),\ \left[ \hat{H}_{\mathrm{el-ph}}(-\tau_2),\ \hat{\rho}_{\mathrm{el}}(t) \otimes \hat{\bar{\sigma}}_\mathrm{ph} \right]\Big] \bigg\}.
    \end{align}
\end{widetext}

\indent
In the following context, we focus on the term $\mathcal{R}_\mathrm{ph} \hat{\rho}_\mathrm{el}(t)$. The electron-phonon coupling $\hat{H}_{\mathrm{el-ph}}$ can be reformulated by Hubbard operators,
\begin{align}
\label{eq:coup_ph_HO}
    \hat{H}_{\mathrm{el-ph}} &= \sum_{i} \sum_{\sigma} \sum_p (n_{i\sigma})_p \hat{X}^{p,p} \otimes g \sum_{\alpha} (\hat{b}_{\alpha}^{\dagger} + \hat{b}_{\alpha}),
\end{align}
where $(n_{i\sigma})_p$ represents the occupation of an electron on site $i$ with spin $\sigma$ in state $\ket*{p}$, and $\alpha$ represents the vibrational modes of phonons. We define the phonon correlation function in Eq.~\eqref{eq:RTph_el} as
\begin{align}
\label{eq:correl_ph}
    \mathrm{C}(\tau_2) \equiv \mathrm{Tr}_{\mathrm{ph}} \Big\{ g^2 &\sum_{\alpha_1} \big[ \hat{b}_{\alpha_1}^\dagger(0) + \hat{b}_{\alpha_1}(0) \big] \nonumber \\
    \times &\sum_{\alpha_2} \big[ \hat{b}_{\alpha_2}^\dagger (-\tau_2) + \hat{b}_{ \alpha_2} (-\tau_2) \big] \hat{\bar{\sigma}}_{\mathrm{ph}} \Big\}.
\end{align}
\begin{figure}[b!]
    \centering
    {\includegraphics[width=0.5\textwidth]{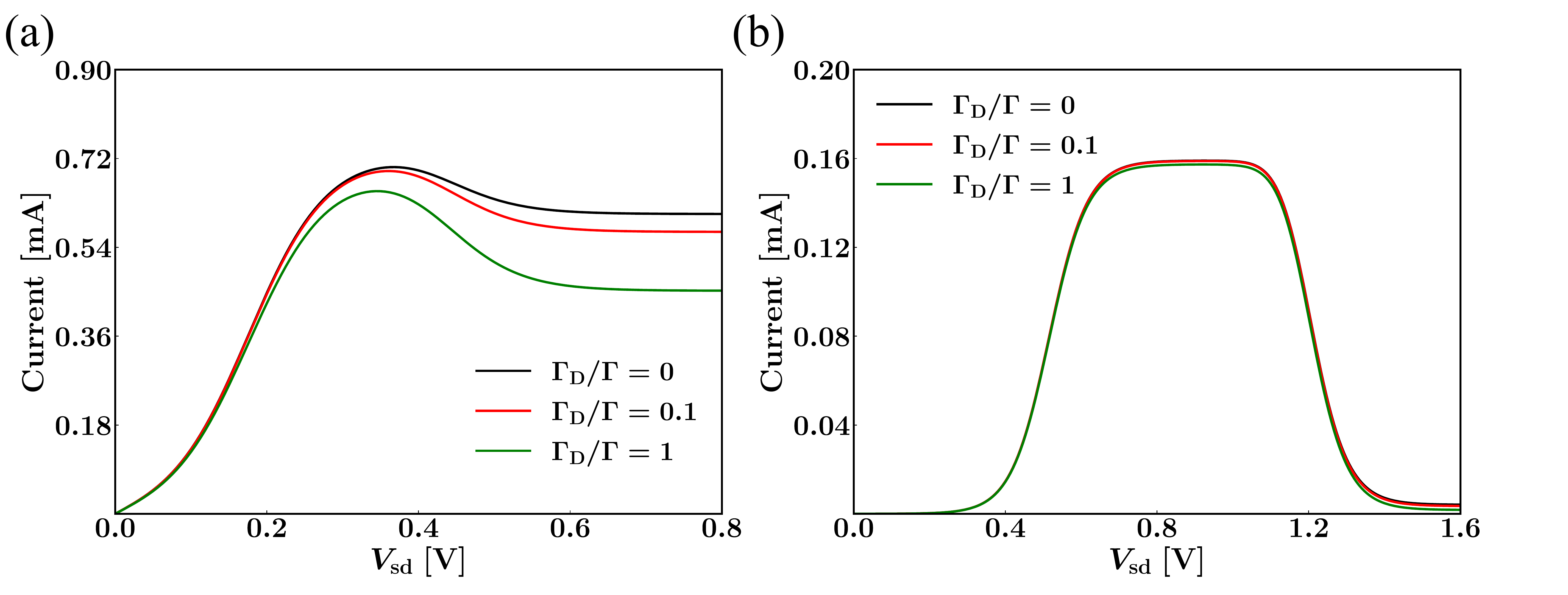}}
    \caption{Current-voltage characteristics with different ratio of $\Gamma_\mathrm{D}/\Gamma$ in an AH system \cite{Koole2016} for (a) $\varepsilon = 0.1\ \mathrm{eV}$, $t = 0.01\ \mathrm{eV}$, $U = 0.08\ \mathrm{eV}$, $\Gamma = 0.005\ \mathrm{eV}$ and in a model system for (b) $\varepsilon = -0.25\ \mathrm{eV}$, $t = 0.005\ \mathrm{eV}$, $U = 0.8\ \mathrm{eV}$, $\Gamma = 0.001\ \mathrm{eV}$. Other parameters are $T=300\ \mathrm{K}$, $V_\mathrm{g}=0\ \mathrm{V}$.\label{fig:1}}
\end{figure}
\indent
In this work, we adopt a phonon spectral density $A_+(\epsilon)$ used in the previous work \cite{Segal2000}. The phonon spectral density originates from the one-sided Fourier transform of the correlation function $\mathrm{C}(\tau_2)$, 
\begin{align}
\label{eq:correl_FT}
    A_+(\epsilon) &\equiv \frac{1}{\hbar} \int_0^\infty \mathrm{d}\tau_2 \mathrm{C}(\tau_2) e^{i\epsilon \tau_2/\hbar} \nonumber \\
    &= \frac{1}{2} \Gamma_\mathrm{D} e^{-\epsilon^2\tau_\mathrm{c}^2/4} e^{-\beta_\mathrm{ph} (\abs{\epsilon}-\epsilon)},
\end{align}
where $\Gamma_\mathrm{D}$ represents a coupling constant, which is proportional to the square of electron-phonon coupling $g^2$. In addition, $\tau_\mathrm{c}$ represents the correlation time in the phonon baths, and $\beta_\mathrm{ph} = 1/kT_\mathrm{ph}$ represents the reciprocal of the thermodynamic temperature of the phonon baths. We then transform the Hubbard operators in Eq.~\eqref{eq:coup_ph_HO} to eigenbasis of the system, and the Redfield tensor for phonons in Eq.~\eqref{eq:RTph_el} can be written as 
\begin{align}
\label{eq:RTphonon}
    \mathcal{R}_{\mathrm{ph}} &\hat{\rho}_{\mathrm{el}}(t) = \nonumber \\
    -&\frac{1}{\hbar} \sum_{p p'} \sum_i \sum_{\sigma \sigma'} \sum_{abcd} \hat{X}^{a,b} \nonumber \\
    \Big\{ &\big[U_{p,d}(n_{i\sigma})_{p}U_{a,p}^{\dagger}\big]\big[U_{d,p'}^{\dagger}(n_{i\sigma'})_{p'}U_{p',c} \big] A_+(\varepsilon_{dc}) \rho_{cb} \nonumber \\
    - &\big[U_{a,p}^{\dagger}(n_{i\sigma})_{p}U_{p,c}\big]\big[U_{p',b}(n_{i\sigma'})_{p'}U_{d,p'}^{\dagger}\big] A_+(\varepsilon_{ac}) \rho_{cd}\Big\} \nonumber \\
    + &\ \mathrm{H.c.}
\end{align}
\indent
The current variations with bias voltage under different ratio of $\Gamma_\mathrm{D}/\Gamma$ are plotted in Fig.~\ref{fig:1}, where $\Gamma$ specifies the system-lead coupling strength as stated in the main text. Both Fig.~\ref{fig:1}a and Fig.~\ref{fig:1}b show evident current blockade (current suppression) when $\Gamma_\mathrm{D}$ is smaller than or equal to $\Gamma$. The results support that the unique phenomenon due to many-body coherence is robust against vibrational relaxation and decoherence.
\section{Analytical Expression for Current-Coherence Relationship and Coherence-Gate Relationship \label{sec:ana}}
\indent
The number of equations in Eq.~\eqref{eq:RME} includes $16 \times 16 = 256$. It is almost impossible to get an analytical solution. 
In order to obtain the analytical expression for current blockade with weak hopping and strong Coulomb repulsion, we need to make two key assumptions: (1) $f_{\rm{L}}(\varepsilon_{ca})=1$ and $f_{\rm{R}}(\varepsilon_{ca})=0$, and (2) we only keep many-body coherence $\rho_{S_{+}^2, T_0^2}$, $\rho_{S_{+}^2, T_{+1}^2}$, $\rho_{S_{+}^2, T_{-1}^2}$ and $\rho_{S_{\mathrm{CS}}^2, S_-^2}$. 
For the first assumption, we consider that $\mathrm{e}V_\mathrm{sd} > U$ in the zero-temperature limit to include Coulomb repulsion. Furthermore, for simplicity, we ignore the influence of $\varepsilon$ and $t$ on the Fermi function. Under this condition, one can approximate  $f_{\rm{L}}(\varepsilon_{ac})=1$ and $f_{\rm{R}}(\varepsilon_{ac})=0$. For the second assumption, it is well-known that coherence can be neglected for a large energy gap between two states. In the case of small $t/U$, we consider these coherence terms, i.e., $\rho_{S^2_+, T^2_0}$, $\rho_{S^2_+, T^2_{+1}}$, $\rho_{S^2_+, T^2_{-1}}$, and $\rho_{S^2_\mathrm{CS}, S^2_-}$ because the energy gap between $\ket*{2,S^2_\mathrm{CS}}$ and $\ket*{2,S^2_-}$ and the energy gap between $\ket*{2,S^2_+}$ and triplet states $\ket*{2,T^2_0}$, $\ket*{2,T^2_{+1}}$, $\ket*{2,T^2_{-1}}$ are the smallest. These two assumptions will reduce $256$ equations to $20$ equations, including $16$ equations for the state populations and $4$ equations for coherence.
The $16$ population terms include $P_{S^0}$, $P_{D^1_{+,\uparrow}}$, $P_{D^1_{+,\downarrow}}$, $P_{D^1_{-,\uparrow}}$, $P_{D^1_{-,\downarrow}}$, $P_{S^2_+}$, $P_{T^2_0}$, $P_{T^2_{+1}}$, $P_{T^2_{-1}}$, $P_{S^2_{\mathrm{CS}}}$, $P_{S^2_-}$, $P_{D^3_{+,\uparrow}}$, $P_{D^3_{+,\downarrow}}$, $P_{D^3_{-,\uparrow}}$, $P_{D^3_{-,\downarrow}}$, and $P_{S^4}$ while the $4$ coherence terms include $\rho_{S^2_+,T^2_0}$, $\rho_{S^2_+,T^2_{+1}}$, $\rho_{S^2_+,T^2_{-1}}$, and $\rho_{S^2_\mathrm{CS},S^2_-}$. We do not list all $20$ equations here because their expressions are so complicated. \\
\newline 
\indent
In the following derivation, we will adopt the index $\tau$ for time in the reduced equations of motion. First, we can easily find that both coherence $\rho_{S^2_+, T^2_{+1}}$ and $\rho_{S^2_+, T^2_{-1}}$ are not affected by steady-state populations, 
\begin{align}
\label{eq:S+T+1}
    \drhodt{\rho_{S^2_+, T^2_{+1}}} &= \frac{i}{\hbar} \times \frac{x-U}{2}  \rho_{S_{+}^{2}, T_{+1}^{2}} - \frac{2\Gamma}{\hbar} \rho_{S_{+}^{2}, T_{+1}^{2}} \\
\label{eq:S+T-1}
    \drhodt{\rho_{S^2_+, T^2_{-1}}} &= \frac{i}{\hbar} \times \frac{x-U}{2}  \rho_{S_{+}^{2}, T_{-1}^{2}} - \frac{2\Gamma}{\hbar} \rho_{S_{+}^{2}, T_{-1}^{2}},
\end{align}
where $x = \sqrt{U^2 + 16t^2}$. Therefore, at steady state, both coherence terms decay to $0$ as a consequence. As for the other coherence between singlet and triplet $\rho_{S^2_+, T^2_0}$, the dynamic equation can be derived as
\begin{widetext}
    \begin{align}
    \label{eq:S+T0}
    \drhodt{\rho_{S^2_+, T^2_0}} &= \frac{i}{\hbar} \times \frac{x-U}{2}  \rho_{S_{+}^{2}, T_{0}^{2}} - \frac{2\Gamma}{\hbar} \rho_{S_{+}^{2}, T_{0}^{2}} \nonumber \\
    &-\ \frac{\Gamma}{4\hbar} \Big\{ \sqrt{1+\frac{4t}{x}} \big(P_{D^1_{+,\uparrow}} -P_{D^1_{+,\downarrow}}\big) + \sqrt{1-\frac{4t}{x}} \big(P_{D^1_{-,\uparrow}} - P_{D^1_{-,\downarrow}}\big) \Big\} \nonumber \\
    &+\ \frac{\Gamma}{4\hbar} \Big\{ \sqrt{1+\frac{4t}{x}} \big(P_{D^3_{-,\uparrow}} - P_{D^3_{-,\downarrow}}\big) + \sqrt{1-\frac{4t}{x}} \big(P_{D^3_{+,\uparrow}} - P_{D^3_{+,\downarrow}}\big) \Big\},
    \end{align}
\end{widetext}
in which $P_a \equiv \rho_{a,a}$ denotes the population of state $\ket*{N_a,a}$. In order to solve Eq.~\eqref{eq:S+T0}, we need several auxiliary equations related to the difference between the time derivative of populations $P_{D^1_{+,\uparrow}}$, $P_{D^1_{+,\downarrow}}$, $P_{D^1_{-,\uparrow}}$, $P_{D^1_{-,\downarrow}}$, $P_{D^3_{+,\uparrow}}$, $P_{D^3_{+,\downarrow}}$, $P_{D^3_{-,\uparrow}}$, and $P_{D^3_{-,\downarrow}}$ at steady state in Eq.~(1) in the main text. The auxiliary equations are listed as follows,
\begin{widetext}
\begin{subequations}
\label{eq:contraction}
    \begin{align}
    \label{eq:1D+u_D+d}
        \drhodt{P_{D^1_{+,\uparrow}}}-\drhodt{P_{D^1_{+,\downarrow}}} = &-\frac{2\Gamma}{\hbar}(P_{D^1_{+,\uparrow}} - P_{D^1_{+,\downarrow}}) + \frac{\Gamma}{2\hbar}(P_{T^2_{+1}} - P_{T^2_{-1}}) + \frac{\Gamma}{\hbar} \sqrt{1+\frac{4t}{x}} \rho_{S_+^2,T_0^2} \\
    \label{eq:1D-u_D-d}
        \drhodt{P_{D^1_{-,\uparrow}}}-\drhodt{P_{D^1_{-,\downarrow}}} = &-\frac{2\Gamma}{\hbar}(P_{D^1_{-,\uparrow}} - P_{D^1_{-,\downarrow}}) + \frac{\Gamma}{2\hbar}(P_{T^2_{+1}} - P_{T^2_{-1}}) + \frac{\Gamma}{\hbar} \sqrt{1-\frac{4t}{x}} \rho_{S_+^2,T_0^2} \\
    \label{eq:3D+u_D+d}
        \drhodt{P_{D^3_{+,\uparrow}}}-\drhodt{P_{D^3_{+,\downarrow}}} = &-\frac{2\Gamma}{\hbar}(P_{D^3_{+,\uparrow}} - P_{D^3_{+,\downarrow}}) + \frac{\Gamma}{2\hbar}(P_{T^2_{+1}} - P_{T^2_{-1}}) - \frac{\Gamma}{\hbar} \sqrt{1-\frac{4t}{x}} \rho_{S_+^2,T_0^2} \\
    \label{eq:3D-u_D-d}
        \drhodt{P_{D^3_{-,\uparrow}}}-\drhodt{P_{D^3_{-,\downarrow}}} = &-\frac{2\Gamma}{\hbar}(P_{D^3_{-,\uparrow}} - P_{D^3_{-,\downarrow}}) + \frac{\Gamma}{2\hbar}(P_{T^2_{+1}} - P_{T^2_{-1}}) - \frac{\Gamma}{\hbar} \sqrt{1+\frac{4t}{x}} \rho_{S_+^2,T_0^2}.
    \end{align}
\end{subequations}
\end{widetext}
\indent
Under the steady-state condition, all the equations in Eq.~\eqref{eq:contraction} equal $0$. Substituting Eq.~\eqref{eq:contraction} into Eq.~\eqref{eq:S+T0}, we derive Eq.~\eqref{eq:S+T0} at steady state as
\begin{align}
\label{eq:S+T0_self}
    \drhodt{\rho_{S^2_+, T^2_0}} =
    &\frac{i}{\hbar} \times \frac{x-U}{2} \rho_{S_{+}^{2}, T_{0}^{2}} -\frac{5\Gamma}{2\hbar} \rho_{S_{+}^{2}, T_{0}^{2}} = 0.
\end{align}
\indent
Obviously, Eq.~\eqref{eq:S+T0_self} indicates that coherence $\rho_{S^2_+, T^2_0} = 0$ under the steady-state situation. By substituting $\rho_{S^2_+, T^2_0} = 0$ into Eq.~\eqref{eq:contraction} and two additional dynamic equations in Eq.~(1) in the main text, i.e.,
\begin{align}
\label{eq:T+1}
    \drhodt{P_{T_{+1}^2}} &= \frac{\Gamma}{2\hbar} (P_{D_{+,\uparrow}^1} + P_{D_{-,\uparrow}^1} + P_{D_{-,\uparrow}^3} + P_{D_{+,\uparrow}^3} - 4P_{T_{+1}^2}) = 0 \\
\label{eq:T-1}
    \drhodt{P_{T_{-1}^2}} &= \frac{\Gamma}{2\hbar} (P_{D_{+,\downarrow}^1} + P_{D_{-,\downarrow}^1} + P_{D_{-,\downarrow}^3} + P_{D_{+,\downarrow}^3} - 4P_{T_{-1}^2}) = 0,
\end{align}
we obtain that population difference in Eq.~\eqref{eq:contraction} as
\begin{align}
\label{eq:equi}
    &P_{D^1_{+,\uparrow}} - P_{D^1_{+,\downarrow}} = P_{D^1_{-,\uparrow}} - P_{D^1_{-,\downarrow}} \nonumber \\
    =~&P_{D^3_{-,\uparrow}} - P_{D^3_{-,\downarrow}} = P_{D^3_{+,\uparrow}} - P_{D^3_{+,\downarrow}} \nonumber \\
    =~&\frac{1}{4}(P_{T^2_{+1}} - P_{T^2_{-1}}) \nonumber \\
    =~&0.
\end{align}
\indent
Up to now, we have already reduced the total number of equations from $20$ to $12$. Among the $8$ vanishing equations, $3$ is from coherence $\rho_{S^2_+,T^2_0}$, $\rho_{S^2_+,T^2_{+1}}$, and $\rho_{S^2_+,T^2_{-1}}$, and $5$ is from Eq.~\eqref{eq:equi}. For simplicity of derivation, we define the following notations,
\begin{subequations}
\begin{align}
    P_{D_{+,\uparrow}^1} &= P_{D_{+,\downarrow}^1} = A \\
    P_{D_{-,\uparrow}^1} &= P_{D_{-,\downarrow}^1} = B \\
    P_{D_{-,\uparrow}^3} &= P_{D_{-,\downarrow}^3} = C \\ 
    P_{D_{+,\uparrow}^3} &= P_{D_{+,\downarrow}^3} = D,
\end{align}    
\end{subequations}
\indent
According to the above notations, one can reduce the total number of equations by $1$ due to Eq.~\eqref{eq:T+1} and the populations $P_{T^2_{+1}}$ and $P_{T^2_{-1}}$ can be expressed as
\begin{equation}
\label{eq:T+1-1}
    P_{T_{+1}^2} = P_{T_{-1}^2} = \frac{1}{4}(A+B+C+D). 
\end{equation}
\indent
From the dynamic equations $\drhodt{P_{S^0}}$, $\drhodt{P_{S^4}}$, 
$\drhodt{P_{S_+^2}}$, and $\drhodt{P_{T_0^2}}$ in Eq.~(1) in the main text, the populations $P_{S^0}$, $P_{S^4}$, $P_{S_+^2}$, and $P_{T_0^2}$ can be solved and then expressed in terms of $A$, $B$, $C$, and $D$ as
\begin{align}
\label{eq:S0_notation}
    P_{S^0} &= \frac{1}{2} (A+B) \\
\label{eq:S4_notation}
    P_{S^4} &= \frac{1}{2} (C+D) \\
\label{eq:S+_notation}
    P_{S_+^2} &= \frac{1}{4} \big[ (1+\frac{4t}{x})(A+C) + (1-\frac{4t}{x})(B+D)\big] \\
\label{eq:T0_notation}
    P_{T_0^2} &= \frac{1}{4} (A+B+C+D).
\end{align}
\indent
In other words, we have reduced the total number of equations from $11$ to $7$.
In addition, we can obtain the relationship among $A$, $B$, $C$, and $D$ by utilizing auxiliary equations related to the difference between the time derivative of populations $P_{D^1_{+,\uparrow}}$, $P_{D^1_{-,\uparrow}}$, $P_{D^3_{+,\uparrow}}$, and $P_{D^3_{-,\uparrow}}$ at steady state in Eq.~(1) in the main text:
\begin{subequations}
\label{eq:D1-D3}
\begin{align}
\label{eq:D1+-D3-}
    \frac{\mathrm{d}P_{D^1_{+,\uparrow}}}{\mathrm{d}\tau} - \frac{\mathrm{d}P_{D^3_{-,\uparrow}}}{\mathrm{d}\tau} &= -\frac{7\Gamma}{4\hbar} (A-C) + \frac{\Gamma}{4\hbar} (B-D) \\
\label{eq:D1--D3+}
    \frac{\mathrm{d}P_{D^1_{-,\uparrow}}}{\mathrm{d}\tau} - \frac{\mathrm{d}P_{D^3_{+,\uparrow}}}{\mathrm{d}\tau} &= \frac{\Gamma}{4\hbar} (A-C) - \frac{7\Gamma}{4\hbar} (B-D).
\end{align}    
\end{subequations}
\indent
By applying the steady-state condition, the two equations in Eq.~\eqref{eq:D1-D3} provide the relations among $A$, $B$, $C$, and $D$: 
\begin{subequations}
\label{eq:ABCD}
\begin{align}
    A &= C \\
    B &= D.
\end{align}
\end{subequations}
\indent
Apparently, the two equality in Eq.~\eqref{eq:ABCD} eliminates $2$ equations, so the total number of equations have been reduced from $7$ to $5$. In other words, if we would like to obtain an analytical expression for current-coherence relationship, we need to solve a system of five equations. \\
\newline 
\indent
To derive a current-coherence relationship, we need an expression for $P_{S^2_\mathrm{CS}}$, $P_{S^2_-}$, and $\mathrm{Re} [\rho_{S^2_\mathrm{CS}, S^2_-}]$, which are defined as
\begin{align}
    &P_{S_{\mathrm{CS}}^2} = a \\ 
    &P_{S_{-}^2} = b \\ 
    &\mathrm{Re}\left[\rho_{S_{\mathrm{CS}}^2, S_{-}^2}\right] = c.
\end{align}
\indent
Next, we can obtain five relations among $A$, $B$, $a$, $b$, and $c$ from the dynamic equations $\drhodt{P_{D^1_{+,\uparrow}}}$, $\drhodt{P_{D^1_{-,\uparrow}}}$, $\drhodt{P_{S^2_\mathrm{CS}}}$, $\drhodt{P_{S^2_-}}$,
$\drhodt{\rho_{S^2_\mathrm{CS}, S^2_-}}$ in Eq.~(1) in the main text as
\begin{widetext}
    \begin{align}
    \label{eq:matrix_ss}
        \begin{bmatrix}
        -5+\frac{8t^2}{x^2}+\frac{4t}{x} & 3-\frac{8t^2}{x^2} & 1 & 1-\frac{4t}{x} & 2\sqrt{1-\frac{4t}{x}} \\
        3-\frac{8t^2}{x^2} & -5+\frac{8t^2}{x^2}-\frac{4t}{x} & 1 & 1+\frac{4t}{x} & 2\sqrt{1+\frac{4t}{x}} \\
        1 & 1 & -2 & 0 & -\sqrt{2}\sqrt{1+\frac{U}{x}} \\
        1-\frac{4t}{x} & 1+\frac{4t}{x} & 0 & -2 & -\sqrt{2}\sqrt{1+\frac{U}{x}} \\
        2\sqrt{1-\frac{4t}{x}} & 2\sqrt{1+\frac{4t}{x}} & \sqrt{2}\cdot\sqrt{1+\frac{U}{x}} & \sqrt{2}\cdot\sqrt{1+\frac{U}{x}} & \left[4+\frac{(x-U)^2}{4\Gamma^2}\right]
        \end{bmatrix}
        \begin{bmatrix}
        A \\
        B \\
        a \\
        b \\
        c
        \end{bmatrix}
        =
        \begin{bmatrix}
        0 \\
        0 \\
        0 \\
        0 \\
        0
        \end{bmatrix}.
    \end{align}
\end{widetext}

\indent
To solve Eq.~(\ref{eq:matrix_ss}), the conservation of probability is required: $\sum_a P_a = 1$. Combining Eq.~(\ref{eq:matrix_ss}) and $\sum_a P_a = 1$, we obtain the solutions as follows,
\begin{subequations}
\label{raw_solutions}
    \begin{align}
    a &= \frac{1}{16} \big[ 1+\frac{7}{2\lambda-3} \big] \\
    b &= \frac{1}{16} \big[ 1+\frac{7-4\eta}{2\lambda-3} \big] \\
    A &= \frac{1}{16} \big[ 1 - (1-\frac{x\eta}{t})\cdot\frac{1}{2\lambda-3} \big] \\
    B &= \frac{1}{16} \big[ 1 - (1+\frac{x\eta}{t})\cdot\frac{1}{2\lambda-3} \big] \\
    \label{eq:ScsS-_para}
    c &= - \sqrt{\frac{\zeta}{2}} \cdot \frac{1}{2\lambda-3},
    \end{align}
\end{subequations}
where $\zeta \equiv x/(x+U)$, $\eta \equiv (x-U)^2/(3x^2+U^2)$, and $\lambda \equiv (\zeta+1/2)\eta + \zeta[4+(x-U)^2/4\Gamma^2]$. Under the weak hopping and strong repulsion, $x \approx U$, $\zeta \approx 1/2$, $\eta \approx 0$, and the populations as well as the magnitude of coherence become

\begin{subequations}
\label{eq:popcoh_limiting_case}
\begin{align}
    &P_{S^2_\mathrm{CS}} = a \approx \frac{1}{2} \\
    &P_{S^2_-} = b \approx \frac{1}{2} \\
    &\mid \rho_{S^2_\mathrm{CS}, S^2_-} \mid = \sqrt{\frac{(x-U)^2}{16\Gamma^2} + 1} \mid c \mid \approx  \frac{1}{2},
\end{align}    
\end{subequations}
where we have utilized the relation from $\drhodt{\rho_{S^2_\mathrm{CS},S^2_-}}$:
\begin{equation}
\label{eq:imag_ScsS-}
    \mathrm{Im}[ \rho_{S_{\mathrm{CS}}^2, S_-^2} ] = \frac{x-U}{4\Gamma} c.
\end{equation}
\newline
\indent
The results in Eq.\eqref{eq:popcoh_limiting_case} can be interpreted as an effective two-level model with states $\ket*{S^2_{\mathrm{CS}}}$ and $\ket*{S^2_-}$, which supports the argument that coherence $\rho_{S^2_\mathrm{CS}, S^2_-}$ is bounded above by $1/2$. 
Recall that our target is to derive the analytical expression for steady-state current. From Eq.~\eqref{eq:I_energy}, the steady-state current can be obtained from dynamic equations of populations:
\begin{align}
\label{eq:I_para}
    I = \frac{\mathrm{e} \Gamma}{\hbar} \cdot \Big[ &1 + \frac{1}{2} \cdot \frac{\eta-2}{2\lambda-3} \Big].
\end{align}

\indent
By using Eq.~\eqref{eq:ScsS-_para}, Eq.~\eqref{eq:imag_ScsS-}, and Eq.~\eqref{eq:I_para}, we can derive the relation between current and the magnitude of coherence $\rho_{S^2_\mathrm{CS}, S^2_-}$. Under the weak hopping and strong repulsion, $x \approx U+8t^2/U$, $\zeta \approx 1/2$, $\eta \approx 0$, and we can derive the current-coherence relationship as follows,
\begin{align}
\label{eq:I_anaI}
    I = \frac{\mathrm{e} \Gamma}{\hbar} \cdot \bigg\{ 1 - 2 \cdot \big[ 1 + \frac{1}{4} \cdot (\frac{4t^2}{U\Gamma})^2\big]^{-1/2} \cdot \abs{\rho_{S_{\mathrm{CS}}^2, S_-^2}} \bigg\}.
\end{align}
\indent
Under the extreme condition, $4t^2/U\Gamma \approx 0$, we can another current-coherence relationship in the main text,
\begin{align}
\label{eq:I_anaII}
    I = \frac{\mathrm{e} \Gamma}{\hbar} \cdot \bigg\{ 1 - 2 \cdot  \abs{\rho_{S_{\mathrm{CS}}^2, S_-^2}} \bigg\}.
\end{align}
\indent
We have mentioned in the main text that the coherence can be tuned by the gate voltage. To obtain the gate dependency of coherence, we rewrite the aforementioned five relations in Eq.~\eqref{eq:matrix_ss} by considering the Fermi function $f_l(E) = \Theta(\mu_l-E)$, where $\Theta(\mu_l-E)$ denotes the Heaviside step function (zero-temperature limit). To simplify the following derivation, we apply the strong Coulomb repulsion and weak hopping condition first, and the matrix in Eq.~\eqref{eq:matrix_ss} can be adapted as
\begin{widetext}
    \begin{align}
    \label{eq:matrix_ss_fermi}
        \begin{bsmallmatrix}
        -4+\frac{1}{2}K(E)+K(E+U) & 2-\frac{1}{2}K(E) & 1-\frac{1}{2}K(E+U) & 1-\frac{1}{2}K(E+U) & L(E+U) \\
        2-\frac{1}{2}K(E) & -4+\frac{1}{2}K(E)+K(E+U) & 1-\frac{1}{2}K(E+U) & 1-\frac{1}{2}K(E+U) & L(E+U) \\
        2-K(E)+K(E+U) & 2-K(E)+K(E+U) & -4+2K(E)-2K(E+U) & 0 & -2L(E)-2L(E+U) \\
        2-K(E)+K(E+U) & 2-K(E)+K(E+U) & 0 & -4+2K(E)-2K(E+U) & -2L(E)-2L(E+U) \\
        -L(E)-L(E+U) & -L(E)-L(E+U) & -L(E)-L(E+U) & -L(E)-L(E+U) & -4+2K(E)-2K(E+U)
        \end{bsmallmatrix}
        \begin{bsmallmatrix}
        A \\
        B \\
        a \\
        b \\
        c
        \end{bsmallmatrix}
        =
        \begin{bsmallmatrix}
        0 \\
        0 \\
        0 \\
        0 \\
        0
        \end{bsmallmatrix},
    \end{align}
\end{widetext}
where $E = \varepsilon-\mathrm{e}V_\mathrm{g}$, $K(E) = \Theta(\mu_\mathrm{L}-E) + \Theta(\mu_\mathrm{R}-E)$, and $L(E) = \Theta(\mu_\mathrm{L}-E) - \Theta(\mu_\mathrm{R}-E)$. By applying the condition $\sum_a P_a = 1$, we derive the real part of coherence $\mathrm{Re}[\rho_{S^2_\mathrm{CS},S^2_-}]$ as a function of the gate voltage $V_\mathrm{g}$:
\begin{widetext}
    \begin{equation}
    \label{eq:coh_fermi}
    c = \frac{2-K(E+U)}{14L(E+U)} \left\{ 1 + \frac{4L(E+U)[L(E)+L(E+U)]+4[2-K(E+U)]\cdot[2-K(E)+K(E+U)]}{3L(E+U)[L(E)+L(E+U)]-4[2-K(E+U)]\cdot[2-K(E)+K(E+U)]} \right\}.
    \end{equation}
\end{widetext}

\indent
In Eq.~\eqref{eq:coh_fermi}, the gate dependence is introduced through step functions in $K(E)$ and $L(E)$. For simplicity, we focus on the case that $\mu_\mathrm{L} \approx E+U$, where current suppression is significant, and make the following simplification: $\Theta(\mu_\mathrm{L}-E) = 1$, $\Theta(\mu_\mathrm{R}-E) = 0$, and $\Theta(\mu_\mathrm{R}-E-U) = 0$. Finally, we obtain the coherence-gate relationship as
\begin{align}
    c &= \frac{1}{2} \times \frac{2-\Theta(\mu_\mathrm{L}-E-U)}{-8+7\Theta(\mu_\mathrm{L}-E-U)},
\end{align}
which specifies the condition $\mathrm{e}V_\mathrm{g} > \varepsilon+U-\mu_\mathrm{L}$ for the magnitude of coherence $\textstyle \mid \rho_{S^2_\mathrm{CS},S^2_-} \mid \approx \mid c \mid$ to reach maximum $1/2$.

\bibliographystyle{apsrev4-2}
\bibliography{references}

\begin{thebibliography}{62}%
\makeatletter
\providecommand \@ifxundefined [1]{%
 \@ifx{#1\undefined}
}%
\providecommand \@ifnum [1]{%
 \ifnum #1\expandafter \@firstoftwo
 \else \expandafter \@secondoftwo
 \fi
}%
\providecommand \@ifx [1]{%
 \ifx #1\expandafter \@firstoftwo
 \else \expandafter \@secondoftwo
 \fi
}%
\providecommand \natexlab [1]{#1}%
\providecommand \enquote  [1]{``#1''}%
\providecommand \bibnamefont  [1]{#1}%
\providecommand \bibfnamefont [1]{#1}%
\providecommand \citenamefont [1]{#1}%
\providecommand \href@noop [0]{\@secondoftwo}%
\providecommand \href [0]{\begingroup \@sanitize@url \@href}%
\providecommand \@href[1]{\@@startlink{#1}\@@href}%
\providecommand \@@href[1]{\endgroup#1\@@endlink}%
\providecommand \@sanitize@url [0]{\catcode `\\12\catcode `\$12\catcode
  `\&12\catcode `\#12\catcode `\^12\catcode `\_12\catcode `\%12\relax}%
\providecommand \@@startlink[1]{}%
\providecommand \@@endlink[0]{}%
\providecommand \url  [0]{\begingroup\@sanitize@url \@url }%
\providecommand \@url [1]{\endgroup\@href {#1}{\urlprefix }}%
\providecommand \urlprefix  [0]{URL }%
\providecommand \Eprint [0]{\href }%
\providecommand \doibase [0]{https://doi.org/}%
\providecommand \selectlanguage [0]{\@gobble}%
\providecommand \bibinfo  [0]{\@secondoftwo}%
\providecommand \bibfield  [0]{\@secondoftwo}%
\providecommand \translation [1]{[#1]}%
\providecommand \BibitemOpen [0]{}%
\providecommand \bibitemStop [0]{}%
\providecommand \bibitemNoStop [0]{.\EOS\space}%
\providecommand \EOS [0]{\spacefactor3000\relax}%
\providecommand \BibitemShut  [1]{\csname bibitem#1\endcsname}%
\let\auto@bib@innerbib\@empty
\bibitem [{\citenamefont {Engel}\ \emph {et~al.}(2007)\citenamefont {Engel},
  \citenamefont {Calhoun}, \citenamefont {Read}, \citenamefont {Ahn},
  \citenamefont {Man{\v{c}}al}, \citenamefont {Cheng}, \citenamefont
  {Blankenship},\ and\ \citenamefont {Fleming}}]{Engel2007}%
  \BibitemOpen
  \bibfield  {author} {\bibinfo {author} {\bibfnamefont {G.~S.}\ \bibnamefont
  {Engel}}, \bibinfo {author} {\bibfnamefont {T.~R.}\ \bibnamefont {Calhoun}},
  \bibinfo {author} {\bibfnamefont {E.~L.}\ \bibnamefont {Read}}, \bibinfo
  {author} {\bibfnamefont {T.-K.}\ \bibnamefont {Ahn}}, \bibinfo {author}
  {\bibfnamefont {T.}~\bibnamefont {Man{\v{c}}al}}, \bibinfo {author}
  {\bibfnamefont {Y.-C.}\ \bibnamefont {Cheng}}, \bibinfo {author}
  {\bibfnamefont {R.~E.}\ \bibnamefont {Blankenship}},\ and\ \bibinfo {author}
  {\bibfnamefont {G.~R.}\ \bibnamefont {Fleming}},\ }\href
  {https://doi.org/10.1038/nature05678} {\bibfield  {journal} {\bibinfo
  {journal} {Nature (London)}\ }\textbf {\bibinfo {volume} {446}},\ \bibinfo
  {pages} {782} (\bibinfo {year} {2007})}\BibitemShut {NoStop}%
\bibitem [{\citenamefont {Panitchayangkoon}\ \emph {et~al.}(2010)\citenamefont
  {Panitchayangkoon}, \citenamefont {Hayes}, \citenamefont {Fransted},
  \citenamefont {Caram}, \citenamefont {Harel}, \citenamefont {Wen},
  \citenamefont {Blankenship},\ and\ \citenamefont
  {Engel}}]{Panitchayangkoon2010}%
  \BibitemOpen
  \bibfield  {author} {\bibinfo {author} {\bibfnamefont {G.}~\bibnamefont
  {Panitchayangkoon}}, \bibinfo {author} {\bibfnamefont {D.}~\bibnamefont
  {Hayes}}, \bibinfo {author} {\bibfnamefont {K.~A.}\ \bibnamefont {Fransted}},
  \bibinfo {author} {\bibfnamefont {J.~R.}\ \bibnamefont {Caram}}, \bibinfo
  {author} {\bibfnamefont {E.}~\bibnamefont {Harel}}, \bibinfo {author}
  {\bibfnamefont {J.}~\bibnamefont {Wen}}, \bibinfo {author} {\bibfnamefont
  {R.~E.}\ \bibnamefont {Blankenship}},\ and\ \bibinfo {author} {\bibfnamefont
  {G.~S.}\ \bibnamefont {Engel}},\ }\href
  {https://doi.org/10.1073/pnas.1005484107} {\bibfield  {journal} {\bibinfo
  {journal} {Proc. Natl. Acad. Sci. U.S.A.}\ }\textbf {\bibinfo {volume}
  {107}},\ \bibinfo {pages} {12766} (\bibinfo {year} {2010})}\BibitemShut
  {NoStop}%
\bibitem [{\citenamefont {Scholes}\ \emph {et~al.}(2017)\citenamefont
  {Scholes}, \citenamefont {Fleming}, \citenamefont {Chen}, \citenamefont
  {Aspuru-Guzik}, \citenamefont {Buchleitner}, \citenamefont {Coker},
  \citenamefont {Engel}, \citenamefont {van Grondelle}, \citenamefont
  {Ishizaki},\ and\ \citenamefont {Jonas~\textit{et al.}}}]{Scholes2017}%
  \BibitemOpen
  \bibfield  {author} {\bibinfo {author} {\bibfnamefont {G.~D.}\ \bibnamefont
  {Scholes}}, \bibinfo {author} {\bibfnamefont {G.~R.}\ \bibnamefont
  {Fleming}}, \bibinfo {author} {\bibfnamefont {L.~X.}\ \bibnamefont {Chen}},
  \bibinfo {author} {\bibfnamefont {A.}~\bibnamefont {Aspuru-Guzik}}, \bibinfo
  {author} {\bibfnamefont {A.}~\bibnamefont {Buchleitner}}, \bibinfo {author}
  {\bibfnamefont {D.~F.}\ \bibnamefont {Coker}}, \bibinfo {author}
  {\bibfnamefont {G.~S.}\ \bibnamefont {Engel}}, \bibinfo {author}
  {\bibfnamefont {R.}~\bibnamefont {van Grondelle}}, \bibinfo {author}
  {\bibfnamefont {A.}~\bibnamefont {Ishizaki}},\ and\ \bibinfo {author}
  {\bibfnamefont {D.~M.}\ \bibnamefont {Jonas~\textit{et al.}}},\ }\href
  {https://doi.org/10.1038/nature21425} {\bibfield  {journal} {\bibinfo
  {journal} {Nature (London)}\ }\textbf {\bibinfo {volume} {543}},\ \bibinfo
  {pages} {647} (\bibinfo {year} {2017})}\BibitemShut {NoStop}%
\bibitem [{\citenamefont {Br{\'{e}}das}\ \emph {et~al.}(2017)\citenamefont
  {Br{\'{e}}das}, \citenamefont {Sargent},\ and\ \citenamefont
  {Scholes}}]{Bredas2017}%
  \BibitemOpen
  \bibfield  {author} {\bibinfo {author} {\bibfnamefont {J.-L.}\ \bibnamefont
  {Br{\'{e}}das}}, \bibinfo {author} {\bibfnamefont {E.~H.}\ \bibnamefont
  {Sargent}},\ and\ \bibinfo {author} {\bibfnamefont {G.~D.}\ \bibnamefont
  {Scholes}},\ }\href {https://doi.org/10.1038/nmat4767} {\bibfield  {journal}
  {\bibinfo  {journal} {Nat. Mater.}\ }\textbf {\bibinfo {volume} {16}},\
  \bibinfo {pages} {35} (\bibinfo {year} {2017})}\BibitemShut {NoStop}%
\bibitem [{\citenamefont {Scully}\ \emph {et~al.}(2011)\citenamefont {Scully},
  \citenamefont {Chapin}, \citenamefont {Dorfman}, \citenamefont {Kim},\ and\
  \citenamefont {Svidzinsky}}]{Scully2011}%
  \BibitemOpen
  \bibfield  {author} {\bibinfo {author} {\bibfnamefont {M.~O.}\ \bibnamefont
  {Scully}}, \bibinfo {author} {\bibfnamefont {K.~R.}\ \bibnamefont {Chapin}},
  \bibinfo {author} {\bibfnamefont {K.~E.}\ \bibnamefont {Dorfman}}, \bibinfo
  {author} {\bibfnamefont {M.~B.}\ \bibnamefont {Kim}},\ and\ \bibinfo {author}
  {\bibfnamefont {A.}~\bibnamefont {Svidzinsky}},\ }\href
  {https://doi.org/10.1073/pnas.1110234108} {\bibfield  {journal} {\bibinfo
  {journal} {Proc. Natl. Acad. Sci. U.S.A.}\ }\textbf {\bibinfo {volume}
  {108}},\ \bibinfo {pages} {15097} (\bibinfo {year} {2011})}\BibitemShut
  {NoStop}%
\bibitem [{\citenamefont {Samuelsson}\ \emph {et~al.}(2017)\citenamefont
  {Samuelsson}, \citenamefont {Kheradsoud},\ and\ \citenamefont
  {Sothmann}}]{Samuelsson2017}%
  \BibitemOpen
  \bibfield  {author} {\bibinfo {author} {\bibfnamefont {P.}~\bibnamefont
  {Samuelsson}}, \bibinfo {author} {\bibfnamefont {S.}~\bibnamefont
  {Kheradsoud}},\ and\ \bibinfo {author} {\bibfnamefont {B.}~\bibnamefont
  {Sothmann}},\ }\href {https://doi.org/10.1103/PhysRevLett.118.256801}
  {\bibfield  {journal} {\bibinfo  {journal} {Phys. Rev. Lett.}\ }\textbf
  {\bibinfo {volume} {118}},\ \bibinfo {pages} {256801} (\bibinfo {year}
  {2017})}\BibitemShut {NoStop}%
\bibitem [{\citenamefont {Saryal}\ \emph {et~al.}(2021)\citenamefont {Saryal},
  \citenamefont {Gerry}, \citenamefont {Khait}, \citenamefont {Segal},\ and\
  \citenamefont {Agarwalla}}]{Saryal2021}%
  \BibitemOpen
  \bibfield  {author} {\bibinfo {author} {\bibfnamefont {S.}~\bibnamefont
  {Saryal}}, \bibinfo {author} {\bibfnamefont {M.}~\bibnamefont {Gerry}},
  \bibinfo {author} {\bibfnamefont {I.}~\bibnamefont {Khait}}, \bibinfo
  {author} {\bibfnamefont {D.}~\bibnamefont {Segal}},\ and\ \bibinfo {author}
  {\bibfnamefont {B.~K.}\ \bibnamefont {Agarwalla}},\ }\href
  {https://doi.org/10.1103/PhysRevLett.127.190603} {\bibfield  {journal}
  {\bibinfo  {journal} {Phys. Rev. Lett.}\ }\textbf {\bibinfo {volume} {127}},\
  \bibinfo {pages} {190603} (\bibinfo {year} {2021})}\BibitemShut {NoStop}%
\bibitem [{\citenamefont {Tajima}\ and\ \citenamefont
  {Funo}(2021)}]{Tajima2021}%
  \BibitemOpen
  \bibfield  {author} {\bibinfo {author} {\bibfnamefont {H.}~\bibnamefont
  {Tajima}}\ and\ \bibinfo {author} {\bibfnamefont {K.}~\bibnamefont {Funo}},\
  }\href {https://doi.org/10.1103/PhysRevLett.127.190604} {\bibfield  {journal}
  {\bibinfo  {journal} {Phys. Rev. Lett.}\ }\textbf {\bibinfo {volume} {127}},\
  \bibinfo {pages} {190604} (\bibinfo {year} {2021})}\BibitemShut {NoStop}%
\bibitem [{\citenamefont {Kamimura}\ \emph {et~al.}(2022)\citenamefont
  {Kamimura}, \citenamefont {Hakoshima}, \citenamefont {Matsuzaki},
  \citenamefont {Yoshida},\ and\ \citenamefont {Tokura}}]{Kamimura2022}%
  \BibitemOpen
  \bibfield  {author} {\bibinfo {author} {\bibfnamefont {S.}~\bibnamefont
  {Kamimura}}, \bibinfo {author} {\bibfnamefont {H.}~\bibnamefont {Hakoshima}},
  \bibinfo {author} {\bibfnamefont {Y.}~\bibnamefont {Matsuzaki}}, \bibinfo
  {author} {\bibfnamefont {K.}~\bibnamefont {Yoshida}},\ and\ \bibinfo {author}
  {\bibfnamefont {Y.}~\bibnamefont {Tokura}},\ }\href
  {https://doi.org/10.1103/PhysRevLett.128.180602} {\bibfield  {journal}
  {\bibinfo  {journal} {Phys. Rev. Lett.}\ }\textbf {\bibinfo {volume} {128}},\
  \bibinfo {pages} {180602} (\bibinfo {year} {2022})}\BibitemShut {NoStop}%
\bibitem [{\citenamefont {Wu}\ \emph {et~al.}(2018)\citenamefont {Wu},
  \citenamefont {Hou}, \citenamefont {Zhao}, \citenamefont {Xiang},
  \citenamefont {Li}, \citenamefont {Guo}, \citenamefont {Ma}, \citenamefont
  {He}, \citenamefont {Thompson},\ and\ \citenamefont {Gu}}]{Wu2018}%
  \BibitemOpen
  \bibfield  {author} {\bibinfo {author} {\bibfnamefont {K.-D.}\ \bibnamefont
  {Wu}}, \bibinfo {author} {\bibfnamefont {Z.}~\bibnamefont {Hou}}, \bibinfo
  {author} {\bibfnamefont {Y.-Y.}\ \bibnamefont {Zhao}}, \bibinfo {author}
  {\bibfnamefont {G.-Y.}\ \bibnamefont {Xiang}}, \bibinfo {author}
  {\bibfnamefont {C.-F.}\ \bibnamefont {Li}}, \bibinfo {author} {\bibfnamefont
  {G.-C.}\ \bibnamefont {Guo}}, \bibinfo {author} {\bibfnamefont
  {J.}~\bibnamefont {Ma}}, \bibinfo {author} {\bibfnamefont {Q.-Y.}\
  \bibnamefont {He}}, \bibinfo {author} {\bibfnamefont {J.}~\bibnamefont
  {Thompson}},\ and\ \bibinfo {author} {\bibfnamefont {M.}~\bibnamefont {Gu}},\
  }\href {https://doi.org/10.1103/PhysRevLett.121.050401} {\bibfield  {journal}
  {\bibinfo  {journal} {Phys. Rev. Lett.}\ }\textbf {\bibinfo {volume} {121}},\
  \bibinfo {pages} {050401} (\bibinfo {year} {2018})}\BibitemShut {NoStop}%
\bibitem [{\citenamefont {Nguyen}\ \emph {et~al.}(2019)\citenamefont {Nguyen},
  \citenamefont {Sukachev}, \citenamefont {Bhaskar}, \citenamefont {Machielse},
  \citenamefont {Levonian}, \citenamefont {Knall}, \citenamefont {Stroganov},
  \citenamefont {Riedinger}, \citenamefont {Park},\ and\ \citenamefont
  {Lon{\v{c}}ar~\textit{et al.}}}]{Nguyen2019}%
  \BibitemOpen
  \bibfield  {author} {\bibinfo {author} {\bibfnamefont {C.~T.}\ \bibnamefont
  {Nguyen}}, \bibinfo {author} {\bibfnamefont {D.~D.}\ \bibnamefont
  {Sukachev}}, \bibinfo {author} {\bibfnamefont {M.~K.}\ \bibnamefont
  {Bhaskar}}, \bibinfo {author} {\bibfnamefont {B.}~\bibnamefont {Machielse}},
  \bibinfo {author} {\bibfnamefont {D.~S.}\ \bibnamefont {Levonian}}, \bibinfo
  {author} {\bibfnamefont {E.~N.}\ \bibnamefont {Knall}}, \bibinfo {author}
  {\bibfnamefont {P.}~\bibnamefont {Stroganov}}, \bibinfo {author}
  {\bibfnamefont {R.}~\bibnamefont {Riedinger}}, \bibinfo {author}
  {\bibfnamefont {H.}~\bibnamefont {Park}},\ and\ \bibinfo {author}
  {\bibfnamefont {M.}~\bibnamefont {Lon{\v{c}}ar~\textit{et al.}}},\ }\href
  {https://doi.org/10.1103/PhysRevLett.123.183602} {\bibfield  {journal}
  {\bibinfo  {journal} {Phys. Rev. Lett.}\ }\textbf {\bibinfo {volume} {123}},\
  \bibinfo {pages} {183602} (\bibinfo {year} {2019})}\BibitemShut {NoStop}%
\bibitem [{\citenamefont {Bhaskar}\ \emph {et~al.}(2020)\citenamefont
  {Bhaskar}, \citenamefont {Riedinger}, \citenamefont {Machielse},
  \citenamefont {Levonian}, \citenamefont {Nguyen}, \citenamefont {Knall},
  \citenamefont {Park}, \citenamefont {Englund}, \citenamefont {Lon{\v{c}}ar},\
  and\ \citenamefont {Sukachev~\textit{et al.}}}]{Bhaskar2020}%
  \BibitemOpen
  \bibfield  {author} {\bibinfo {author} {\bibfnamefont {M.~K.}\ \bibnamefont
  {Bhaskar}}, \bibinfo {author} {\bibfnamefont {R.}~\bibnamefont {Riedinger}},
  \bibinfo {author} {\bibfnamefont {B.}~\bibnamefont {Machielse}}, \bibinfo
  {author} {\bibfnamefont {D.~S.}\ \bibnamefont {Levonian}}, \bibinfo {author}
  {\bibfnamefont {C.~T.}\ \bibnamefont {Nguyen}}, \bibinfo {author}
  {\bibfnamefont {E.~N.}\ \bibnamefont {Knall}}, \bibinfo {author}
  {\bibfnamefont {H.}~\bibnamefont {Park}}, \bibinfo {author} {\bibfnamefont
  {D.}~\bibnamefont {Englund}}, \bibinfo {author} {\bibfnamefont
  {M.}~\bibnamefont {Lon{\v{c}}ar}},\ and\ \bibinfo {author} {\bibfnamefont
  {D.~D.}\ \bibnamefont {Sukachev~\textit{et al.}}},\ }\href
  {https://doi.org/10.1038/s41586-020-2103-5} {\bibfield  {journal} {\bibinfo
  {journal} {Nature (London)}\ }\textbf {\bibinfo {volume} {580}},\ \bibinfo
  {pages} {60} (\bibinfo {year} {2020})}\BibitemShut {NoStop}%
\bibitem [{\citenamefont {Zhai}\ \emph {et~al.}(2022)\citenamefont {Zhai},
  \citenamefont {Nguyen}, \citenamefont {Spinnler}, \citenamefont {Ritzmann},
  \citenamefont {L{\"{o}}bl}, \citenamefont {Wieck}, \citenamefont {Ludwig},
  \citenamefont {Javadi},\ and\ \citenamefont {Warburton}}]{Zhai2022}%
  \BibitemOpen
  \bibfield  {author} {\bibinfo {author} {\bibfnamefont {L.}~\bibnamefont
  {Zhai}}, \bibinfo {author} {\bibfnamefont {G.~N.}\ \bibnamefont {Nguyen}},
  \bibinfo {author} {\bibfnamefont {C.}~\bibnamefont {Spinnler}}, \bibinfo
  {author} {\bibfnamefont {J.}~\bibnamefont {Ritzmann}}, \bibinfo {author}
  {\bibfnamefont {M.~C.}\ \bibnamefont {L{\"{o}}bl}}, \bibinfo {author}
  {\bibfnamefont {A.~D.}\ \bibnamefont {Wieck}}, \bibinfo {author}
  {\bibfnamefont {A.}~\bibnamefont {Ludwig}}, \bibinfo {author} {\bibfnamefont
  {A.}~\bibnamefont {Javadi}},\ and\ \bibinfo {author} {\bibfnamefont {R.~J.}\
  \bibnamefont {Warburton}},\ }\href
  {https://doi.org/10.1038/s41565-022-01131-2} {\bibfield  {journal} {\bibinfo
  {journal} {Nat. Nanotechnol.}\ }\textbf {\bibinfo {volume} {17}},\ \bibinfo
  {pages} {829} (\bibinfo {year} {2022})}\BibitemShut {NoStop}%
\bibitem [{\citenamefont {Gu{\'{e}}don}\ \emph {et~al.}(2012)\citenamefont
  {Gu{\'{e}}don}, \citenamefont {Valkenier}, \citenamefont {Markussen},
  \citenamefont {Thygesen}, \citenamefont {Hummelen},\ and\ \citenamefont
  {van~der Molen}}]{Guedon2012}%
  \BibitemOpen
  \bibfield  {author} {\bibinfo {author} {\bibfnamefont {C.~M.}\ \bibnamefont
  {Gu{\'{e}}don}}, \bibinfo {author} {\bibfnamefont {H.}~\bibnamefont
  {Valkenier}}, \bibinfo {author} {\bibfnamefont {T.}~\bibnamefont
  {Markussen}}, \bibinfo {author} {\bibfnamefont {K.~S.}\ \bibnamefont
  {Thygesen}}, \bibinfo {author} {\bibfnamefont {J.~C.}\ \bibnamefont
  {Hummelen}},\ and\ \bibinfo {author} {\bibfnamefont {S.~J.}\ \bibnamefont
  {van~der Molen}},\ }\href {https://doi.org/10.1038/nnano.2012.37} {\bibfield
  {journal} {\bibinfo  {journal} {Nat. Nanotechnol.}\ }\textbf {\bibinfo
  {volume} {7}},\ \bibinfo {pages} {305} (\bibinfo {year} {2012})}\BibitemShut
  {NoStop}%
\bibitem [{\citenamefont {Ballmann}\ \emph {et~al.}(2012)\citenamefont
  {Ballmann}, \citenamefont {H{\"{a}}rtle}, \citenamefont {Coto}, \citenamefont
  {Elbing}, \citenamefont {Mayor}, \citenamefont {Bryce}, \citenamefont
  {Thoss},\ and\ \citenamefont {Weber}}]{Ballmann2012}%
  \BibitemOpen
  \bibfield  {author} {\bibinfo {author} {\bibfnamefont {S.}~\bibnamefont
  {Ballmann}}, \bibinfo {author} {\bibfnamefont {R.}~\bibnamefont
  {H{\"{a}}rtle}}, \bibinfo {author} {\bibfnamefont {P.~B.}\ \bibnamefont
  {Coto}}, \bibinfo {author} {\bibfnamefont {M.}~\bibnamefont {Elbing}},
  \bibinfo {author} {\bibfnamefont {M.}~\bibnamefont {Mayor}}, \bibinfo
  {author} {\bibfnamefont {M.~R.}\ \bibnamefont {Bryce}}, \bibinfo {author}
  {\bibfnamefont {M.}~\bibnamefont {Thoss}},\ and\ \bibinfo {author}
  {\bibfnamefont {H.~B.}\ \bibnamefont {Weber}},\ }\href
  {https://doi.org/10.1103/PhysRevLett.109.056801} {\bibfield  {journal}
  {\bibinfo  {journal} {Phys. Rev. Lett.}\ }\textbf {\bibinfo {volume} {109}},\
  \bibinfo {pages} {056801} (\bibinfo {year} {2012})}\BibitemShut {NoStop}%
\bibitem [{\citenamefont {Hsu}\ and\ \citenamefont {Rabitz}(2012)}]{Hsu2012}%
  \BibitemOpen
  \bibfield  {author} {\bibinfo {author} {\bibfnamefont {L.-Y.}\ \bibnamefont
  {Hsu}}\ and\ \bibinfo {author} {\bibfnamefont {H.}~\bibnamefont {Rabitz}},\
  }\href {https://doi.org/10.1103/PhysRevLett.109.186801} {\bibfield  {journal}
  {\bibinfo  {journal} {Phys. Rev. Lett.}\ }\textbf {\bibinfo {volume} {109}},\
  \bibinfo {pages} {186801} (\bibinfo {year} {2012})}\BibitemShut {NoStop}%
\bibitem [{\citenamefont {Li}\ \emph {et~al.}(2019)\citenamefont {Li},
  \citenamefont {Buerkle}, \citenamefont {Li}, \citenamefont {Rostamian},
  \citenamefont {Wang}, \citenamefont {Wang}, \citenamefont {Bowler},
  \citenamefont {Miyazaki}, \citenamefont {Xiang},\ and\ \citenamefont
  {Asai~\textit{et al.}}}]{Li2019}%
  \BibitemOpen
  \bibfield  {author} {\bibinfo {author} {\bibfnamefont {Y.}~\bibnamefont
  {Li}}, \bibinfo {author} {\bibfnamefont {M.}~\bibnamefont {Buerkle}},
  \bibinfo {author} {\bibfnamefont {G.}~\bibnamefont {Li}}, \bibinfo {author}
  {\bibfnamefont {A.}~\bibnamefont {Rostamian}}, \bibinfo {author}
  {\bibfnamefont {H.}~\bibnamefont {Wang}}, \bibinfo {author} {\bibfnamefont
  {Z.}~\bibnamefont {Wang}}, \bibinfo {author} {\bibfnamefont {D.~R.}\
  \bibnamefont {Bowler}}, \bibinfo {author} {\bibfnamefont {T.}~\bibnamefont
  {Miyazaki}}, \bibinfo {author} {\bibfnamefont {L.}~\bibnamefont {Xiang}},\
  and\ \bibinfo {author} {\bibfnamefont {Y.}~\bibnamefont {Asai~\textit{et
  al.}}},\ }\href {https://doi.org/10.1038/s41563-018-0280-5} {\bibfield
  {journal} {\bibinfo  {journal} {Nat. Mater.}\ }\textbf {\bibinfo {volume}
  {18}},\ \bibinfo {pages} {357} (\bibinfo {year} {2019})}\BibitemShut
  {NoStop}%
\bibitem [{\citenamefont {Bai}\ \emph {et~al.}(2019)\citenamefont {Bai},
  \citenamefont {Daaoub}, \citenamefont {Sangtarash}, \citenamefont {Li},
  \citenamefont {Tang}, \citenamefont {Zou}, \citenamefont {Sadeghi},
  \citenamefont {Liu}, \citenamefont {Huang},\ and\ \citenamefont
  {Tan~\textit{et al.}}}]{Bai2019}%
  \BibitemOpen
  \bibfield  {author} {\bibinfo {author} {\bibfnamefont {J.}~\bibnamefont
  {Bai}}, \bibinfo {author} {\bibfnamefont {A.}~\bibnamefont {Daaoub}},
  \bibinfo {author} {\bibfnamefont {S.}~\bibnamefont {Sangtarash}}, \bibinfo
  {author} {\bibfnamefont {X.}~\bibnamefont {Li}}, \bibinfo {author}
  {\bibfnamefont {Y.}~\bibnamefont {Tang}}, \bibinfo {author} {\bibfnamefont
  {Q.}~\bibnamefont {Zou}}, \bibinfo {author} {\bibfnamefont {H.}~\bibnamefont
  {Sadeghi}}, \bibinfo {author} {\bibfnamefont {S.}~\bibnamefont {Liu}},
  \bibinfo {author} {\bibfnamefont {X.}~\bibnamefont {Huang}},\ and\ \bibinfo
  {author} {\bibfnamefont {Z.}~\bibnamefont {Tan~\textit{et al.}}},\ }\href
  {https://doi.org/10.1038/s41563-018-0265-4} {\bibfield  {journal} {\bibinfo
  {journal} {Nat. Mater.}\ }\textbf {\bibinfo {volume} {18}},\ \bibinfo {pages}
  {364} (\bibinfo {year} {2019})}\BibitemShut {NoStop}%
\bibitem [{\citenamefont {Greenwald}\ \emph {et~al.}(2021)\citenamefont
  {Greenwald}, \citenamefont {Cameron}, \citenamefont {Findlay}, \citenamefont
  {Fu}, \citenamefont {Gunasekaran}, \citenamefont {Skabara},\ and\
  \citenamefont {Venkataraman}}]{Greenwald2021}%
  \BibitemOpen
  \bibfield  {author} {\bibinfo {author} {\bibfnamefont {J.~E.}\ \bibnamefont
  {Greenwald}}, \bibinfo {author} {\bibfnamefont {J.}~\bibnamefont {Cameron}},
  \bibinfo {author} {\bibfnamefont {N.~J.}\ \bibnamefont {Findlay}}, \bibinfo
  {author} {\bibfnamefont {T.}~\bibnamefont {Fu}}, \bibinfo {author}
  {\bibfnamefont {S.}~\bibnamefont {Gunasekaran}}, \bibinfo {author}
  {\bibfnamefont {P.~J.}\ \bibnamefont {Skabara}},\ and\ \bibinfo {author}
  {\bibfnamefont {L.}~\bibnamefont {Venkataraman}},\ }\href
  {https://doi.org/10.1038/s41565-020-00807-x} {\bibfield  {journal} {\bibinfo
  {journal} {Nat. Nanotechnol.}\ }\textbf {\bibinfo {volume} {16}},\ \bibinfo
  {pages} {313} (\bibinfo {year} {2021})}\BibitemShut {NoStop}%
\bibitem [{\citenamefont {Galperin}\ \emph {et~al.}(2007)\citenamefont
  {Galperin}, \citenamefont {Ratner},\ and\ \citenamefont
  {Nitzan}}]{Galperin2007}%
  \BibitemOpen
  \bibfield  {author} {\bibinfo {author} {\bibfnamefont {M.}~\bibnamefont
  {Galperin}}, \bibinfo {author} {\bibfnamefont {M.~A.}\ \bibnamefont
  {Ratner}},\ and\ \bibinfo {author} {\bibfnamefont {A.}~\bibnamefont
  {Nitzan}},\ }\href {https://doi.org/10.1088/0953-8984/19/10/103201}
  {\bibfield  {journal} {\bibinfo  {journal} {J. Phys. Condens. Matter}\
  }\textbf {\bibinfo {volume} {19}},\ \bibinfo {pages} {103201} (\bibinfo
  {year} {2007})}\BibitemShut {NoStop}%
\bibitem [{\citenamefont {Mitchell}\ \emph {et~al.}(2017)\citenamefont
  {Mitchell}, \citenamefont {Pedersen}, \citenamefont {Hedeg{\aa}rd},\ and\
  \citenamefont {Paaske}}]{Mitchell2017}%
  \BibitemOpen
  \bibfield  {author} {\bibinfo {author} {\bibfnamefont {A.~K.}\ \bibnamefont
  {Mitchell}}, \bibinfo {author} {\bibfnamefont {K.~G.~L.}\ \bibnamefont
  {Pedersen}}, \bibinfo {author} {\bibfnamefont {P.}~\bibnamefont
  {Hedeg{\aa}rd}},\ and\ \bibinfo {author} {\bibfnamefont {J.}~\bibnamefont
  {Paaske}},\ }\href {https://doi.org/10.1038/ncomms15210} {\bibfield
  {journal} {\bibinfo  {journal} {Nat. Commun.}\ }\textbf {\bibinfo {volume}
  {8}},\ \bibinfo {pages} {15210} (\bibinfo {year} {2017})}\BibitemShut
  {NoStop}%
\bibitem [{\citenamefont {Yu}\ \emph {et~al.}(2017)\citenamefont {Yu},
  \citenamefont {Koci{\'{c}}}, \citenamefont {Repp}, \citenamefont {Siegert},\
  and\ \citenamefont {Donarini}}]{Yu2017}%
  \BibitemOpen
  \bibfield  {author} {\bibinfo {author} {\bibfnamefont {P.}~\bibnamefont
  {Yu}}, \bibinfo {author} {\bibfnamefont {N.}~\bibnamefont {Koci{\'{c}}}},
  \bibinfo {author} {\bibfnamefont {J.}~\bibnamefont {Repp}}, \bibinfo {author}
  {\bibfnamefont {B.}~\bibnamefont {Siegert}},\ and\ \bibinfo {author}
  {\bibfnamefont {A.}~\bibnamefont {Donarini}},\ }\href
  {https://doi.org/10.1103/PhysRevLett.119.056801} {\bibfield  {journal}
  {\bibinfo  {journal} {Phys. Rev. Lett.}\ }\textbf {\bibinfo {volume} {119}},\
  \bibinfo {pages} {056801} (\bibinfo {year} {2017})}\BibitemShut {NoStop}%
\bibitem [{\citenamefont {Fu}\ \emph {et~al.}(2018)\citenamefont {Fu},
  \citenamefont {Mosquera}, \citenamefont {Schatz}, \citenamefont {Ratner},\
  and\ \citenamefont {Hsu}}]{Fu2018}%
  \BibitemOpen
  \bibfield  {author} {\bibinfo {author} {\bibfnamefont {B.}~\bibnamefont
  {Fu}}, \bibinfo {author} {\bibfnamefont {M.~A.}\ \bibnamefont {Mosquera}},
  \bibinfo {author} {\bibfnamefont {G.~C.}\ \bibnamefont {Schatz}}, \bibinfo
  {author} {\bibfnamefont {M.~A.}\ \bibnamefont {Ratner}},\ and\ \bibinfo
  {author} {\bibfnamefont {L.-Y.}\ \bibnamefont {Hsu}},\ }\href
  {https://doi.org/10.1021/acs.nanolett.8b01838} {\bibfield  {journal}
  {\bibinfo  {journal} {Nano Lett.}\ }\textbf {\bibinfo {volume} {18}},\
  \bibinfo {pages} {5015} (\bibinfo {year} {2018})}\BibitemShut {NoStop}%
\bibitem [{\citenamefont {Kimura}\ \emph {et~al.}(2019)\citenamefont {Kimura},
  \citenamefont {Miwa}, \citenamefont {Imada}, \citenamefont {Imai-Imada},
  \citenamefont {Kawahara}, \citenamefont {Takeya}, \citenamefont {Kawai},
  \citenamefont {Galperin},\ and\ \citenamefont {Kim}}]{Kimura2019}%
  \BibitemOpen
  \bibfield  {author} {\bibinfo {author} {\bibfnamefont {K.}~\bibnamefont
  {Kimura}}, \bibinfo {author} {\bibfnamefont {K.}~\bibnamefont {Miwa}},
  \bibinfo {author} {\bibfnamefont {H.}~\bibnamefont {Imada}}, \bibinfo
  {author} {\bibfnamefont {M.}~\bibnamefont {Imai-Imada}}, \bibinfo {author}
  {\bibfnamefont {S.}~\bibnamefont {Kawahara}}, \bibinfo {author}
  {\bibfnamefont {J.}~\bibnamefont {Takeya}}, \bibinfo {author} {\bibfnamefont
  {M.}~\bibnamefont {Kawai}}, \bibinfo {author} {\bibfnamefont
  {M.}~\bibnamefont {Galperin}},\ and\ \bibinfo {author} {\bibfnamefont
  {Y.}~\bibnamefont {Kim}},\ }\href {https://doi.org/10.1038/s41586-019-1284-2}
  {\bibfield  {journal} {\bibinfo  {journal} {Nature (London)}\ }\textbf
  {\bibinfo {volume} {570}},\ \bibinfo {pages} {210} (\bibinfo {year}
  {2019})}\BibitemShut {NoStop}%
\bibitem [{\citenamefont {Fetherolf}\ \emph {et~al.}(2020)\citenamefont
  {Fetherolf}, \citenamefont {Gole{\v{z}}},\ and\ \citenamefont
  {Berkelbach}}]{Fetherolf2020}%
  \BibitemOpen
  \bibfield  {author} {\bibinfo {author} {\bibfnamefont {J.~H.}\ \bibnamefont
  {Fetherolf}}, \bibinfo {author} {\bibfnamefont {D.}~\bibnamefont
  {Gole{\v{z}}}},\ and\ \bibinfo {author} {\bibfnamefont {T.~C.}\ \bibnamefont
  {Berkelbach}},\ }\href {https://doi.org/10.1103/PhysRevX.10.021062}
  {\bibfield  {journal} {\bibinfo  {journal} {Phys. Rev. X}\ }\textbf {\bibinfo
  {volume} {10}},\ \bibinfo {pages} {021062} (\bibinfo {year}
  {2020})}\BibitemShut {NoStop}%
\bibitem [{\citenamefont {Shein-Lumbroso}\ \emph {et~al.}(2022)\citenamefont
  {Shein-Lumbroso}, \citenamefont {Liu}, \citenamefont {Shastry}, \citenamefont
  {Segal},\ and\ \citenamefont {Tal}}]{Shein-Lumbroso2022}%
  \BibitemOpen
  \bibfield  {author} {\bibinfo {author} {\bibfnamefont {O.}~\bibnamefont
  {Shein-Lumbroso}}, \bibinfo {author} {\bibfnamefont {J.}~\bibnamefont {Liu}},
  \bibinfo {author} {\bibfnamefont {A.}~\bibnamefont {Shastry}}, \bibinfo
  {author} {\bibfnamefont {D.}~\bibnamefont {Segal}},\ and\ \bibinfo {author}
  {\bibfnamefont {O.}~\bibnamefont {Tal}},\ }\href
  {https://doi.org/10.1103/PhysRevLett.128.237701} {\bibfield  {journal}
  {\bibinfo  {journal} {Phys. Rev. Lett.}\ }\textbf {\bibinfo {volume} {128}},\
  \bibinfo {pages} {237701} (\bibinfo {year} {2022})}\BibitemShut {NoStop}%
\bibitem [{\citenamefont {Park}\ \emph {et~al.}(2002)\citenamefont {Park},
  \citenamefont {Pasupathy}, \citenamefont {Goldsmith}, \citenamefont {Chang},
  \citenamefont {Yalsh}, \citenamefont {Petta}, \citenamefont {Rinkoski},
  \citenamefont {Sethna}, \citenamefont {Abru{\~{n}}a}, \citenamefont
  {McEuen},\ and\ \citenamefont {Ralph}}]{Park2002}%
  \BibitemOpen
  \bibfield  {author} {\bibinfo {author} {\bibfnamefont {J.}~\bibnamefont
  {Park}}, \bibinfo {author} {\bibfnamefont {A.~N.}\ \bibnamefont {Pasupathy}},
  \bibinfo {author} {\bibfnamefont {J.~I.}\ \bibnamefont {Goldsmith}}, \bibinfo
  {author} {\bibfnamefont {C.}~\bibnamefont {Chang}}, \bibinfo {author}
  {\bibfnamefont {Y.}~\bibnamefont {Yalsh}}, \bibinfo {author} {\bibfnamefont
  {J.~R.}\ \bibnamefont {Petta}}, \bibinfo {author} {\bibfnamefont
  {M.}~\bibnamefont {Rinkoski}}, \bibinfo {author} {\bibfnamefont {J.~P.}\
  \bibnamefont {Sethna}}, \bibinfo {author} {\bibfnamefont {H.~D.}\
  \bibnamefont {Abru{\~{n}}a}}, \bibinfo {author} {\bibfnamefont {P.~L.}\
  \bibnamefont {McEuen}},\ and\ \bibinfo {author} {\bibfnamefont {D.~C.}\
  \bibnamefont {Ralph}},\ }\href {https://doi.org/10.1038/nature00791}
  {\bibfield  {journal} {\bibinfo  {journal} {Nature (London)}\ }\textbf
  {\bibinfo {volume} {417}},\ \bibinfo {pages} {722} (\bibinfo {year}
  {2002})}\BibitemShut {NoStop}%
\bibitem [{\citenamefont {Brotons-Gisbert}\ \emph {et~al.}(2019)\citenamefont
  {Brotons-Gisbert}, \citenamefont {Branny}, \citenamefont {Kumar},
  \citenamefont {Picard}, \citenamefont {Proux}, \citenamefont {Gray},
  \citenamefont {Burch}, \citenamefont {Watanabe}, \citenamefont {Taniguchi},\
  and\ \citenamefont {Gerardot}}]{Brotons-Gisbert2019}%
  \BibitemOpen
  \bibfield  {author} {\bibinfo {author} {\bibfnamefont {M.}~\bibnamefont
  {Brotons-Gisbert}}, \bibinfo {author} {\bibfnamefont {A.}~\bibnamefont
  {Branny}}, \bibinfo {author} {\bibfnamefont {S.}~\bibnamefont {Kumar}},
  \bibinfo {author} {\bibfnamefont {R.}~\bibnamefont {Picard}}, \bibinfo
  {author} {\bibfnamefont {R.}~\bibnamefont {Proux}}, \bibinfo {author}
  {\bibfnamefont {M.}~\bibnamefont {Gray}}, \bibinfo {author} {\bibfnamefont
  {K.~S.}\ \bibnamefont {Burch}}, \bibinfo {author} {\bibfnamefont
  {K.}~\bibnamefont {Watanabe}}, \bibinfo {author} {\bibfnamefont
  {T.}~\bibnamefont {Taniguchi}},\ and\ \bibinfo {author} {\bibfnamefont
  {B.~D.}\ \bibnamefont {Gerardot}},\ }\href
  {https://doi.org/10.1038/s41565-019-0402-5} {\bibfield  {journal} {\bibinfo
  {journal} {Nat. Nanotechnol.}\ }\textbf {\bibinfo {volume} {14}},\ \bibinfo
  {pages} {442} (\bibinfo {year} {2019})}\BibitemShut {NoStop}%
\bibitem [{\citenamefont {Liang}\ \emph {et~al.}(2002)\citenamefont {Liang},
  \citenamefont {Shores}, \citenamefont {Bockrath}, \citenamefont {Long},\ and\
  \citenamefont {Park}}]{Liang2002}%
  \BibitemOpen
  \bibfield  {author} {\bibinfo {author} {\bibfnamefont {W.}~\bibnamefont
  {Liang}}, \bibinfo {author} {\bibfnamefont {M.~P.}\ \bibnamefont {Shores}},
  \bibinfo {author} {\bibfnamefont {M.}~\bibnamefont {Bockrath}}, \bibinfo
  {author} {\bibfnamefont {J.~R.}\ \bibnamefont {Long}},\ and\ \bibinfo
  {author} {\bibfnamefont {H.}~\bibnamefont {Park}},\ }\href
  {https://doi.org/10.1038/nature00790} {\bibfield  {journal} {\bibinfo
  {journal} {Nature (London)}\ }\textbf {\bibinfo {volume} {417}},\ \bibinfo
  {pages} {725} (\bibinfo {year} {2002})}\BibitemShut {NoStop}%
\bibitem [{\citenamefont {Kurzmann}\ \emph {et~al.}(2021)\citenamefont
  {Kurzmann}, \citenamefont {Kleeorin}, \citenamefont {Tong}, \citenamefont
  {Garreis}, \citenamefont {Knothe}, \citenamefont {Eich}, \citenamefont
  {Mittag}, \citenamefont {Gold}, \citenamefont {de~Vries},\ and\ \citenamefont
  {Watanabe~\textit{et al.}}}]{Kurzmann2021}%
  \BibitemOpen
  \bibfield  {author} {\bibinfo {author} {\bibfnamefont {A.}~\bibnamefont
  {Kurzmann}}, \bibinfo {author} {\bibfnamefont {Y.}~\bibnamefont {Kleeorin}},
  \bibinfo {author} {\bibfnamefont {C.}~\bibnamefont {Tong}}, \bibinfo {author}
  {\bibfnamefont {R.}~\bibnamefont {Garreis}}, \bibinfo {author} {\bibfnamefont
  {A.}~\bibnamefont {Knothe}}, \bibinfo {author} {\bibfnamefont
  {M.}~\bibnamefont {Eich}}, \bibinfo {author} {\bibfnamefont {C.}~\bibnamefont
  {Mittag}}, \bibinfo {author} {\bibfnamefont {C.}~\bibnamefont {Gold}},
  \bibinfo {author} {\bibfnamefont {F.~K.}\ \bibnamefont {de~Vries}},\ and\
  \bibinfo {author} {\bibfnamefont {K.}~\bibnamefont {Watanabe~\textit{et
  al.}}},\ }\href {https://doi.org/10.1038/s41467-021-26149-3} {\bibfield
  {journal} {\bibinfo  {journal} {Nat. Commun.}\ }\textbf {\bibinfo {volume}
  {12}},\ \bibinfo {pages} {6004} (\bibinfo {year} {2021})}\BibitemShut
  {NoStop}%
\bibitem [{\citenamefont {Koch}\ and\ \citenamefont {von
  Oppen}(2005)}]{Koch2005}%
  \BibitemOpen
  \bibfield  {author} {\bibinfo {author} {\bibfnamefont {J.}~\bibnamefont
  {Koch}}\ and\ \bibinfo {author} {\bibfnamefont {F.}~\bibnamefont {von
  Oppen}},\ }\href {https://doi.org/10.1103/PhysRevLett.94.206804} {\bibfield
  {journal} {\bibinfo  {journal} {Phys. Rev. Lett.}\ }\textbf {\bibinfo
  {volume} {94}},\ \bibinfo {pages} {206804} (\bibinfo {year}
  {2005})}\BibitemShut {NoStop}%
\bibitem [{\citenamefont {Burzur{\'{i}}}\ \emph {et~al.}(2014)\citenamefont
  {Burzur{\'{i}}}, \citenamefont {Yamamoto}, \citenamefont {Warnock},
  \citenamefont {Zhong}, \citenamefont {Park}, \citenamefont {Cornia},\ and\
  \citenamefont {{Van Der Zant}}}]{Burzuri2014}%
  \BibitemOpen
  \bibfield  {author} {\bibinfo {author} {\bibfnamefont {E.}~\bibnamefont
  {Burzur{\'{i}}}}, \bibinfo {author} {\bibfnamefont {Y.}~\bibnamefont
  {Yamamoto}}, \bibinfo {author} {\bibfnamefont {M.}~\bibnamefont {Warnock}},
  \bibinfo {author} {\bibfnamefont {X.}~\bibnamefont {Zhong}}, \bibinfo
  {author} {\bibfnamefont {K.}~\bibnamefont {Park}}, \bibinfo {author}
  {\bibfnamefont {A.}~\bibnamefont {Cornia}},\ and\ \bibinfo {author}
  {\bibfnamefont {H.~S.}\ \bibnamefont {{Van Der Zant}}},\ }\href
  {https://doi.org/10.1021/nl500524w} {\bibfield  {journal} {\bibinfo
  {journal} {Nano Lett.}\ }\textbf {\bibinfo {volume} {14}},\ \bibinfo {pages}
  {3191} (\bibinfo {year} {2014})}\BibitemShut {NoStop}%
\bibitem [{\citenamefont {Du}\ \emph {et~al.}(2021)\citenamefont {Du},
  \citenamefont {Hashikawa}, \citenamefont {Ito}, \citenamefont {Hashimoto},
  \citenamefont {Murata}, \citenamefont {Hirayama},\ and\ \citenamefont
  {Hirakawa}}]{Du2021}%
  \BibitemOpen
  \bibfield  {author} {\bibinfo {author} {\bibfnamefont {S.}~\bibnamefont
  {Du}}, \bibinfo {author} {\bibfnamefont {Y.}~\bibnamefont {Hashikawa}},
  \bibinfo {author} {\bibfnamefont {H.}~\bibnamefont {Ito}}, \bibinfo {author}
  {\bibfnamefont {K.}~\bibnamefont {Hashimoto}}, \bibinfo {author}
  {\bibfnamefont {Y.}~\bibnamefont {Murata}}, \bibinfo {author} {\bibfnamefont
  {Y.}~\bibnamefont {Hirayama}},\ and\ \bibinfo {author} {\bibfnamefont
  {K.}~\bibnamefont {Hirakawa}},\ }\href
  {https://doi.org/10.1021/acs.nanolett.1c03604} {\bibfield  {journal}
  {\bibinfo  {journal} {Nano Lett.}\ }\textbf {\bibinfo {volume} {21}},\
  \bibinfo {pages} {10346} (\bibinfo {year} {2021})}\BibitemShut {NoStop}%
\bibitem [{\citenamefont {Galperin}\ \emph {et~al.}(2005)\citenamefont
  {Galperin}, \citenamefont {Ratner},\ and\ \citenamefont
  {Nitzan}}]{Galperin2005}%
  \BibitemOpen
  \bibfield  {author} {\bibinfo {author} {\bibfnamefont {M.}~\bibnamefont
  {Galperin}}, \bibinfo {author} {\bibfnamefont {M.~A.}\ \bibnamefont
  {Ratner}},\ and\ \bibinfo {author} {\bibfnamefont {A.}~\bibnamefont
  {Nitzan}},\ }\href {https://doi.org/10.1021/nl048216c} {\bibfield  {journal}
  {\bibinfo  {journal} {Nano Lett.}\ }\textbf {\bibinfo {volume} {5}},\
  \bibinfo {pages} {125} (\bibinfo {year} {2005})}\BibitemShut {NoStop}%
\bibitem [{\citenamefont {Schwarz}\ \emph {et~al.}(2016)\citenamefont
  {Schwarz}, \citenamefont {Kastlunger}, \citenamefont {Lissel}, \citenamefont
  {Egler-Lucas}, \citenamefont {Semenov}, \citenamefont {Venkatesan},
  \citenamefont {Berke}, \citenamefont {Stadler},\ and\ \citenamefont
  {L{\"{o}}rtscher}}]{Schwarz2016}%
  \BibitemOpen
  \bibfield  {author} {\bibinfo {author} {\bibfnamefont {F.}~\bibnamefont
  {Schwarz}}, \bibinfo {author} {\bibfnamefont {G.}~\bibnamefont {Kastlunger}},
  \bibinfo {author} {\bibfnamefont {F.}~\bibnamefont {Lissel}}, \bibinfo
  {author} {\bibfnamefont {C.}~\bibnamefont {Egler-Lucas}}, \bibinfo {author}
  {\bibfnamefont {S.~N.}\ \bibnamefont {Semenov}}, \bibinfo {author}
  {\bibfnamefont {K.}~\bibnamefont {Venkatesan}}, \bibinfo {author}
  {\bibfnamefont {H.}~\bibnamefont {Berke}}, \bibinfo {author} {\bibfnamefont
  {R.}~\bibnamefont {Stadler}},\ and\ \bibinfo {author} {\bibfnamefont
  {E.}~\bibnamefont {L{\"{o}}rtscher}},\ }\href
  {https://doi.org/10.1038/nnano.2015.255} {\bibfield  {journal} {\bibinfo
  {journal} {Nat. Nanotechnol.}\ }\textbf {\bibinfo {volume} {11}},\ \bibinfo
  {pages} {170} (\bibinfo {year} {2016})}\BibitemShut {NoStop}%
\bibitem [{\citenamefont {Brandes}(2005)}]{Brandes2005}%
  \BibitemOpen
  \bibfield  {author} {\bibinfo {author} {\bibfnamefont {T.}~\bibnamefont
  {Brandes}},\ }\href {https://doi.org/10.1016/j.physrep.2004.12.002}
  {\bibfield  {journal} {\bibinfo  {journal} {Phys. Rep.}\ }\textbf {\bibinfo
  {volume} {408}},\ \bibinfo {pages} {315} (\bibinfo {year}
  {2005})}\BibitemShut {NoStop}%
\bibitem [{\citenamefont {Landi}\ \emph {et~al.}(2022)\citenamefont {Landi},
  \citenamefont {Poletti},\ and\ \citenamefont {Schaller}}]{Landi2022}%
  \BibitemOpen
  \bibfield  {author} {\bibinfo {author} {\bibfnamefont {G.~T.}\ \bibnamefont
  {Landi}}, \bibinfo {author} {\bibfnamefont {D.}~\bibnamefont {Poletti}},\
  and\ \bibinfo {author} {\bibfnamefont {G.}~\bibnamefont {Schaller}},\ }\href
  {https://doi.org/10.1103/RevModPhys.94.045006} {\bibfield  {journal}
  {\bibinfo  {journal} {Rev. Mod. Phys.}\ }\textbf {\bibinfo {volume} {94}},\
  \bibinfo {pages} {45006} (\bibinfo {year} {2022})}\BibitemShut {NoStop}%
\bibitem [{\citenamefont {Dani}\ \emph {et~al.}(2022)\citenamefont {Dani},
  \citenamefont {Hussein}, \citenamefont {Bayer}, \citenamefont {Kohler},\ and\
  \citenamefont {Haug}}]{Dani2022}%
  \BibitemOpen
  \bibfield  {author} {\bibinfo {author} {\bibfnamefont {O.}~\bibnamefont
  {Dani}}, \bibinfo {author} {\bibfnamefont {R.}~\bibnamefont {Hussein}},
  \bibinfo {author} {\bibfnamefont {J.~C.}\ \bibnamefont {Bayer}}, \bibinfo
  {author} {\bibfnamefont {S.}~\bibnamefont {Kohler}},\ and\ \bibinfo {author}
  {\bibfnamefont {R.~J.}\ \bibnamefont {Haug}},\ }\href
  {https://doi.org/10.1038/s42005-022-01074-z} {\bibfield  {journal} {\bibinfo
  {journal} {Commun. Phys.}\ }\textbf {\bibinfo {volume} {5}},\ \bibinfo
  {pages} {1} (\bibinfo {year} {2022})}\BibitemShut {NoStop}%
\bibitem [{\citenamefont {Segal}\ \emph {et~al.}(2000)\citenamefont {Segal},
  \citenamefont {Nitzan}, \citenamefont {Davis}, \citenamefont {Wasielewski},\
  and\ \citenamefont {Ratner}}]{Segal2000}%
  \BibitemOpen
  \bibfield  {author} {\bibinfo {author} {\bibfnamefont {D.}~\bibnamefont
  {Segal}}, \bibinfo {author} {\bibfnamefont {A.}~\bibnamefont {Nitzan}},
  \bibinfo {author} {\bibfnamefont {W.~B.}\ \bibnamefont {Davis}}, \bibinfo
  {author} {\bibfnamefont {M.~R.}\ \bibnamefont {Wasielewski}},\ and\ \bibinfo
  {author} {\bibfnamefont {M.~A.}\ \bibnamefont {Ratner}},\ }\href
  {https://doi.org/10.1021/jp993260f} {\bibfield  {journal} {\bibinfo
  {journal} {J. Phys. Chem. B}\ }\textbf {\bibinfo {volume} {104}},\ \bibinfo
  {pages} {3817} (\bibinfo {year} {2000})}\BibitemShut {NoStop}%
\bibitem [{\citenamefont {Hsu}\ \emph {et~al.}(2014)\citenamefont {Hsu},
  \citenamefont {Wu},\ and\ \citenamefont {Rabitz}}]{Hsu2014}%
  \BibitemOpen
  \bibfield  {author} {\bibinfo {author} {\bibfnamefont {L.-Y.}\ \bibnamefont
  {Hsu}}, \bibinfo {author} {\bibfnamefont {N.}~\bibnamefont {Wu}},\ and\
  \bibinfo {author} {\bibfnamefont {H.}~\bibnamefont {Rabitz}},\ }\href
  {https://doi.org/10.1021/jz5005818} {\bibfield  {journal} {\bibinfo
  {journal} {J. Phys. Chem. Lett.}\ }\textbf {\bibinfo {volume} {5}},\ \bibinfo
  {pages} {1831} (\bibinfo {year} {2014})}\BibitemShut {NoStop}%
\bibitem [{\citenamefont {Agarwalla}\ \emph {et~al.}(2015)\citenamefont
  {Agarwalla}, \citenamefont {Jiang},\ and\ \citenamefont
  {Segal}}]{Agarwalla2015}%
  \BibitemOpen
  \bibfield  {author} {\bibinfo {author} {\bibfnamefont {B.~K.}\ \bibnamefont
  {Agarwalla}}, \bibinfo {author} {\bibfnamefont {J.-H.}\ \bibnamefont
  {Jiang}},\ and\ \bibinfo {author} {\bibfnamefont {D.}~\bibnamefont {Segal}},\
  }\href {https://doi.org/10.1103/PhysRevB.92.245418} {\bibfield  {journal}
  {\bibinfo  {journal} {Phys. Rev. B}\ }\textbf {\bibinfo {volume} {92}},\
  \bibinfo {pages} {245418} (\bibinfo {year} {2015})}\BibitemShut {NoStop}%
\bibitem [{\citenamefont {Anto-Sztrikacs}\ \emph {et~al.}(2023)\citenamefont
  {Anto-Sztrikacs}, \citenamefont {Nazir},\ and\ \citenamefont
  {Segal}}]{Anto-Sztrikacs2023}%
  \BibitemOpen
  \bibfield  {author} {\bibinfo {author} {\bibfnamefont {N.}~\bibnamefont
  {Anto-Sztrikacs}}, \bibinfo {author} {\bibfnamefont {A.}~\bibnamefont
  {Nazir}},\ and\ \bibinfo {author} {\bibfnamefont {D.}~\bibnamefont {Segal}},\
  }\href {https://doi.org/10.1103/PRXQuantum.4.020307} {\bibfield  {journal}
  {\bibinfo  {journal} {PRX Quantum}\ }\textbf {\bibinfo {volume} {4}},\
  \bibinfo {pages} {020307} (\bibinfo {year} {2023})}\BibitemShut {NoStop}%
\bibitem [{\citenamefont {Li}\ \emph {et~al.}(2014)\citenamefont {Li},
  \citenamefont {Miller}, \citenamefont {Levy},\ and\ \citenamefont
  {Rabani}}]{Li2014}%
  \BibitemOpen
  \bibfield  {author} {\bibinfo {author} {\bibfnamefont {B.}~\bibnamefont
  {Li}}, \bibinfo {author} {\bibfnamefont {W.~H.}\ \bibnamefont {Miller}},
  \bibinfo {author} {\bibfnamefont {T.~J.}\ \bibnamefont {Levy}},\ and\
  \bibinfo {author} {\bibfnamefont {E.}~\bibnamefont {Rabani}},\ }\href
  {https://doi.org/10.1063/1.4878736} {\bibfield  {journal} {\bibinfo
  {journal} {J. Chem. Phys.}\ }\textbf {\bibinfo {volume} {140}},\ \bibinfo
  {pages} {204106} (\bibinfo {year} {2014})}\BibitemShut {NoStop}%
\bibitem [{\citenamefont {Koole}\ \emph {et~al.}(2016)\citenamefont {Koole},
  \citenamefont {Hummelen},\ and\ \citenamefont {van~der Zant}}]{Koole2016}%
  \BibitemOpen
  \bibfield  {author} {\bibinfo {author} {\bibfnamefont {M.}~\bibnamefont
  {Koole}}, \bibinfo {author} {\bibfnamefont {J.~C.}\ \bibnamefont
  {Hummelen}},\ and\ \bibinfo {author} {\bibfnamefont {H.~S.~J.}\ \bibnamefont
  {van~der Zant}},\ }\href {https://doi.org/10.1103/PhysRevB.94.165414}
  {\bibfield  {journal} {\bibinfo  {journal} {Phys. Rev. B}\ }\textbf {\bibinfo
  {volume} {94}},\ \bibinfo {pages} {165414} (\bibinfo {year}
  {2016})}\BibitemShut {NoStop}%
\bibitem [{\citenamefont {Esposito}\ and\ \citenamefont
  {Galperin}(2009)}]{Esposito2009}%
  \BibitemOpen
  \bibfield  {author} {\bibinfo {author} {\bibfnamefont {M.}~\bibnamefont
  {Esposito}}\ and\ \bibinfo {author} {\bibfnamefont {M.}~\bibnamefont
  {Galperin}},\ }\href {https://doi.org/10.1103/PhysRevB.79.205303} {\bibfield
  {journal} {\bibinfo  {journal} {Phys. Rev. B}\ }\textbf {\bibinfo {volume}
  {79}},\ \bibinfo {pages} {205303} (\bibinfo {year} {2009})}\BibitemShut
  {NoStop}%
\bibitem [{\citenamefont {Wunsch}\ \emph {et~al.}(2005)\citenamefont {Wunsch},
  \citenamefont {Braun}, \citenamefont {K{\"{o}}nig},\ and\ \citenamefont
  {Pfannkuche}}]{Wunsch2005}%
  \BibitemOpen
  \bibfield  {author} {\bibinfo {author} {\bibfnamefont {B.}~\bibnamefont
  {Wunsch}}, \bibinfo {author} {\bibfnamefont {M.}~\bibnamefont {Braun}},
  \bibinfo {author} {\bibfnamefont {J.}~\bibnamefont {K{\"{o}}nig}},\ and\
  \bibinfo {author} {\bibfnamefont {D.}~\bibnamefont {Pfannkuche}},\ }\href
  {https://doi.org/10.1103/PhysRevB.72.205319} {\bibfield  {journal} {\bibinfo
  {journal} {Phys. Rev. B}\ }\textbf {\bibinfo {volume} {72}},\ \bibinfo
  {pages} {205319} (\bibinfo {year} {2005})}\BibitemShut {NoStop}%
\bibitem [{\citenamefont {Luo}\ \emph {et~al.}(2011)\citenamefont {Luo},
  \citenamefont {Jiao}, \citenamefont {Shen}, \citenamefont {Cen},
  \citenamefont {He},\ and\ \citenamefont {Wang}}]{Luo2011}%
  \BibitemOpen
  \bibfield  {author} {\bibinfo {author} {\bibfnamefont {J.}~\bibnamefont
  {Luo}}, \bibinfo {author} {\bibfnamefont {H.}~\bibnamefont {Jiao}}, \bibinfo
  {author} {\bibfnamefont {Y.}~\bibnamefont {Shen}}, \bibinfo {author}
  {\bibfnamefont {G.}~\bibnamefont {Cen}}, \bibinfo {author} {\bibfnamefont
  {X.-L.}\ \bibnamefont {He}},\ and\ \bibinfo {author} {\bibfnamefont
  {C.}~\bibnamefont {Wang}},\ }\href
  {https://doi.org/10.1088/0953-8984/23/14/145301} {\bibfield  {journal}
  {\bibinfo  {journal} {J. Phys. Condens. Matter}\ }\textbf {\bibinfo {volume}
  {23}},\ \bibinfo {pages} {145301} (\bibinfo {year} {2011})}\BibitemShut
  {NoStop}%
\bibitem [{\citenamefont {Thomas}\ \emph {et~al.}(2021)\citenamefont {Thomas},
  \citenamefont {Sowa}, \citenamefont {Limburg}, \citenamefont {Bian},
  \citenamefont {Evangeli}, \citenamefont {Swett}, \citenamefont {Tewari},
  \citenamefont {Baugh}, \citenamefont {Schatz},\ and\ \citenamefont
  {Briggs~\textit{et al.}}}]{Thomas2021}%
  \BibitemOpen
  \bibfield  {author} {\bibinfo {author} {\bibfnamefont {J.~O.}\ \bibnamefont
  {Thomas}}, \bibinfo {author} {\bibfnamefont {J.~K.}\ \bibnamefont {Sowa}},
  \bibinfo {author} {\bibfnamefont {B.}~\bibnamefont {Limburg}}, \bibinfo
  {author} {\bibfnamefont {X.}~\bibnamefont {Bian}}, \bibinfo {author}
  {\bibfnamefont {C.}~\bibnamefont {Evangeli}}, \bibinfo {author}
  {\bibfnamefont {J.~L.}\ \bibnamefont {Swett}}, \bibinfo {author}
  {\bibfnamefont {S.}~\bibnamefont {Tewari}}, \bibinfo {author} {\bibfnamefont
  {J.}~\bibnamefont {Baugh}}, \bibinfo {author} {\bibfnamefont {G.~C.}\
  \bibnamefont {Schatz}},\ and\ \bibinfo {author} {\bibfnamefont {G.~A.~D.}\
  \bibnamefont {Briggs~\textit{et al.}}},\ }\href
  {https://doi.org/10.1039/D1SC03050G} {\bibfield  {journal} {\bibinfo
  {journal} {Chem. Sci.}\ }\textbf {\bibinfo {volume} {12}},\ \bibinfo {pages}
  {11121} (\bibinfo {year} {2021})}\BibitemShut {NoStop}%
\bibitem [{\citenamefont {Ueda}\ \emph {et~al.}(2010)\citenamefont {Ueda},
  \citenamefont {Entin-Wohlman}, \citenamefont {Eto},\ and\ \citenamefont
  {Aharony}}]{Ueda2010}%
  \BibitemOpen
  \bibfield  {author} {\bibinfo {author} {\bibfnamefont {A.}~\bibnamefont
  {Ueda}}, \bibinfo {author} {\bibfnamefont {O.}~\bibnamefont {Entin-Wohlman}},
  \bibinfo {author} {\bibfnamefont {M.}~\bibnamefont {Eto}},\ and\ \bibinfo
  {author} {\bibfnamefont {A.}~\bibnamefont {Aharony}},\ }\href
  {https://doi.org/10.1103/PhysRevB.82.245317} {\bibfield  {journal} {\bibinfo
  {journal} {Phys. Rev. B}\ }\textbf {\bibinfo {volume} {82}},\ \bibinfo
  {pages} {245317} (\bibinfo {year} {2010})}\BibitemShut {NoStop}%
\bibitem [{\citenamefont {H{\"{a}}rtle}\ \emph {et~al.}(2011)\citenamefont
  {H{\"{a}}rtle}, \citenamefont {Butzin}, \citenamefont {Rubio-Pons},\ and\
  \citenamefont {Thoss}}]{Hartle2011}%
  \BibitemOpen
  \bibfield  {author} {\bibinfo {author} {\bibfnamefont {R.}~\bibnamefont
  {H{\"{a}}rtle}}, \bibinfo {author} {\bibfnamefont {M.}~\bibnamefont
  {Butzin}}, \bibinfo {author} {\bibfnamefont {O.}~\bibnamefont {Rubio-Pons}},\
  and\ \bibinfo {author} {\bibfnamefont {M.}~\bibnamefont {Thoss}},\ }\href
  {https://doi.org/10.1103/PhysRevLett.107.046802} {\bibfield  {journal}
  {\bibinfo  {journal} {Phys. Rev. Lett.}\ }\textbf {\bibinfo {volume} {107}},\
  \bibinfo {pages} {046802} (\bibinfo {year} {2011})}\BibitemShut {NoStop}%
\bibitem [{\citenamefont {Kilgour}\ and\ \citenamefont
  {Segal}(2015)}]{Kilgour2015}%
  \BibitemOpen
  \bibfield  {author} {\bibinfo {author} {\bibfnamefont {M.}~\bibnamefont
  {Kilgour}}\ and\ \bibinfo {author} {\bibfnamefont {D.}~\bibnamefont
  {Segal}},\ }\href {https://doi.org/10.1063/1.4926395} {\bibfield  {journal}
  {\bibinfo  {journal} {J. Chem. Phys.}\ }\textbf {\bibinfo {volume} {143}},\
  \bibinfo {pages} {024111} (\bibinfo {year} {2015})}\BibitemShut {NoStop}%
\bibitem [{\citenamefont {Covito}\ \emph {et~al.}(2018)\citenamefont {Covito},
  \citenamefont {Eich}, \citenamefont {Tuovinen}, \citenamefont {Sentef},\ and\
  \citenamefont {Rubio}}]{Covito2018}%
  \BibitemOpen
  \bibfield  {author} {\bibinfo {author} {\bibfnamefont {F.}~\bibnamefont
  {Covito}}, \bibinfo {author} {\bibfnamefont {F.~G.}\ \bibnamefont {Eich}},
  \bibinfo {author} {\bibfnamefont {R.}~\bibnamefont {Tuovinen}}, \bibinfo
  {author} {\bibfnamefont {M.~A.}\ \bibnamefont {Sentef}},\ and\ \bibinfo
  {author} {\bibfnamefont {A.}~\bibnamefont {Rubio}},\ }\href
  {https://doi.org/10.1021/acs.jctc.8b00077} {\bibfield  {journal} {\bibinfo
  {journal} {J. Chem. Theory Comput.}\ }\textbf {\bibinfo {volume} {14}},\
  \bibinfo {pages} {2495} (\bibinfo {year} {2018})}\BibitemShut {NoStop}%
\bibitem [{\citenamefont {Haug}\ and\ \citenamefont
  {Jauho}(2008)}]{haug2008quantum}%
  \BibitemOpen
  \bibfield  {author} {\bibinfo {author} {\bibfnamefont {H.}~\bibnamefont
  {Haug}}\ and\ \bibinfo {author} {\bibfnamefont {A.-P.}\ \bibnamefont
  {Jauho}},\ }\href@noop {} {\emph {\bibinfo {title} {Quantum Kinetics in
  Transport and Optics of Semiconductors}}}\ (\bibinfo  {publisher} {Springer,
  Berlin, Heidelberg},\ \bibinfo {year} {2008})\BibitemShut {NoStop}%
\bibitem [{\citenamefont {Boyle}\ \emph {et~al.}(2019)\citenamefont {Boyle},
  \citenamefont {Upadhyaya}, \citenamefont {Wang}, \citenamefont {Renna},
  \citenamefont {Lu-D{\'{i}}az}, \citenamefont {{Pyo Jeong}}, \citenamefont
  {Hight-Huf}, \citenamefont {Korugic-Karasz}, \citenamefont {Barnes},\ and\
  \citenamefont {Aksamija~\textit{et al.}}}]{Boyle2019}%
  \BibitemOpen
  \bibfield  {author} {\bibinfo {author} {\bibfnamefont {C.~J.}\ \bibnamefont
  {Boyle}}, \bibinfo {author} {\bibfnamefont {M.}~\bibnamefont {Upadhyaya}},
  \bibinfo {author} {\bibfnamefont {P.}~\bibnamefont {Wang}}, \bibinfo {author}
  {\bibfnamefont {L.~A.}\ \bibnamefont {Renna}}, \bibinfo {author}
  {\bibfnamefont {M.}~\bibnamefont {Lu-D{\'{i}}az}}, \bibinfo {author}
  {\bibfnamefont {S.}~\bibnamefont {{Pyo Jeong}}}, \bibinfo {author}
  {\bibfnamefont {N.}~\bibnamefont {Hight-Huf}}, \bibinfo {author}
  {\bibfnamefont {L.}~\bibnamefont {Korugic-Karasz}}, \bibinfo {author}
  {\bibfnamefont {M.~D.}\ \bibnamefont {Barnes}},\ and\ \bibinfo {author}
  {\bibfnamefont {Z.}~\bibnamefont {Aksamija~\textit{et al.}}},\ }\href
  {https://doi.org/10.1038/s41467-019-10567-5} {\bibfield  {journal} {\bibinfo
  {journal} {Nat. Commun.}\ }\textbf {\bibinfo {volume} {10}},\ \bibinfo
  {pages} {2827} (\bibinfo {year} {2019})}\BibitemShut {NoStop}%
\bibitem [{\citenamefont {Vyas}\ \emph {et~al.}(2020)\citenamefont {Vyas},
  \citenamefont {{Van de Put}},\ and\ \citenamefont {Fischetti}}]{Vyas2020}%
  \BibitemOpen
  \bibfield  {author} {\bibinfo {author} {\bibfnamefont {P.~B.}\ \bibnamefont
  {Vyas}}, \bibinfo {author} {\bibfnamefont {M.~L.}\ \bibnamefont {{Van de
  Put}}},\ and\ \bibinfo {author} {\bibfnamefont {M.~V.}\ \bibnamefont
  {Fischetti}},\ }\href {https://doi.org/10.1103/PhysRevApplied.13.014067}
  {\bibfield  {journal} {\bibinfo  {journal} {Phys. Rev. Appl.}\ }\textbf
  {\bibinfo {volume} {13}},\ \bibinfo {pages} {014067} (\bibinfo {year}
  {2020})}\BibitemShut {NoStop}%
\bibitem [{\citenamefont {Grifoni}\ and\ \citenamefont
  {H{\"{a}}nggi}(1998)}]{Grifoni1998}%
  \BibitemOpen
  \bibfield  {author} {\bibinfo {author} {\bibfnamefont {M.}~\bibnamefont
  {Grifoni}}\ and\ \bibinfo {author} {\bibfnamefont {P.}~\bibnamefont
  {H{\"{a}}nggi}},\ }\href {https://doi.org/10.1016/S0370-1573(98)00022-2}
  {\bibfield  {journal} {\bibinfo  {journal} {Phys. Rep.}\ }\textbf {\bibinfo
  {volume} {304}},\ \bibinfo {pages} {229} (\bibinfo {year}
  {1998})}\BibitemShut {NoStop}%
\bibitem [{\citenamefont {Hsu}\ \emph {et~al.}(2013)\citenamefont {Hsu},
  \citenamefont {Li},\ and\ \citenamefont {Rabitz}}]{Hsu2013}%
  \BibitemOpen
  \bibfield  {author} {\bibinfo {author} {\bibfnamefont {L.-Y.}\ \bibnamefont
  {Hsu}}, \bibinfo {author} {\bibfnamefont {E.~Y.}\ \bibnamefont {Li}},\ and\
  \bibinfo {author} {\bibfnamefont {H.}~\bibnamefont {Rabitz}},\ }\href
  {https://doi.org/10.1021/nl401340c} {\bibfield  {journal} {\bibinfo
  {journal} {Nano Lett.}\ }\textbf {\bibinfo {volume} {13}},\ \bibinfo {pages}
  {5020} (\bibinfo {year} {2013})}\BibitemShut {NoStop}%
\bibitem [{\citenamefont {White}\ \emph {et~al.}(2013)\citenamefont {White},
  \citenamefont {Peskin},\ and\ \citenamefont {Galperin}}]{White2013}%
  \BibitemOpen
  \bibfield  {author} {\bibinfo {author} {\bibfnamefont {A.~J.}\ \bibnamefont
  {White}}, \bibinfo {author} {\bibfnamefont {U.}~\bibnamefont {Peskin}},\ and\
  \bibinfo {author} {\bibfnamefont {M.}~\bibnamefont {Galperin}},\ }\href
  {https://doi.org/10.1103/PhysRevB.88.205424} {\bibfield  {journal} {\bibinfo
  {journal} {Phys. Rev. B}\ }\textbf {\bibinfo {volume} {88}},\ \bibinfo
  {pages} {205424} (\bibinfo {year} {2013})}\BibitemShut {NoStop}%
\bibitem [{\citenamefont {Damanet}\ \emph {et~al.}(2019)\citenamefont
  {Damanet}, \citenamefont {Mascarenhas}, \citenamefont {Pekker},\ and\
  \citenamefont {Daley}}]{Damanet2019}%
  \BibitemOpen
  \bibfield  {author} {\bibinfo {author} {\bibfnamefont {F.}~\bibnamefont
  {Damanet}}, \bibinfo {author} {\bibfnamefont {E.}~\bibnamefont
  {Mascarenhas}}, \bibinfo {author} {\bibfnamefont {D.}~\bibnamefont
  {Pekker}},\ and\ \bibinfo {author} {\bibfnamefont {A.~J.}\ \bibnamefont
  {Daley}},\ }\href {https://doi.org/10.1103/PhysRevLett.123.180402} {\bibfield
   {journal} {\bibinfo  {journal} {Phys. Rev. Lett.}\ }\textbf {\bibinfo
  {volume} {123}},\ \bibinfo {pages} {180402} (\bibinfo {year}
  {2019})}\BibitemShut {NoStop}%
\bibitem [{\citenamefont {Datta}(2005)}]{Datta2005}%
  \BibitemOpen
  \bibfield  {author} {\bibinfo {author} {\bibfnamefont {S.}~\bibnamefont
  {Datta}},\ }\href {https://doi.org/10.1017/CBO9781139164313} {\emph {\bibinfo
  {title} {{Quantum Transport: Atom to Transistor}}}}\ (\bibinfo  {publisher}
  {Cambridge University Press},\ \bibinfo {address} {Cambridge},\ \bibinfo
  {year} {2005})\BibitemShut {NoStop}%
\bibitem [{\citenamefont {Toyozawa}(1981)}]{Toyozawa1981}%
  \BibitemOpen
  \bibfield  {author} {\bibinfo {author} {\bibfnamefont {Y.}~\bibnamefont
  {Toyozawa}},\ }\href {https://doi.org/10.1143/JPSJ.50.1861} {\bibfield
  {journal} {\bibinfo  {journal} {J. Phys. Soc. Japan}\ }\textbf {\bibinfo
  {volume} {50}},\ \bibinfo {pages} {1861} (\bibinfo {year}
  {1981})}\BibitemShut {NoStop}%
\bibitem [{\citenamefont {Hsu}\ \emph {et~al.}(2010)\citenamefont {Hsu},
  \citenamefont {Tsai},\ and\ \citenamefont {Jin}}]{Hsu2010}%
  \BibitemOpen
  \bibfield  {author} {\bibinfo {author} {\bibfnamefont {L.-Y.}\ \bibnamefont
  {Hsu}}, \bibinfo {author} {\bibfnamefont {T.-W.}\ \bibnamefont {Tsai}},\ and\
  \bibinfo {author} {\bibfnamefont {B.-Y.}\ \bibnamefont {Jin}},\ }\href
  {https://doi.org/10.1063/1.3499746} {\bibfield  {journal} {\bibinfo
  {journal} {J. Chem. Phys.}\ }\textbf {\bibinfo {volume} {133}},\ \bibinfo
  {pages} {144705} (\bibinfo {year} {2010})}\BibitemShut {NoStop}%
\end{thebibliography}%

\end{document}